\documentclass[sigconf, nonacm]{acmart}
\usepackage{syntax}
\usepackage{xcolor,listings}
\usepackage{fancyvrb}
\usepackage{enumitem}
\usepackage{alltt}
\usepackage[skip=0pt]{caption}
\usepackage[aboveskip=-2pt,belowskip=0pt]{subcaption}
\usepackage[linesnumbered,ruled,noend]{algorithm2e}
\usepackage{multirow}
\setlist[enumerate]{topsep=0pt,itemsep=-1ex,partopsep=1ex,parsep=1ex}

\colorlet{RED}{red}
\colorlet{BLUE}{blue}

\newcommand\vldbdoi{XX.XX/XXX.XX}

\newcommand\vldbavailabilityurl{http://vldb.org/pvldb/format_vol14.html}
\begin{document}

\title{AWESOME: Empowering Scalable Data Science on \\Social Media Data with an Optimized Tri-Store Data System}


 \author{Xiuwen Zheng, Subhasis Dasgupta, Arun Kumar, Amarnath Gupta} 
    \affiliation{ 
    \institution{University of California, San Diego}
    }
    \email{xiz675@eng.ucsd.edu, sudasgupta@ucsd.edu, arunkk@eng.ucsd.edu, a1gupta@ucsd.edu}
    



\begin{abstract}
Modern data science applications increasingly use  heterogeneous data sources and analytics. 
This has led to growing interest in polystore systems, especially analytical polystores. 
In this work, we focus on emerging multi-data model analytics workloads over social media data that fluidly straddle relational, graph, and text analytics. 
Instead of a generic polystore, we build a ``tri-store'' system that is more aware of the underlying data models to better optimize execution to improve scalability and runtime efficiency.
We name our system AWESOME (Analytics WorkbEnch for SOcial MEdia). It features a powerful domain-specific language named ADIL. ADIL builds on top of underlying query engines (e.g., SQL and Cypher) and features native data types for succinctly specifying cross-engine queries and NLP operations, as well as automatic in-memory and query optimizations. 
Using real-world tri-model analytical workloads and datasets, we empirically demonstrate the functionalities of AWESOME for scalable data science over social media data and evaluate its efficiency.
\end{abstract}

\maketitle


\section{Introduction}\label{sec:intro}

The rise of data science over large-scale social media data has been transforming several fields, including many social sciences, digital humanities, and cybersecurity~\cite{wolff2021information, moustaka2018systematic, lu2018internet}. For instance, in our own ongoing multi-year collaboration with political scientists at UC San Diego, unified analyses of large volumes of data from Twitter, microblogs, and news corpora are enabling a deeper understanding of pressing sociopolitical phenomena observed in social media such as conspiratorial disinformation spreading during elections and changes in criminal justice ~\cite{wu2019constructing,zheng2019social}. 

A defining characteristic of such emerging data science workloads is that they are \textit{multi-model}, more specifically \textit{tri-model}, spanning the canonical logical data models of relational, graph, and text data. This is largely a consequence of the end-users (e.g., political scientists or security policy experts) naturally thinking at the level of abstraction offered by all three of tables, graphs, and natural language text. 
Furthermore, apart from the content of the social media data (viz., tweets or other microblogs such as Sina Weibo), large text corpora in the form of news articles, court documents, public records, etc., and relational corpora in the form of entity dictionaries, census data, geographic data, etc., are also commonly used in such analytics tasks.
Naturally, it is increasingly common at the software level to handle such workloads using a mix of graph DBMSs (e.g., Neo4j~\cite{neo4j}), relational DBMSs (e.g., PostgreSQL~\cite{postgres}), and full-text search systems (e.g., Solr~\cite{solr}).

\begin{figure}[t]
\centering
     \includegraphics[width=0.48\textwidth]{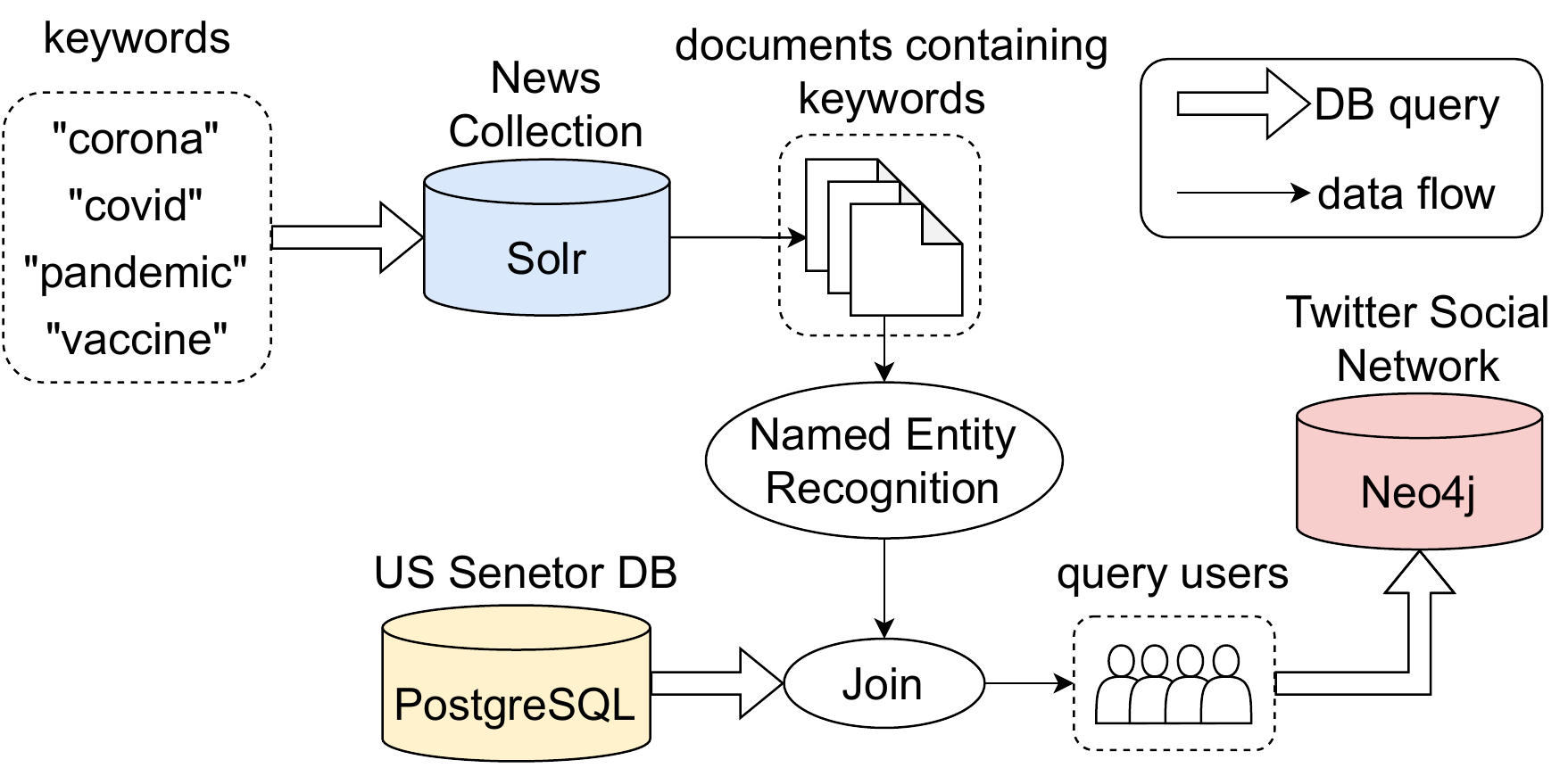}
     \vspace{2mm}
    \caption{Illustration of \textit{PoliSci} workload.}
    \vspace{2mm}
     \label{fig:poliscidiagram}
\end{figure}

\paragraph*{\textbf{Example: PoliSci Workload.}}
Figure~\ref{fig:poliscidiagram} illustrates a simplified political science workload on tweet data from the last few years. 
Given a set of keywords about COVID-19, recent news articles containing any of them are found out through text queries against a Solr document database. 
Then, a named entity recognition (NER) algorithm is invoked on the collected documents to retrieve named entities (e.g., ``President Trump'').
The returned entity list is then joined with a table with Twitter handles of US Senators, which is stored in a PostgreSQL relational database, to obtain the Twitter users for named entities who are Senators.
Finally, the Twitter social network, stored in a Neo4j graph, is queried to retrieve all the users who mentioned any of these Twitter users and all tweets that contain any of these senators’ names.

\vspace{2mm}
While libraries in Python/R suffice for small datasets, handling such data science workloads at scale is a key database systems research challenge. Naturally, there is a growing body of work on \textit{polystore systems}, e.g., \cite{duggan2015bigdawg2,khan2019one,lu2019multi, shrestha2020survey, guo2020multi}.
In a polystore, a query processing \textit{middleware} is built to access multiple underlying data stores, giving users the illusion of a single engine. However, as the complexity of social media analytics workloads grow, it is increasingly crucial to support not just cross-store queries for retrieval or simple analysis but also \textit{complex analytical operations}. Such operations could involve the DBMS invoking specific external libraries, e.g., an NLP library for NER in our PoliSci example.
Other examples include machine learning-based classification tasks on relational tuples and graph analytics tasks such as centrality computation.

\begin{figure*}[ht]
    \centering
    \includegraphics[width=0.85\textwidth]{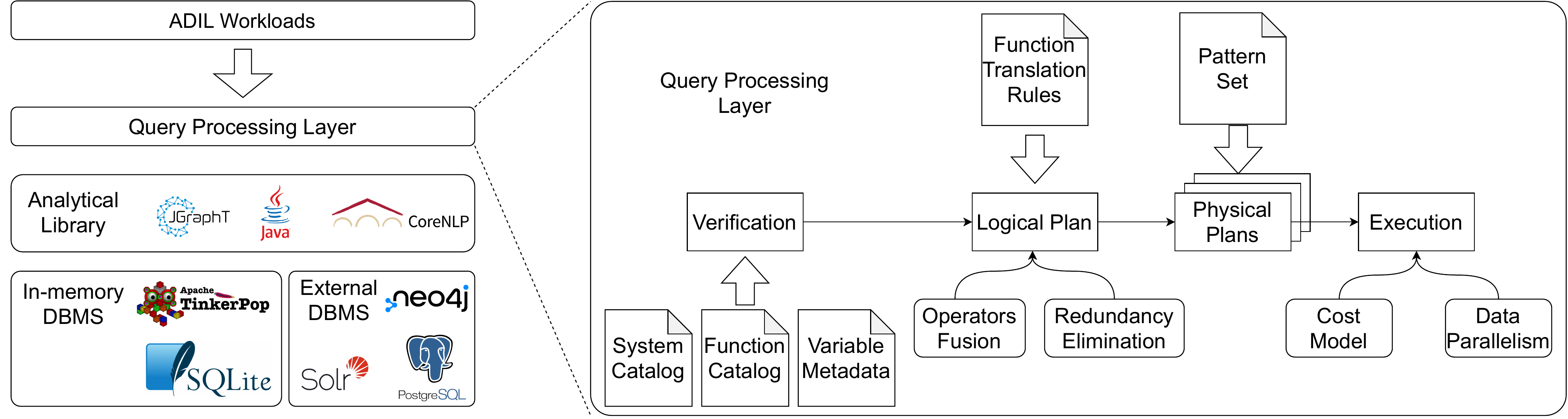}
    \vspace{2mm}
    \caption{Illustration of the AWESOME system architecture.}
    \vspace{2mm}
    \label{fig:sysoverview}
\end{figure*}

Unfortunately, we find that most prior polystores do not support such rich analytics across these three data models. Furthermore, much prior art also aim to be highly general in supporting many stores, which often significantly restricts their ability to look into the specifics of the cross-engine dataflows to optimize their runtime performance at scale. In this work, we mitigate these issues by proposing a ``tri-store'' system focused specifically on relational, graph, and text analytics and elevating cross-engine dataflow optimization to the middleware. We call our system AWESOME: Analytics WorkbEnch for SOcial MEdia data. Table~\ref{tab:relatedwork} summarizes the differences between AWESOME and key prior polystores (elaborated further in Section 8). We now explain the key desiderata and design decisions for our system.

\vspace{-4mm}
\subsection{System Desiderata and Design Decisions}
\begin{itemize}[leftmargin=*]
\item \textbf{Tri-Model Dataflow Language.} AWESOME must offer a \textit{unified} high-level language to enable users to succinctly express tri-store analytics spanning relations, graph, and text. It should support ``unistore-native'' queries (e.g., SQL over RDBMS), basic control flows such as iteration and conditionals, and basic data types such as \texttt{List} and \texttt{String} to handle intermediates and pass arguments to unistore-native queries.

\item \textbf{Tri-Store Middleware.} AWESOME must work \textit{transparently} with underlying uni-stores to handle relational, graph, and text data. We use PostgreSQL, SQLite, Neo4j, Tinkerpop, and Solr as exemplars. Intermediate data must be handled automatically without manual exports/imports.
In our \textit{PoliSci} example, the named entities are joined with a PostgreSQL table and the join result must be passed to a Cypher query in Neo4j.

\item \textbf{Complex Analytical Functions.} AWESOME must provide a set of popular complex analytical operators to support the tri-store analytics, including NLP algorithms on text and graph analytics algorithms. Extensibility via user-defined functions is desired.
In our \textit{PoliSci} example, the Solr query result is sent to an \texttt{NER} operator implemented by a popular NLP library.

\item \textbf{Execution Optimizations.} AWESOME must support \textit{transparent} physical query optimizations such as splitting computations across underlying unistores carefully and automatically caching (intermediate) data in memory when possible to reduce runtime.
In our \textit{PoliSci} example, \texttt{NER} returns a table that is much larger than the senators table in PostgreSQL. So, loading the senators table into memory to process the join in memory can be faster than pushing down the NER result to PostgreSQL.

\item \textbf{Strict Validation Mechanism.}
Some analytical operations such as \texttt{PageRank} are time-consuming and run-time errors/exceptions (e.g., type mismatch) lead to high overheads. So, a rigorous semantics check mechanism at compile time is needed to avoid runtime errors as much as possible.
\end{itemize}

\vspace{-1.5ex}
\subsection{Technical Contributions}
In summary, this paper makes the following technical contributions.

\begin{itemize}[leftmargin=*]

    \item We present a formal description of ADIL, a dataflow language straddling relations, graphs, and text data models with rigorous semantics that enables the expression of complex workloads. An early version of ADIL was briefly introduced but not formalized in~\cite{gupta:awesome:2016}.
    
    \item We present the architecture of AWESOME, the first optimized tri-store system to enable rich data science over social media data at scale, spanning relations, graphs, and text data. 
    Figure~\ref{fig:sysoverview} illustrates our system architecture. AWESOME automates handling of intermediate data to enable seamless user experience.
    
    \item We formalize the logical and physical levels of query planning in AWESOME. We present a suite of transparent optimizations to reduce runtime, including reducing data movement, placing computations in memory, careful apportioning of resources, and a cost model-based selection of execution plans.

    \item We present an extensive set of experiments using real-world datasets to demonstrate AWESOME's support for rich social media analytics, while also enabling higher scalability and runtime efficiency than baseline approaches based only on an RDBMS or in-memory Python.
\end{itemize}

\vspace{-1.5ex}
\begin{small}
\begin{table*}[ht!]
\centering
\caption{Features of existing polystore systems and AWESOME.}\label{tab:relatedwork}
\begin{tabular}{|c|ccccc|ccccc|}
\hline
\multicolumn{1}{|l|}{\multirow{2}{*}{}} & \multicolumn{5}{c|}{Language}& \multicolumn{5}{c|}{System Design} \\ \cline{2-11} 
\multicolumn{1}{|l|}{}& \multicolumn{1}{c|}{\begin{tabular}[c]{@{}c@{}}Native DBMS \\ Query\end{tabular}} & \multicolumn{1}{c|}{\begin{tabular}[c]{@{}c@{}}Function \end{tabular}} & \multicolumn{1}{c|}{\begin{tabular}[c]{@{}c@{}}Graph  \\ Analytics\end{tabular}} & \multicolumn{1}{c|}{\begin{tabular}[c]{@{}c@{}}Text  \\ Analytics\end{tabular}}  & \multicolumn{1}{c|}{\begin{tabular}[c]{@{}c@{}}Control \\ Flow\end{tabular}} & \multicolumn{1}{c|}{\begin{tabular}[c]{@{}c@{}} Native Tri-\\Data Model\end{tabular}} & \multicolumn{1}{c|}{\begin{tabular}[c]{@{}c@{}}RDBMS \\ Support\end{tabular}} & \multicolumn{1}{c|}{\begin{tabular}[c]{@{}c@{}} Graph DBMS \\ Support\end{tabular}} & \multicolumn{1}{c|}{\begin{tabular}[c]{@{}c@{}}Text DBMS \\ Support\end{tabular}} & \begin{tabular}[c]{@{}c@{}}In-memory\\  DBMS Support\end{tabular} \\ \hline
BigDAWG~\cite{she2016bigdawg}  & \multicolumn{1}{c|}{\checkmark} & \multicolumn{1}{c|}{}&\multicolumn{1}{c|}{}&
\multicolumn{1}{c|}{}& & \multicolumn{1}{c|}{} & \multicolumn{1}{c|}{\checkmark} & \multicolumn{1}{c|}{} & \multicolumn{1}{c|}{\checkmark}& \\ \hline
Rheemix~\cite{kruse2020rheemix,agrawal2018rheem} & \multicolumn{1}{c|}{}& \multicolumn{1}{c|}{\checkmark}& \multicolumn{1}{c|}{\checkmark}& \multicolumn{1}{c|}{\checkmark}& \checkmark& \multicolumn{1}{c|}{}& \multicolumn{1}{c|}{\checkmark}& \multicolumn{1}{c|}{}  &\multicolumn{1}{c|}{}& \\ \hline
Estocada~\cite{alotaibi2020estocada} & \multicolumn{1}{c|}{\checkmark}& \multicolumn{1}{c|}{}&\multicolumn{1}{c|}{}&\multicolumn{1}{c|}{}& & \multicolumn{1}{c|}{}& \multicolumn{1}{c|}{\checkmark}& \multicolumn{1}{c|}{}&  \multicolumn{1}{c|}{\checkmark} &  \\ \hline
Tatooine~\cite{bonaque2016mixed} & \multicolumn{1}{c|}{\checkmark}& \multicolumn{1}{c|}{}&\multicolumn{1}{c|}{}&\multicolumn{1}{c|}{}& & \multicolumn{1}{c|}{}& \multicolumn{1}{c|}{\checkmark}& \multicolumn{1}{c|}{\checkmark}& \multicolumn{1}{c|}{\checkmark} & \\ \hline
Myria~\cite{wang2017myria}& \multicolumn{1}{c|}{\checkmark}& \multicolumn{1}{c|}{\checkmark}&\multicolumn{1}{c|}{\checkmark}&\multicolumn{1}{c|}{}& \multicolumn{1}{c|}{\checkmark} & \multicolumn{1}{c|}{}& \multicolumn{1}{c|}{\checkmark}& \multicolumn{1}{c|}{}& \multicolumn{1}{c|}{}& \\ \hline
Hybrid~\cite{simmons2017hybrid,podkorytov2019hybrid}& 
\multicolumn{1}{c|}{\checkmark}& \multicolumn{1}{c|}{\checkmark}& \multicolumn{1}{c|}{} & \multicolumn{1}{c|}{}& \multicolumn{1}{c|}{}& \multicolumn{1}{c|}{}& \multicolumn{1}{c|}{\checkmark}&\multicolumn{1}{c|}{}& \multicolumn{1}{c|}{}& \multicolumn{1}{c|}{\checkmark}\\ \hline
\textbf{AWESOME}& \multicolumn{1}{c|}{\checkmark}&\multicolumn{1}{c|}{\checkmark}&\multicolumn{1}{c|}{\checkmark}&\multicolumn{1}{c|}{\checkmark}&\multicolumn{1}{c|}{\checkmark}& \multicolumn{1}{c|}{\checkmark}& \multicolumn{1}{c|}{\checkmark} & \multicolumn{1}{c|}{\checkmark}& \multicolumn{1}{c|}{\checkmark}& \multicolumn{1}{c|}{\checkmark}\\ \hline
\end{tabular}
\end{table*}
\end{small}

\section{ADIL: A Dataflow Language}
\label{sec: language}
\begin{figure}[t]
     \centering
     \includegraphics[width=0.46\textwidth]{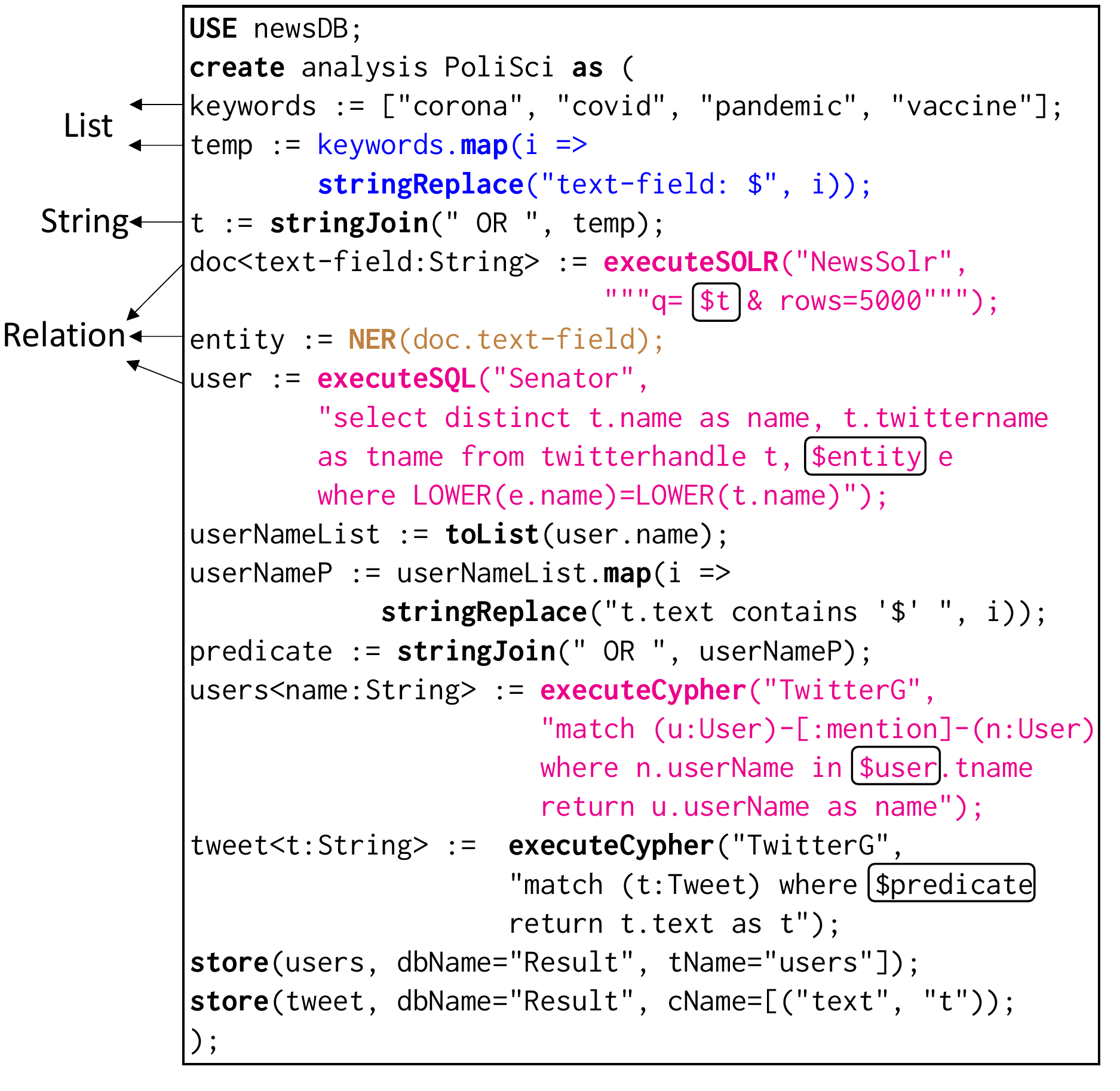}
     \caption{\textit{PoliSci} represented in ADIL.}
     \label{fig:polisciscript}
\end{figure}

ADIL, the surface language for AWESOME, is designed as a dataflow language. The user expresses an analysis workload in ADIL as a sequence of assignment statements where the LHS of the assignment is a variable or multiple variables and the RHS  is an expression.  Figure~\ref{fig:polisciscript} presents the ADIL script for the \textit{PoliSci} workload. 

\subsection{Data Types}
\label{sec:datatype}
ADIL  supports the following data types in native. We annotate the data types for some variables in Figure~\ref{fig:polisciscript}.
\begin{itemize}[leftmargin=*]
    \item \textbf{Primitive types}: \textsf{Integer}, \textsf{Double}, \textsf{String}, and \textsf{Boolean}. 
    \item \textbf{Relation and Record}: A \textsf{Relation}  variable represents a relational table and a \textsf{Record}  variable  is a single tuple of a relation. 
    \item \textbf{Property Graph and Graph Element}:  Users can  construct,  query against, or apply analytical functions (e.g.,  PageRank) on property graphs. A \textsf{GraphElement} variable can be either a node or an edge with labels and properties. 
    \item \textbf{Corpus and Document}: A \textsf{Corpus} is a collection of documents, and each \textsf{document} consists of document content (\textsf{String}), a document identifier (\textsf{Integer}) and tokens (\textsf{List<String>}).
    \item \textbf{Matrix}: We support \textsf{Matrix} data type and commonly-used matrix operators such as dot products on matrix-valued variables. In addition, an AWESOME matrix has  optional \textsf{row map} and \textsf{column map} properties which are semantic mappings from matrix row (resp. column) indices  to values in another  data type. For example, for a document term matrix, the row map is a mapping from row indices to the document ids and the column map is a mapping from column indices to terms (i.e., tokens).
    \item \textbf{Collection}: A \textsf{List} is a collection of indexed elements with homogeneous type; a \textsf{Tuple} is a finite ordered sequence of elements with any type.    \textsf{List} data type is strictly homogeneous: each element should have the same type. However,  there can be  heterogeneous objects in a  \textsf{Tuple} variable. For example, the following tuple $T$ contains a relation, a graph, a list and constant values. 

\begin{Verbatim}[commandchars=\&\~\!]
R := &textbf~executeSQL!(..., ...); //produces relation R
G := &textbf~BuildGraphFromRelation!(...); //produces graph G
T := {R, G, [1, 2, 3], "string", 2 }; 
\end{Verbatim}
\end{itemize}
In this paper, \textsf{Relation}, \textsf{PropertyGraph} and \textsf{Corpus} types are collectively referred to as the ``constituent data models'' because they correspond to the data models of underlying stores. 


\subsection{ADIL Workload Structure} 
\label{subsec:adil-syntax}
 An ADIL script starts by declaring a polystore instance registered in  AWESOME system  catalog:
\begin{Verbatim}[commandchars=+\[\]]
+textbf[USE] newsDB;
+textbf[create] analysis NewsAnalysis +textbf[as] {/*main code block*/}
\end{Verbatim}
AWESOME system  catalog is a file that maintains the metadata for each user-defined polystore instance including  the alias, connection detail, and schema of data stores in this instance. For underlying data store which admits a schema (e.g., PostgreSQL, Solr), a copy of the schema is maintained in the catalog. For stores that do not admit a schema (e.g., Neo4j), a set of schema-like information (e.g., node/edge labels/properties) is maintained. In the above example,  the  metadata of polystore instance \textit{newsDB} will be retrieved from the system catalog 
which contains the information of all DBMSs used in the workload named \textit{NewsAnalysis}. 

The main code block contains a sequence of assignment statements (Section \ref{subsubsec:assignment}) and store statements (Section \ref{sec:store}). 
\subsection{Assignment Statement}\label{subsubsec:assignment}
An ADIL assignment statement evaluates an RHS expression and assigns the result to one or more LHS variables. The grammar for assignment statement is shown as follows.

\begin{grammar}
<assignment-statement> ::= <var1> `, ' <var2> `, ' $\cdots$  `:=' <assign>

\vspace{-8pt}
<assign> ::= <basic-expr> | <ho-expr>
\end{grammar}
\vspace{-3pt}
The RHS expression (<assign>)  can be ``basic'' or ``higher-order'' explained  by the following grammar fragments,  
\begin{grammar} 
\label{grammar:expr}
<basic-expr> ::= <const>|<query>|<func> 

\vspace{-8pt}
<ho-expr> ::= <assign> `>' | `==' | `<' <assign>
\alt <var>`.map(' <lVar> `->' <assign> `)'
\alt <var>`.reduce((' <lVar1> `,' <lVar2> `) ->' <assign> `)'
\alt <var> ` where ' <assign> 
\end{grammar}


\subsubsection{Basic Expression}
<basic-expr> includes three types:

\noindent \textbf{Constant Expression (<const>):} A constant expression evaluates to a constant of any allowed data type. The expression can itself be a constant, e.g., \texttt{['x', 'y', 'z']},  or a prior constant variable, or an element of a prior collection  variable, e.g., \texttt{a[1]}.


\noindent \textbf{Query Expression (<query>):} A query expression executes a query against a data store  or against an AWESOME variable with a constituent data model. It uses standard query languages: SQL-93 for relational queries, OpenCypher~\cite{francis2018cypher} for property graph queries, and Lucene~\cite{mccandless2010lucene} for retrieval from text indices. In Figure~\ref{fig:polisciscript}, three query expressions are marked in pink and they use \texttt{executeSOLR}, \texttt{executeSQL} and \texttt{executeCypher} keywords respectively. The first argument of a query expression is the alias of target DBMS registered in the polystore instance.
If the query is  against a variable  created in prior statements, the first argument is left empty. 
The second argument is a standard Lucene/SQL/Cypher query with the exception of the \texttt{\$} followed by a variable name (highlighted by the rounded rectangles in the figure). ADIL uses \texttt{\$} as a prefix of the variable  passed as a parameter to a  query. 

\noindent \textbf{Function Expression (<func>):} 
AWESOME supports a rich native library for common data analytical tasks. The   expression   includes  function name with required positional parameters followed by optional  and  named parameters. A parameter can be a constant  or a variable. The expression can return a single or multiple variables. The \texttt{NER} function expression marked as brown in Figure~\ref{fig:polisciscript} takes a relation variable as parameter and returns a relation variable. 


\subsubsection{Higher-Order Expression.}\label{sec:highorder}
A higher-order  expression is recursively defined where another  expression  serves as its sub-expression.
The following  snippet from \textit{NewsAnalysis} workload shows an example statement where the RHS is a nested higher-order  expression:
\begin{Verbatim} [commandchars=+\[\]]
wtmPerTopic := topicID.+textbf[map](i =>
        WTM +textbf[where] getValue(_:Row, i) +textbf[>] 0.00);
\end{Verbatim}
\texttt{topicID} is a list of Integers and \texttt{WTM} is word-topic matrix where each row presents a word's weights on all topics. For each topic, it produces a word-topic matrix consisting of words with weights higher than 0 on this topic.
This snippet contains map, filter and binary comparison which are explained as follows. 



\noindent \textbf{Map Expression:} 
A map expression operates on a collection variable, evaluates a sub-expression for each element in the collection, and returns a new collection object. 
The sub-expression can be a constant, a query, a function or another higher-order expression. In this snippet, it takes a list of integers (\textit{topicID}) as input, and for each, applies another higher-order expression (a filter expression) on the \texttt{WTM} matrix to generate a matrix. Thus the returned variable (\textit{wtmPerTopic}) is a list of matrices. 

\noindent \textbf{Filter Expression:} The filter expression is indicated by the \texttt{where} clause -- its sub-expression is a predicate; it returns a new collection with values which satisfy the given predicate.  Since a  matrix can be  iterated by rows or  by columns, users need to specify the iteration mode: the underscore sign ($\_$) is used to represent  every  single element in the matrix, and the colon ($:$) followed by the type specify the element type. In the example snippet, it applies a binary comparison predicate on each row of the matrix and returns a new matrix consists of the rows satisfying the predicate.  

\noindent \textbf{Binary Comparison and Logical Operations:} A binary comparison accepts two expressions and  compares their values to return a Boolean value. 
In the example above, \begin{verbatim}
getValue(_:Row, i) > 0.00
\end{verbatim}
checks whether the $i$-th element of a row vector is positive.  More generally, ADIL supports any binary logical operators such as AND, OR and NOT over predicates. 

\noindent \textbf{Reduce Expression:} A \texttt{reduce} operation  aggregates results from a collection by passing a commutative and associative binary operator as its sub-expression. 
For example, the following snippet
\begin{Verbatim} [commandchars=+\[\]]
R := relations.+textbf[reduce]((r1,r2) => join(r1,r2,on="id"))
\end{Verbatim}
takes a list of relations as input and then joins each two tables and returns a new table at the end.




\subsection{Store Statement}
\label{sec:store}
A store statement specifies the variables to be stored to a persistent storage, which can be an underlying DBMS registered in the system catalog or the AWESOME file system; it also includes the instructions for how to store the variable. In Figure~\ref{fig:polisciscript}, the last two lines store \texttt{users} and \texttt{tweet} variables to   relational DBMS,   and specifies the DBMS alias (\texttt{dbName} parameter), table name (\texttt{tName} optional parameter) and mapping between  the targeted  column names to the relational variables' column names (\texttt{cName} optional parameter). 



\subsection{Some Properties of ADIL} 
\label{sec:expressiveness}
A full discussion of the formal properties of  ADIL  is beyond the scope of this paper. Here we provide a few  properties that will be useful in validating and developing logical plans from ADIL scripts.
\begin{enumerate}[leftmargin=*]
    \item ADIL does not have a \textsf{for loop} or a \textsf{while} operation. Instead, it uses the \textsf{map} operation to iterate over a collection and apply function over each element, the \textsf{filter} operation to select out elements from a collection that satisfies predicates, the \textsf{reduce} operation to compute an aggregate function on  a collection. In ADIL, the collection must be \textit{completely constructed} before the \textsf{map} (resp. filter or reduce) operation can be performed. Therefore, these operations are guaranteed to terminate.
    \item ADIL is strongly typed.
    \item In an assignment where the RHS expression is a query in a schemaless language like OpenCypher, the user must specify a schema for the LHS variable in the current system.
    \item The data type and some metadata information of any LHS variable can be uniquely and correctly determined by analyzing the RHS expression (see Section \ref{sec: check}).
\end{enumerate}


\vspace{-1ex}
\section{System Architecture}\label{sec:architecture}
The system architecture of  AWESOME polystore is shown in Figure \ref{fig:sysoverview}. We summarize some primary architectural components:

\noindent \textbf{(a) Data Stores.} It supports on-disk DBMSs (Neo4j, Postgres and Solr)  and in-memory DBMSs (Tinkerpop and SQLite).

\noindent \textbf{(b) Analytical Capability.} It incorporates existing analytical libraries for NLP and graph algorithms   such as CoreNLP and JGraphT, and AWESOME native functions written in Java. 

\noindent \textbf{(c) Query Processing.} The ``query'' in AWESOME is essentially a multi-statement analysis plan consisting of data retrieval, transformation, storage, function execution and management of intermediate results.  The query processor verifies an ADIL script, creates the optimal logical plan, generates a set of physical plans and then applies cost model and data parallelism mechanism to create an optimal execution plan.

\vspace{-1.5ex}
\section{Validating ADIL Scripts}\label{sec: check}
An ADIL script is  complex with many expensive operations. 
To reduce the  avoidable run-time errors, AWESOME implements a strict compile-time semantics check mechanism which consists of two parts:  1) \textit{Validation}  refers to  determining the semantic correctness of each expression, 2) \textit{Inference} refers to inferring the data type and metadata of the variables generated from each expression.
 

\vspace{-1ex}
\subsection{Validation}
For different RHS expressions, the validation process is different.


\noindent\textbf{System catalog based validation.}
To validate a  query expression (<query>) against an external DBMS, the system catalog is used to get the schema information. For example, for a SQL query,  it checks if the relations and columns  in the query exist in the database.

\noindent\textbf{Function catalog based validation.} For a function expression, AWESOME checks if  the data types of the input variables/constant values are consistent with the parameters information registered in the function catalog. 

\noindent\textbf{Validation with Variable Metadata.}
Variable metadata map stores the key properties of  variables and is built through inference process. It is looked up for every expression containing a variable.
For a query expression, if it queries on  relation-valued variables, their schema is found from the variable metadata map instead of the system catalog. For a function expression, if an input parameter is a variable, its data type will  be  found in the map. 

\noindent\textbf{Validation Example.} Usually, more than one types of validation need to be used. 
We use the example snippet from Sec.~\ref{sec:highorder}  to show how to validate a nested higher-order expression. To validate the Map expression,  it gets the data type and element type of  \texttt{topicID} from the variable metadata map, then it checks 
if the variable has a collection  type and the element type will be used to validate the sub-expression which is a Filter expression; to validate the Filter expression, similar to the Map expression, the data type of   \texttt{WTM} is checked and the element type is used to validate the sub-expression which is a binary comparison expression, besides, it also checks if the return type of the sub-expression is a  Boolean; to validate the binary comparison expression, it validates if the two operands have the same data type and the data type  is comparable: in this example, the type of the left operand can be inferred based on the function catalog; At the end, it checks the \texttt{getValue} function using the element   type information of \texttt{WTM} and \texttt{topicID}.


\vspace{-1ex}
\subsection{Inference}\label{sec:infer}
Inference refers to  building variable metadata map. Table~2  in the technical report shows the variable types,  and their corresponding metadata properties. 
For each statement in an  analysis plan, the RHS expression is validated and then the type and metadata of the LHS variables are inferred as much as possible and be stored in the  map.

\begin{table}[t]
\caption{Metadata for different data types.}
\label{tab:meta}
\footnotesize
    \begin{tabular}{ll}
    \toprule
    \textbf{Data Type}      & \textbf{Metadata}  \\ \midrule
    \texttt{Relation}       & Schema $S = \{\textit{ColName}:\textit{Type}\}$ \\ \hline
    \texttt{Property Graph} & \begin{tabular}[l]{@{}l@{}}Node labels set $NL$\\ Node properties map $NP =\{\textit{PropName}:\textit{Type}\}$
    \\ Edge labels set $EL$ \\ Edge properties map $EP=\{\textit{PropName}:\textit{Type}\}$ \end{tabular} \\ \hline
    \texttt{List}           & Element type, Element metadata, Size \\ \hline
    \texttt{Tuple}          & Each element's type and metadata, Size \\ \hline
    \texttt{Map} & \begin{tabular}[l]{@{}l@{}}
    Key type, Key metadata, Value type,\\ Value metadata, Size
    \end{tabular} \\ \hline
    \texttt{Matrix}         & Row (and column) count, Element type \\ \bottomrule
    \end{tabular}
\end{table}
The  inference mechanisms are different for different   expressions. For a SQL query expression, the schema of the returned relation is inferred by parsing the SELECT clause and looking up the system catalog/variable metadata map to get column types. 
For function expressions, the return types reside in the function catalog. 
For example, the following expression invokes function \texttt{lda}. 
By querying the function catalog, we know that it outputs two matrix variables.
Thus the data types of \texttt{DTM} and \texttt{WTM} will be set as Matrix. 
\begin{Verbatim}[commandchars=+\[\]]
DTM, WTM := +textbf[lda](processedNews,
              docid=true, topic=numTopic);
\end{Verbatim}
For nested expressions, the inference is handled from the innermost expression to the  outermost expression. Taking  the  snippet shown in Sec.~\ref{sec:highorder} as an example,   the LHS variable's type and  metadata is inferred by the following steps: 1) the Filter expression returns a matrix since \texttt{WTM} is a matrix; and 2) Map expression will return a list of matrices since its sub-expression returns a matrix.

\vspace{-1.5ex}
\section{Logical Plan}
\label{sec:logicalplan}
After validating the correctness of an ADIL script, a logical plan will be constructed. A logical plan is a DAG where each node represents a logical operator. 
\begin{figure}[t]
    \centering
    \includegraphics[width=0.33\textwidth]{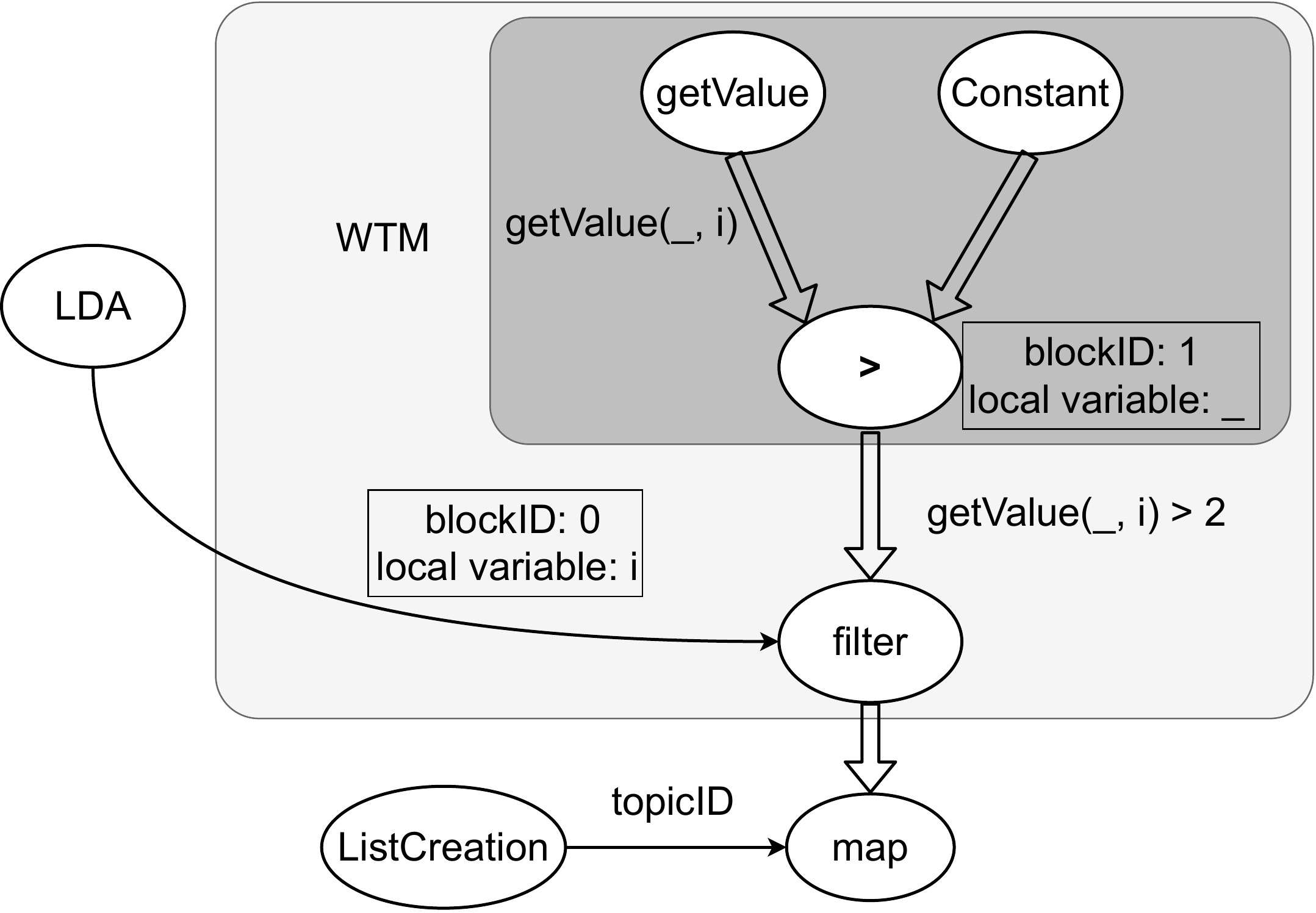}
    \caption{Logical plan of higher-order expressions.}
    \label{fig:logical-plan-map}
\end{figure}
\vspace{-1.5ex}
\subsection{Logical Plan Creation}
The initial logical plan is directly translated from the parsing results. 
In most cases, each expression in the ADIL script corresponds to a single  logical operator. 
For example, an \texttt{ExecuteSQL} query expression will be mapped to an \textit{ExecuteSQL} logical operator. 
However, for specific functions expressions or higher-order expressions, extra processing steps are required to generate the initial logical plan. 

\noindent\textbf{Input-based Function Translation.}  For analytical functions, the corresponding logical operators can vary based on different function inputs. 
Table~\ref{tab:logic-operator} presents some functions. For example, the function \texttt{LDA} can take either a \texttt{Matrix} variable or a \texttt{Corpus} variable as input, which corresponds to logical operators \textit{LDAOnTextMatrix} and \textit{LDAOnCorpus} respectively.

\noindent\textbf{Higher-order Expressions to Sub-plans.}
For higher-order expressions (e.g., map expressions), a single expression will be translated to a sub-plan since it contains sub-expressions. For the nested higher order expression shown in Section~\ref{sec:highorder}, the logical plan is given in Figure~\ref{fig:logical-plan-map}. 
In Figure~\ref{fig:logical-plan-map}, there are two types of edges denoting data flow and sub-operator consumption, respectively. The \textit{Filter} operator takes data from  \textit{LDA} and applies a binary comparison sub-operator. The \textit{Map} operator takes data from \textit{ListCreation} and applies the \textit{Filter} sub-operator on each element of the data. 
Both \textit{Map} and \textit{Filter} create a local variable to denote each element of a collection, and the scope of such local variable is the  sub-expression block of that higher-order expression. 

\begin{table}[ht] 
\centering
\footnotesize
\caption{Some of ADIL functions and logical operators.}\label{tab:logic-operator}
\begin{tabular}{lll}
\toprule
\multicolumn{1}{l}{\textbf{ADIL Function}} & \textbf{Input Parameter}                                                                                               & \textbf{Logical Operator(s)}         \\ \midrule
Preprocess         & \begin{tabular}[c]{@{}l@{}}Column\\ List\textless{}String\textgreater\\ Corpus\end{tabular}        &\begin{tabular}[c]{@{}l@{}}CreateCorpusFromColumn\\ CreateCorpusFromList\\ NLPAnnotator(tokenize)\\ NLPAnnotator(ssplit)\\ NLPAnnotator(pos)\\ NLPAnnotator(lemma)\\ FilterStopWords\end{tabular}                               \\ \hline
NER                                 & \begin{tabular}[c]{@{}l@{}}Column\\ List\textless{}String\textgreater\\ Corpus\\ AnnotatedCorpus\end{tabular} & \begin{tabular}[c]{@{}l@{}}CreateCorpusFromColumn\\ CreateCorpusFromList\\ NLPAnnotator(tokenize)\\ NLPAnnotator(ssplit)\\ NLPAnnotator(pos)\\ NLPAnnotator(lemma)\\ NLPAnnotator(ner)\end{tabular}            \\ \hline
\multirow{2}{*}{TopicModel}         & TextMatrix                                                                                                    & TopicModelOnTextMatrix                                                     \\ 
                                    & Corpus                                                                                                        & TopicModelOnCorpus                                 \\ \hline
\multirow{2}{*}{LDA}                & Matrix                                                                                                    & LDAOnTextMatrix                                                                                                                                                                                                                    \\ 
                                    & Corpus                                                                                                        & LDAOnCorpus                                                                                                                                                                                                                        \\ \hline
\multirow{2}{*}{SVD}                & Matrix                                                                                                    & \multirow{2}{*}{\begin{tabular}[c]{@{}l@{}}CreateTextMatrix\\ SVDOnTextMatrix\end{tabular}}                                                                                                                                        \\ 
                                    & Corpus &                                \\ \hline
\multirow{4}{*}{Sum}                & List                                                                                                          & \multirow{4}{*}{\begin{tabular}[c]{@{}l@{}}Column2List\\ GetVector\\ SumList\\ SumVector\end{tabular}}                                                                                                                             \\
                                    & Column            &                                                                          \\ & Vector  &   \\   & Matrix, Index     & \\\bottomrule
\end{tabular}
\end{table}

\vspace{-1ex}
\subsection{Logical rewriting}
After creating the initial logical plan, a set of rewriting rules will be applied to generate an optimized logical plan.   

\noindent\textbf{Rule 1: Function decomposition.}
Some functions can be decomposed to several logical operators to achieve a deeper level of optimization. For example, for \texttt{NER} function which recognizes named entities in corpus, it will be translated to a series of \texttt{CoreNLPAnnotator} operators with different annotation sub-operators. 

\noindent\textbf{Rule 2: Redundancy elimination.}
The same operators which take the same input data  will only be executed once.  As Table~\ref{tab:logic-operator} shows, some  functions may share common logical operators, and these common operators will be merged.

\begin{figure}[t]
    \centering
    \includegraphics[width=0.45\textwidth]{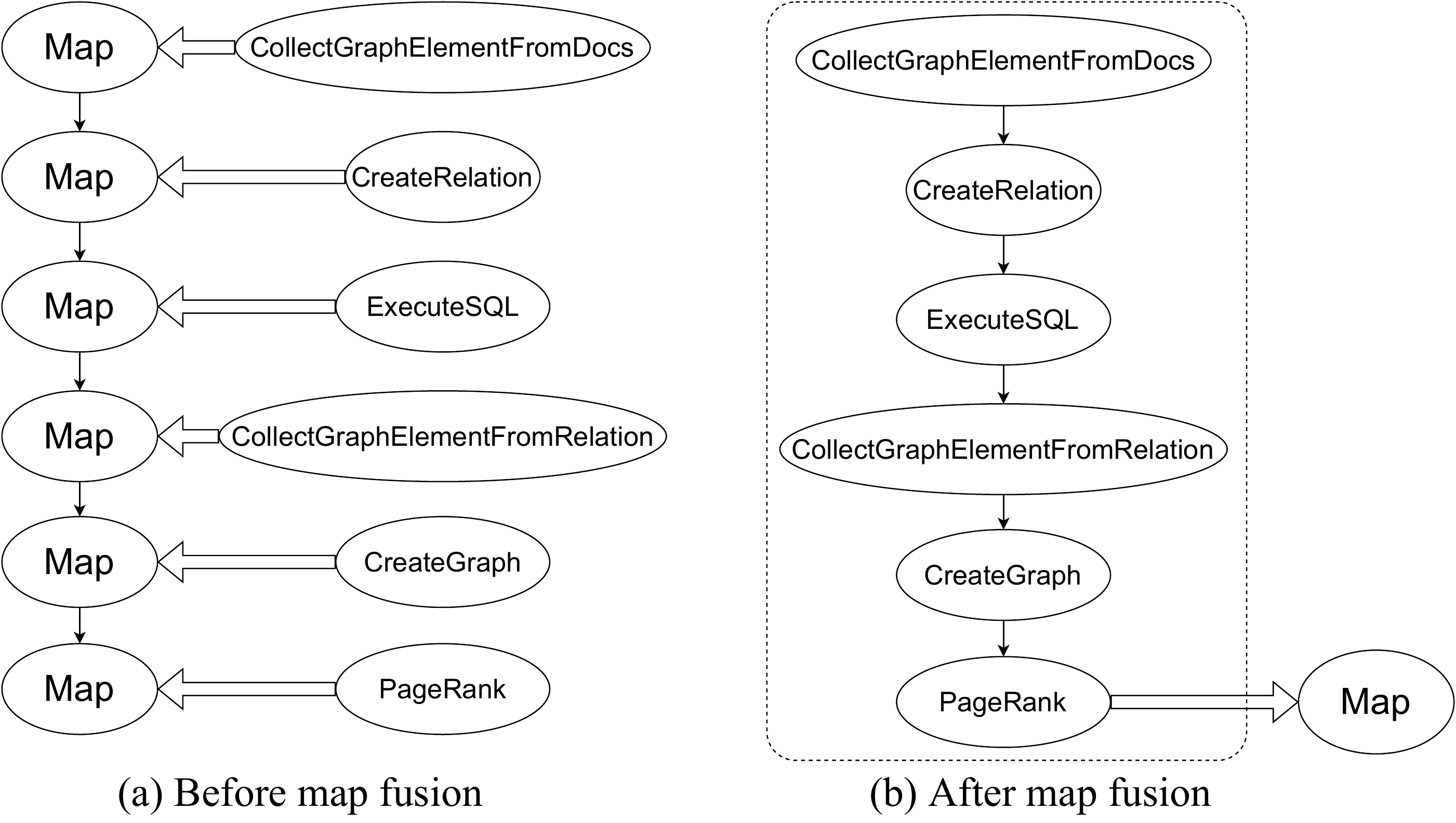}
    \caption{Illustration of map fusion.}
    \label{fig:mapfusion}
\end{figure}

\noindent\textbf{Rule 3: Operators fusion.}
There are two special operators which apply a sub-operator on each single element of a collection variable: \textit{Map} and \textit{NLPAnnotator}. For a series of \textit{Map} or \textit{NLPAnnotator}, they will be fused and the sub-operators of them will be connected, which are termed as Map Fusion and NLP annotation pipeline. Figure~\ref{fig:mapfusion} shows an example that corresponds to a snippet of  workload \textit{NewsAnalysis}, the left plot is the initial logical plan, and the right one applies map fusion. 
This rewriting has two advantages: 1) the intermediate results will not be materialized which saves  memory. 2) it will benefit the candidate physical plans generation which will be discussed in detail in the physical planning section (Section~\ref{sec:physical}). 

\begin{figure}[t]
    \centering
    \includegraphics[width=0.49\textwidth]{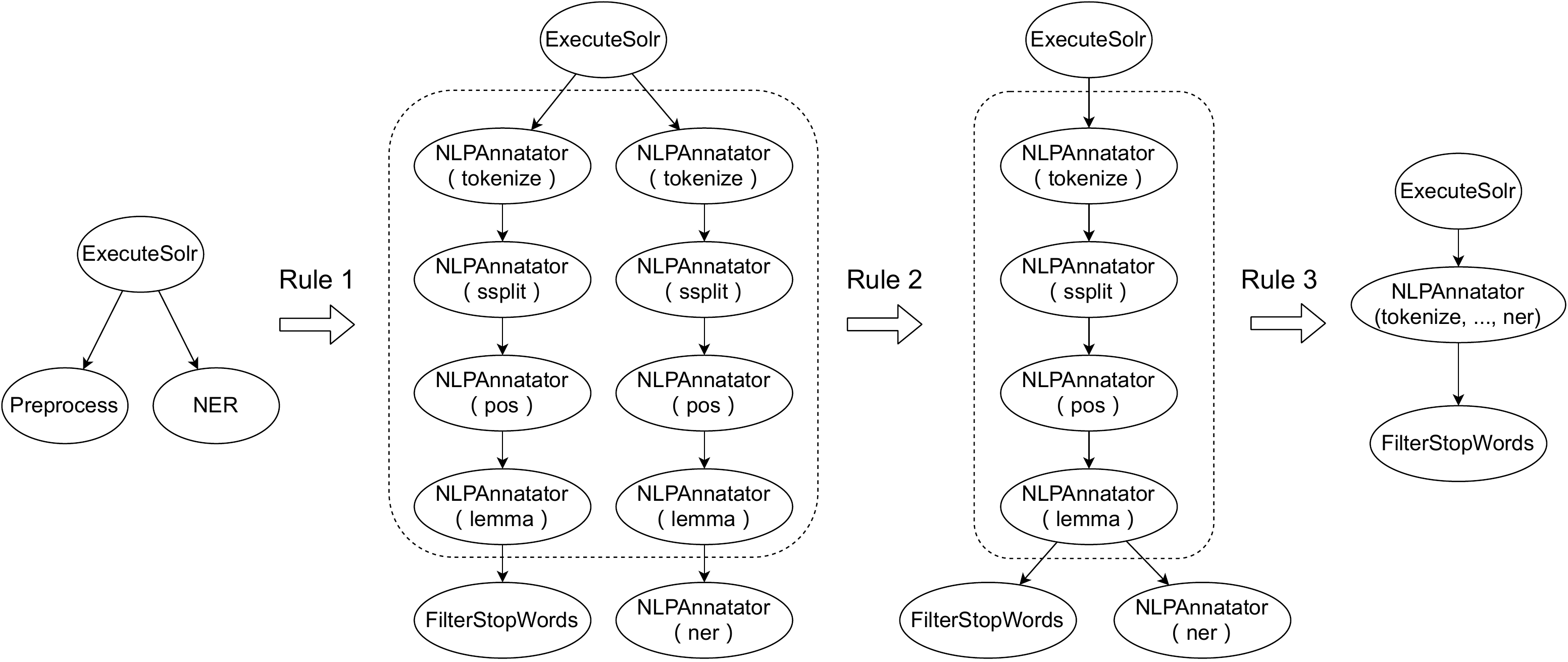}
    \caption{An Illustration of logical plan rewriting rules.}
    \label{fig:logicalrewrite}
\end{figure}

Supposing the following ADIL snippet, 
\begin{Verbatim} [commandchars=+\[\]]
    processedDoc:= +textbf[Preprocess](doc);
    entity:= +textbf[NER](doc);
\end{Verbatim}
Figure~\ref{fig:logicalrewrite} illustrates the aforementioned rewriting rules. 
The first part is the initial logical plan, and the second part shows the plan after applying the function decomposition rule. 
For the third part, functions \textit{Preprocess} and \textit{NER} share a series of common logical operators which are merged based on Rule 2. The final part applies Rule 3 to generate an NLP annotator pipeline, which is a common practice in NLP toolkits such as Standford coreNLP. 

\vspace{-1.5ex}
\section{Physical Plan}\label{sec:physical}
Based on the optimized logical plan, we introduce the physical planning details of AWESOME. 
As shown in Algorithm~\ref{algo:physicalplan}, there are mainly three steps to generate the candidate physical plans, each of which  will be introduced in the following sections.   

\begin{algorithm}[t]
\small
\caption{\emph{Physical Plan Generation}} \label{algo:physicalplan}
\SetKwFunction{CandidatePhsicalPlanGen}{CandidatePhsicalPlanGen}
\SetKwFunction{AddDataParallelism}{AddDataParallelism}
\SetKwFunction{AddBuffering}{AddBuffering}
\SetKwProg{Fn}{Function}{:}{}
\KwIn{A logical plan $G = (V, E)$, a boolean flag $buffer$.}
\KwOut{Candidate physical plans: $candPlans$}
$candPlans \gets$ \CandidatePhsicalPlanGen($G$)\; 
$candPlans \gets$ \AddDataParallelism($candPlans$)\;
\lIf{$buffer$}{$candPlans \gets$ \AddBuffering($candPlans$)}
\end{algorithm}

\subsection{Candidate Physical Plans Generation}
We introduce the pattern based transform algorithm to generate a set of candidate physical plans from a logical plan DAG. To begin with, we provide some definitions  as follows.

\begin{definition}[Pattern Set]
A pattern set $Pat: \{\{OP^l, E^l\} \to \{ \{OP^p, E\} \}\}$ is a  mapping where a key is a logical sub-DAG and a value is a set of physical sub-plans. The pattern set is ordered by the sizes of keys (i.e., the numbers of nodes in the logical sub-DAGs) to make sure that in the subsequent procedures the larger patterns in a logical plan are matched earlier. 
\end{definition}

\begin{definition}[Candidate Physical Plans]
Candidate physical plans consist of a DAG $PG = \{OP^p, E\}$ which contains some virtual nodes, and a map $PM: I \to \{OP^p, E\}$ where a key is a  virtual node id and a value is a set of physical sub-plans.
\end{definition}

\begin{algorithm}[b]
\footnotesize
\caption{\emph{Candidate Physical Plan Generation}} \label{algo:candplan}
\KwIn{An ordered  pattern set $Pat$; An optimal logical plan DAG $G = (V, E)$. 
}
\KwOut{Candidate physical plans: $PG$ and $PM$.}
\SetKwFunction{FindMatchPattern}{FindMatchPattern}
\SetKwFunction{SingleOperatorTransform}{SingleOperatorTransform}
\SetKwFunction{PatternTransform}{PatternTransform}
\SetKwProg{Fn}{Function}{:}{}
$PG \gets G, PM \gets \{\}$\;
\tcc{match patterns from the largest to the  smallest.}
\For{$pat \in Pat$} {
    $pSubs \gets Pat[pat]$\;
    $lSubs \gets$ \FindMatchPattern($PG, pat$) \;
    \For {$sub \in lSubs$} {
        \tcc{for a pattern which only has one physical sub-plan, directly replace the pattern with the DAG.}
        \If {$pSubs.size == 1$}{
            $PG \gets$\SingleOperatorTransform($PG, sub, pSubs$)\;
        }
        \tcc{for a pattern with several candidate physical sub-plans, transform $sub$ to a virtual node and add the node id and physical sub-plans to map $PM$.}
        \Else{
            $PG, PM \gets$ \PatternTransform($PG, PM, sub, pSubs$)\;
        }
    } 
}
\end{algorithm}

\begin{figure}[t]
    \centering
    \includegraphics[width=0.48\textwidth]{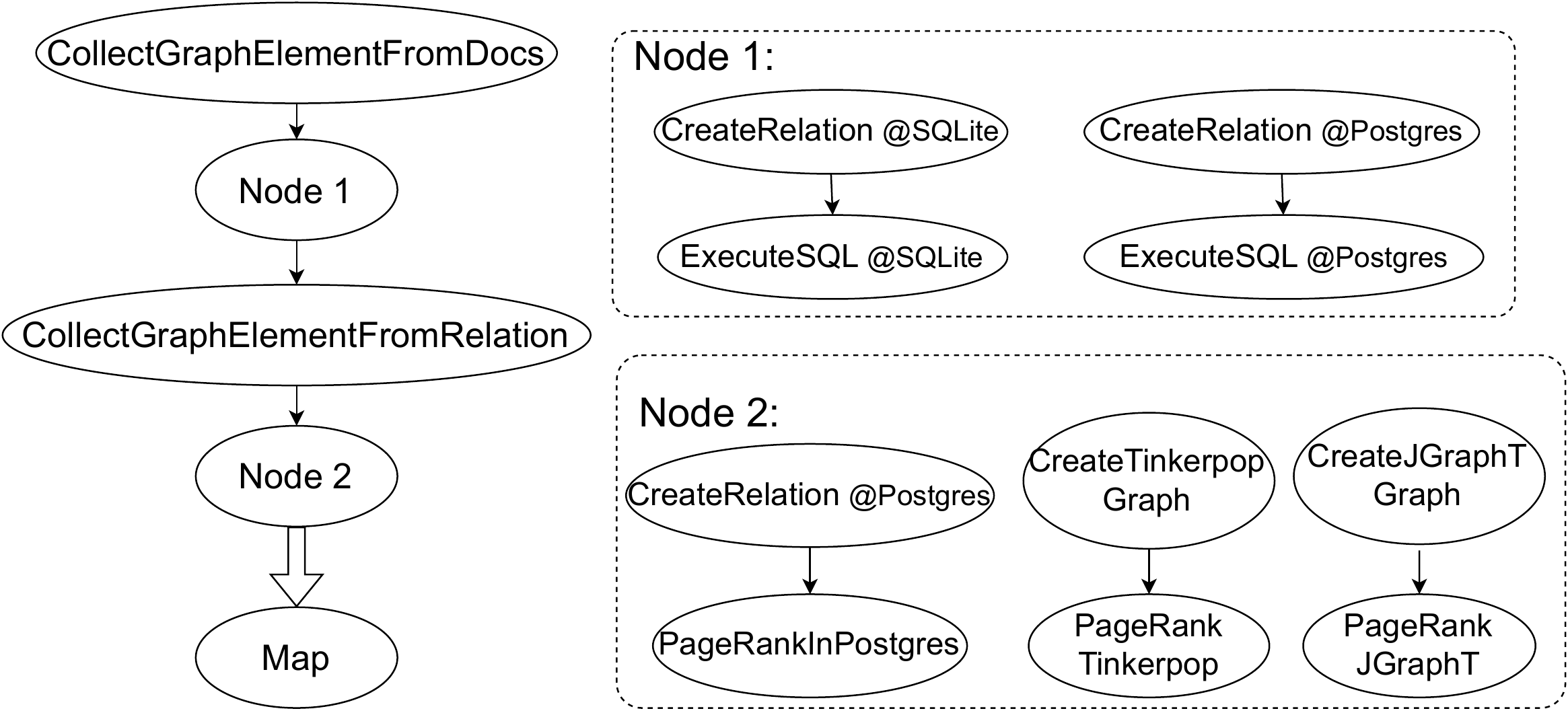}
    \caption{Candidate physical plans for Figure \ref{fig:mapfusion} logical plan.}
    \label{fig:physicalplan}
\end{figure}

We propose Algorithm~\ref{algo:candplan} to generate the candidate physical plans. 
Table~\ref{tab:operator} lists some logical operators and their corresponding physical operators. 
For some logical operators like \textit{LDAOnCorpus} or  some logical  sub-DAGs, each  corresponds to only one candidate physical operator or sub-plan.
In this case, each operator or sub-DAG will be directly replaced by the physical operator or sub-plan (lines 6-7). 
On the other hand, there exist logical operators or  sub-DAGs corresponding to multiple candidate physical operators or sub-plans. For example, in Figure \ref{fig:physicalplan}, a logical sub-DAG \textit{CreateGraph} $\to$ \textit{ExecuteCypher} can be transformed to multiple different physical sub-plans shown in the dashed rectangle marked by Node 1. In this case, the operator or sub-DAG will be replaced by a virtual node, and the virtual node ID with its corresponding physical sub-plans will be stored in the map $PM$ (lines 8-9).


\begin{table*} 
\centering
\caption{Summary of AWESOME logical and physical operators.}\label{tab:operator}
\footnotesize
\begin{tabular}{lllll} 
\toprule
\textbf{Types} & \textbf{Logical Operators} & \textbf{Physical Operators}  & \textbf{DataParallelCap}  & \textbf{BufferingCap} \\ 
\midrule
Query & \begin{tabular}[c]{@{}l@{}}ExecuteCypher \\ExecuteSQL \\ExecuteSolr\end{tabular} &   \begin{tabular}[c]{@{}l@{}} ExecuteCypherInNeo4j~\\ExecuteCypherInTinkerpop~\\ExecuteSQLInPostgres~\\ExecuteSQLInSQLite~\\ExecuteSolr~\end{tabular}  & \begin{tabular}[c]{@{}l@{}}ST \\EX \\ST \\EX \\ ST \end{tabular} & \begin{tabular}[c]{@{}l@{}}SO \\SO \\SO \\SO \\SO \end{tabular} 
\\ 
\midrule
Graph Operations 
& \begin{tabular}[c]{@{}l@{}}BuildWordNeighborGraph \\BuildGraphFromRelation\\PageRank\end{tabular} 
& \begin{tabular}[c]{@{}l@{}}CollectGraphElementsFromDocs ~\\CollectGraphElementsFromRelation ~\\CreateTinkerpopGraph~\\CreateNeo4jGraph ~\\PageRankTinkerpop~\\ \end{tabular}  
&  \begin{tabular}[c]{@{}l@{}}PR \\PR \\ST \\ST\\EX\end{tabular} 
& \begin{tabular}[c]{@{}l@{}} SS \\SS \\SI\\SI\\ SO \end{tabular} \\
\midrule
Text Operations       & \begin{tabular}[c]{@{}l@{}}NLPAnnotator\\LDAOnCorpus\\TopicModel\\\end{tabular} & \begin{tabular}[c]{@{}l@{}}CreatDocumentsFromRecords\\CreatDocumentsFromList\\CreateTextMatrix\\ LDAOnCorpus\\SVD\end{tabular}   
& \begin{tabular}[c]{@{}l@{}}PR\\PR\\PR\\EX\\EX\end{tabular} 
&\begin{tabular}[c]{@{}l@{}}SS\\SS\\SS\\SI\\ B\end{tabular} 
\\ 
\bottomrule
\end{tabular}
\end{table*}

Figure \ref{fig:physicalplan} shows the candidate physical plans for the logical plan in Figure \ref{fig:mapfusion}. 
There are two virtual nodes and each corresponds to several physical sub-plans. This figure also shows the advantage of Map Fusion which connects all the sub-operators of a series of Maps so that the sub-DAG matched with a pattern in the pattern set could have larger size and thus leads to more efficient candidate plans. For example, after map fusion, the  \textit{CreateGraph}$\to$\textit{ExecuteCypher} pattern will be translated to three physical sub-plans. However, without map fusion, the  \texttt{CreateGraph} itself will be translated to three candidate operators: \textit{CreateTinkerpopGraph},  \textit{CreateNeo4jGraph} and \textit{CreateJgrapht},  since the JgraphT library does not support Cypher queries, if \textit{CreateJgrapht} is chosen  during execution,  then it will initiate a set of  typecasting transformations before executing \textit{ExecuteCypher}, which increases the overall cost.

\vspace{-1ex}
\subsection{Partitioned Data Parallelism}
AWESOME exploits data parallelism to take advantage of modern multi-core systems. 
Table~\ref{tab:operator} presents some physical operators with their data parallel capabilities. \textbf{ST} means single-threaded operators which can not be executed in a data parallel fashion,  \textbf{PR} means data parallelizable operators,  and \textbf{EX} means operators provided by external libraries. 
The execution of  \textbf{EX} operators is fully supported by external libraries and can utilize multi-core feature in their native implementation, and thus  they are excluded from the subsequent AWESOME optimizations which are based on data parallelism.

\begin{figure}[t]
    \centering
    \includegraphics[width=0.36\textwidth]{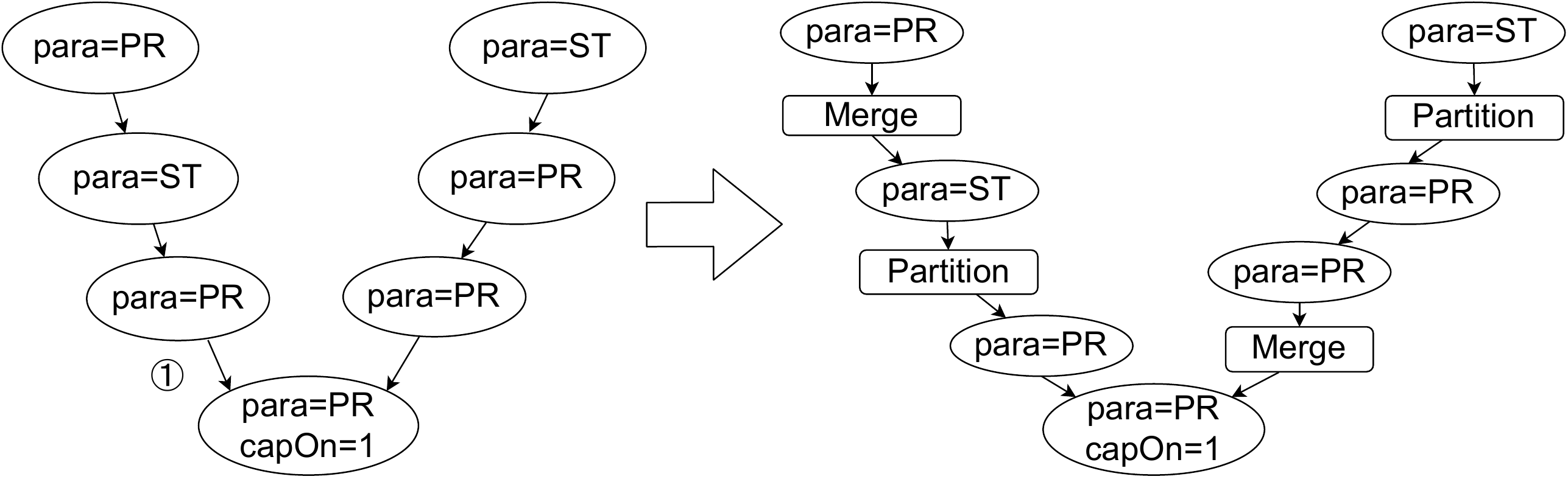}
    \caption{Illustration of data parallel execution.}
    \label{fig:parallelexe}
\end{figure}

For a \textbf{PR} operator with multiple inputs, it is associated with a $capOn$ attribute specifying the input on which it has data parallelism capability. 
For example, the \textit{FilterStopWords} operator takes a corpus and also a list of stop-words as input, and it can be executed in parallel by partitioning the corpus input. In this case, $capOn$ will be set as the ID of the corpus variable. 
Every \textbf{PR} operator will be executed in parallel by partitioning the $capOn$ input data. 
Figure \ref{fig:parallelexe} shows an illustration. 
The left sub-figure shows the original physical plan DAG and the right sub-figure shows the plan DAG after considering data-parallelism.  
When an operator with \textbf{PR} capability gets its input: if its $capOn$ input was not partitioned, a Partition step will be added which generates partitioned result; if a non-$capOn$ input was partitioned, then  a Merge step will be  added to collect data from multi-threads to a single collection; When an operator with \textbf{ST} capability gets data from an  operator with \textbf{PR} capability,  a Merge step will be added. 

\subsection{Buffering Mechanism}
AWESOME employs a buffering mechanism to avoid storing unnecessary intermediate results in memory. Different from pipeline, buffering mechanism does not utilize multiple cores to execute different operators simultaneously.
Some operators can process input in a batch-by-batch manner, and some can generate output in a batch-by-batch manner. 
We refer data with this manner as stream hereafter.  There are four types of buffering capabilities: 
\begin{enumerate}[leftmargin=*]
    \item $SI$ (Stream-Input): the input can be passed as stream to the operator, but it produces a whole inseparable result at once;
    \item $SO$ (Stream-Output): the operator takes an inseparable input but can produce result progressively as stream;
    \item $B$ (Blocking): both the input and output need to be a whole;
    \item $SS$ (Stream-Stream): both the input and output can be a stream.
\end{enumerate}

Each physical operator is associated  with its buffering capability. Table \ref{tab:operator} presents it for some physical operators.   Similar to data parallelism capability, there is another $capOn$ attribute associated if the operator has more than one input. The physical DAG will be partitioned to a collection of chains. Inside each chain, the intermediate result is  not stored in memory; the upstream operator produces stream output to be consumed by the downstream operator. 
The data across chains has to be stored in memory.

\begin{figure}
    \centering
    \includegraphics[width=0.25\textwidth]{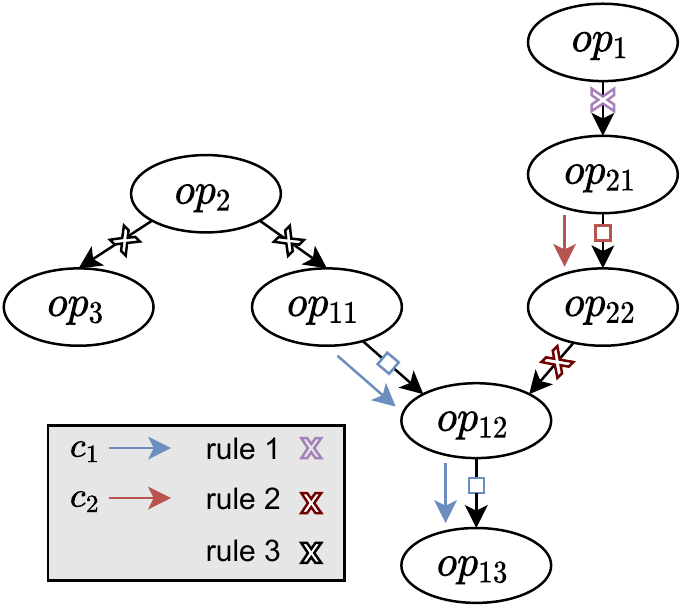}
    \caption{Illustration of buffering rules.}
    \label{fig:cutgraph}
\end{figure}

The  collection of chains is collected from the physical DAG by partitioning it based on the  partition rules which are shown below and also illustrated in Fig.~\ref{fig:cutgraph}:
\begin{itemize}[leftmargin=*]
    \item For an edge $e = (op_{e1}, op_{e2})$, if $op_{e1}$ can't generate stream result or $op_{e2}$ can't take stream input, $e$ will be cut. For example, in Fig.~\ref{fig:cutgraph}, the edge between $op_1$ and $op_{21}$  is cut. 
    \item For an edge $e = (op_{e1}, op_{e2})$, if data from $op_{e1}$ to  $op_{e2}$  is not the $capOn$ input of $op_{e2}$, $e$ will be cut. In Fig.~\ref{fig:cutgraph}, the edge between $op_{22}$ and $op_{12}$ is cut. 
    \item For an operator $op$, if it has more than one outgoing edges, then all outgoing edges will be cut. In Fig.~\ref{fig:cutgraph}, the outgoing edges from $op_2$ are all cut. 
\end{itemize}

AWESOME users have the option to turn on buffering mechanism.  The buffering mechanism would be very helpful if : 1) the analysis plan contains many  data flow edges which buffer can be added to; or 2) AWESOME is running on a memory-limited machine; or 3) users care about the responsiveness of the system and expect to get some initial results back soon  without waiting for the complete results. 


\vspace{-1.5ex}
\subsection{Failed Attempt: Pipeline + Data Parallelism}
We built a framework that hybridizes pipeline (i.e., task parallelism) and data parallelism, however, the experimental results reveal that such framework is not suitable for AWESOME. We briefly introduce this framework and explain why this technique did not boost performance to provide some insights for future researches. 

Similar to the buffering mechanism, an AWESOME physical DAG is partitioned into a list of  chains based on the partition rules. Then each chain will form a pipeline where operators can be executed simultaneously using multi-cores. Once the upstream operator produces a batch of results, the downstream operator will be executed on that batch immediately and simultaneously. Both pipeline and data parallelism utilizes multi-cores to increase resources utilization, thus we define a scheduling problem to allocate a specific amount of cores (the number of cores in an OS) to operators in each pipeline chain. A simple solution is to allocate cores to match the produce and consume rates of data.  

However, from the experimental results, this framework is not more efficient than  data parallelism framework even under the best allocation strategy due to two properties of AWESOME operators. We theoretically explain the reason why this framework does not outperfrom data parallelism framework.  For a simple pipeline chain with two operators: $op_1 \to op_2$, suppose that there are a total of $n$ cores and it costs $t_1$ for $op_1$ to produce a batch of data and $t_2$ for $op_2$ to consume the batch, then there will be $t_1n/(t_2+t_1)$ cores assigned to $op_1$ and the rest of cores assigned to $op_2$.

Suppose that $op_1$ will produce $m$ batches in total, then the execution time of applying data parallelism solely $T_1$ and of applying pipeline + data parallelism $T_2$ can be computed as, 
\begin{equation}
    \begin{split}
    T_1 & = \frac{(t_1 + t_2) m}{n} + agg*n\\
    T_2 & = \max\{\frac{t_1 m}{n_1}, \frac{t_2 m}{n - n_1}\} + agg * n_1, 
    \end{split}
\end{equation}
where $n_1$ is the number of cores assigned to $op_1$, and $agg * \#core$ is the sequential aggregation cost of data parallelism. Since for AWESOME aggregation operators such as \textit{SUM}, the aggregation cost is usually very small and can be  negligible comparing to other time-consuming analytical functions, we can  prove that $T_1 	\approx \frac{(t_1 + t_2) m}{n} \leq \max\{\frac{t_1 m}{n_1}, \frac{t_2 m}{n - n_1}\} \approx T_2$ always holds where the equality is achieved when the above optimal allocation solution is applied. Thus, the pipeline and data parallelism framework can't outperform data parallelism if all operators in a chain are data parallel-able.


Appendix D  presents the data parallel capability and buffering capabilities for most AWESOME operators. In the future, when there are more operators with different properties are added to AWESOME, this framework may have chance to outperform the solely data parallelism framework.  

\vspace{-1.5ex}
\section{Learned Cost model} \label{sec:costmodel}
The query planning stage generates multiple candidate physical plans, and in the execution stage, the optimal one will be chosen at run-time based on a learned cost model. 

For each virtual node which corresponds to multiple candidate sub-plans,  the cost model is applied to each sub-plan to estimate the execution cost and the sub-plan with the lowest cost will be chosen. 
We use a learned cost model instead of a rule-based model based on two reasons:
\begin{itemize}[leftmargin=*]
    \item Cost should be decided at sub-plan level instead of operator-level, which makes rule-based optimization hard to design. 
    For a single logical operator with different physical implementations, it is easy to design rules to decide which implementation should be chosen under what circumstance. 
    However, in the pattern set, each logical sub-plan may consist of several logical operators and each of them may be transferred to multiple different physical operators, leading to a large size of rules space. 
    \item The cost of a physical operator may depend on several features and a rule-based model is too simple to represent the complex relationship. 
\end{itemize}

\subsection{Cost Model}
We provide a learned cost model to estimate the execution time for each candidate physical sub-plan denoted as $S$. 
Suppose that $S$ consists of multiple operators $\{op_1, \cdots, op_n\}$, 
the overall cost estimation is given as the sum of the estimated cost of all operators since AWESOME does not apply task parallelism, i.e.,
\begin{equation}\label{equ:cost}
    est_{S} = Cost(op_1) + \cdots + Cost(op_n),
\end{equation}
where $Cost(\cdot)$ is a trained linear regression model with the polynomial of raw features (degree 2) as variables that predicts the execution cost of a physical operator, i.e.,
\begin{equation}
\begin{split}
Cost(op) = w_0 + w_1 f_1 + \cdots + w_n f_n + w'_1 f^2_1 +  \cdots + w'_n f^2_n \\
+ w_{(1,2)} f_1f_2 +  \cdots + w_{(n-1, n)} f_{n-1} f_n,
\end{split}
\end{equation}
where $f_{1}, \cdots, f_n$ are the raw  features for $op$. $Cost(op)$ is trained based on training data collected from  calibration for   operator $op$. For relation-related operators, the raw features include the sizes of tables; for graph-related operators, node count or  edge count is selected as a raw feature and for some graph queries, the predicate size can also be a raw feature.

\vspace{-1.5ex}
\subsection{Calibration}
To train the individual cost model $Cost(\cdot)$, we design a set of synthetic datasets which vary at some parameters, and  run each   operator on different datasets to collect a set of execution time.

\noindent\textbf{Operators and features.} We mainly train cost model for operators  which are graph-related or relation-related. 

For graph operators, we evaluate common operators such as \textit{CreateGraph} and \textit{PageRank}. The graph size serves as  one feature for the  cost estimation. For \textit{ExecuteCypher}, there are various types of Cypher queries and we  evaluate two typical types of queries: 
\noindent \textbf{Type I}: Queries with a series of  node or edge property predicates. For example,
\textit{Match (n)-[]-(m) where n.value in L and m.value in L} where $L$ is a list of strings. The size of $L$ is another raw feature  that decides the query cost.  

\noindent\textbf{Type II}: Full text search queries. In this kind of queries, there is a node/edge property which contains long text and the queries will find out nodes/edges whose text property contains specific strings. For example, \textit{Match (n)-[]-(m) where n.value contains string1 or n.value contains string2 or .....}. The number of the OR predicates is another raw feature of the cost model.

\begin{figure}
    \begin{subfigure}{.5\textwidth}
      \centering
      \includegraphics[width=0.85\linewidth]{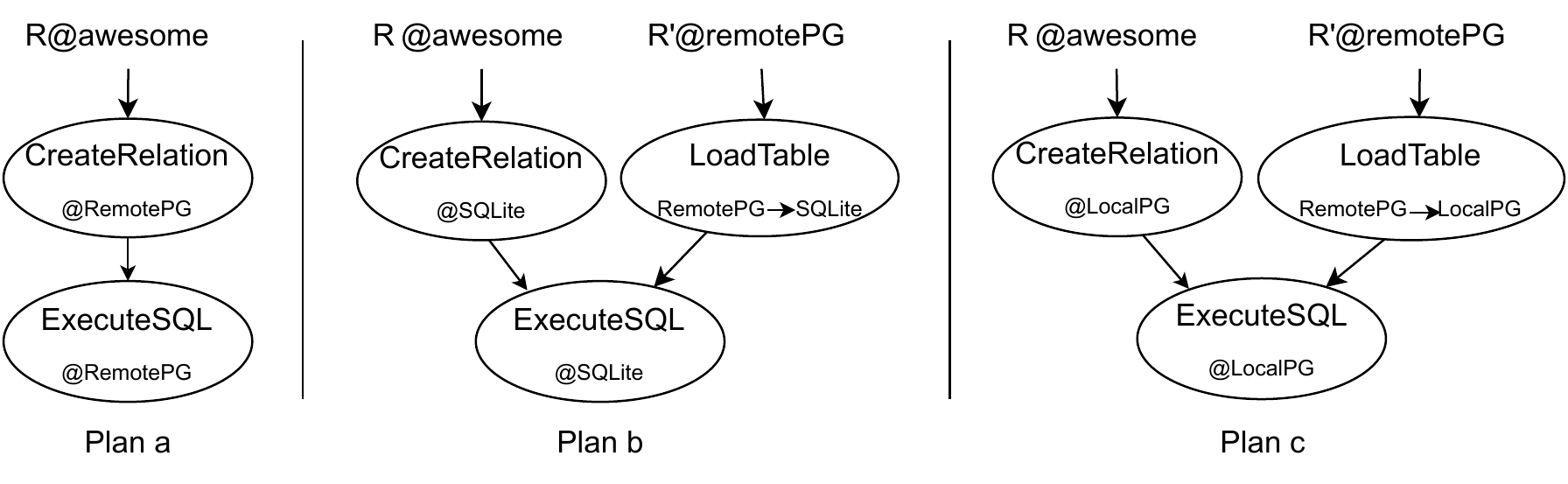}
      \caption{Execution Sub-plans}\label{fig:executeSQL}
    \end{subfigure}%

    \begin{subfigure}{.25\textwidth}
      \centering
      \includegraphics[width=\linewidth]{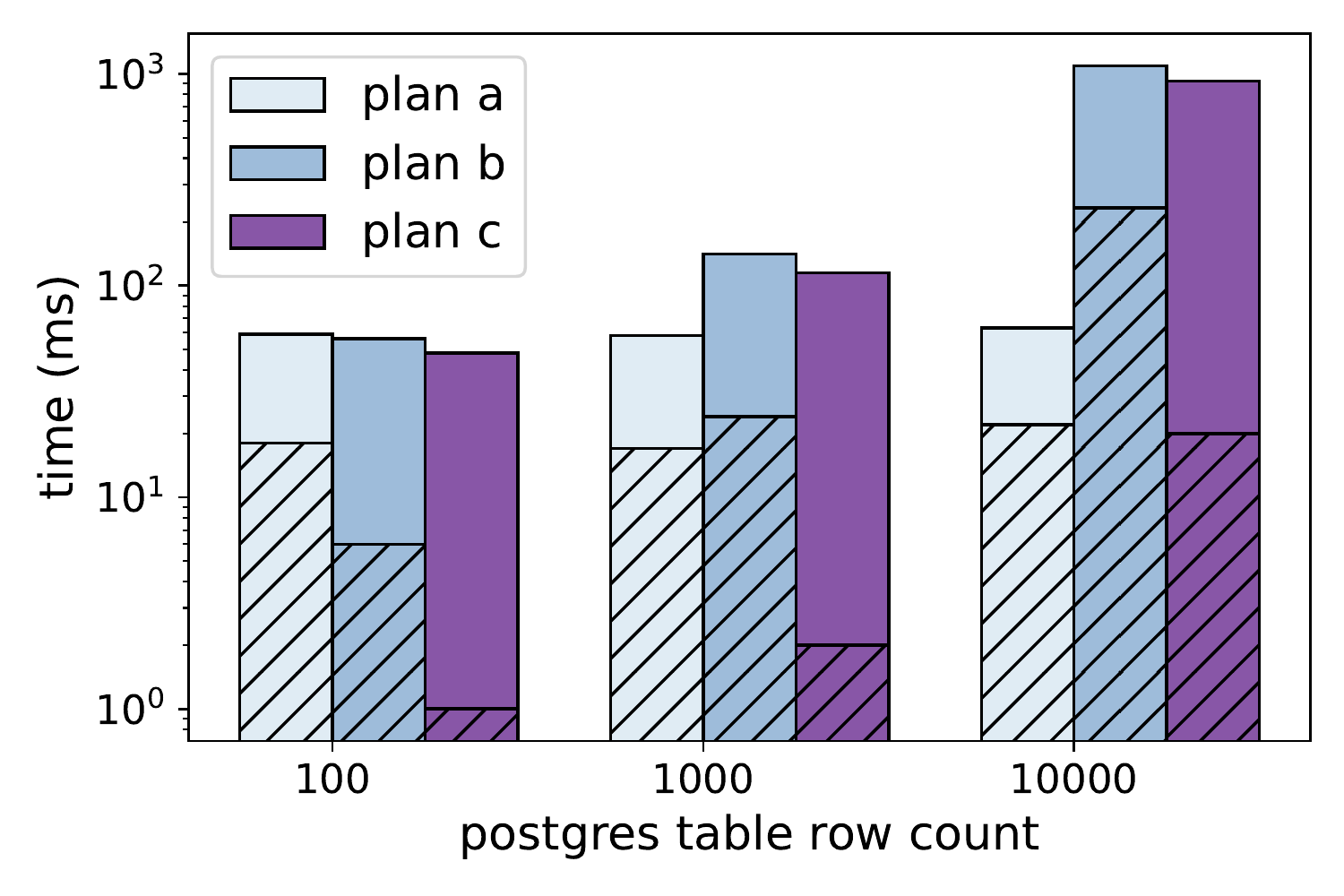}
      \caption{Row count of R: 100.}\label{fig:tablejoin1}
    \end{subfigure}%
    \begin{subfigure}{.25\textwidth}
      \centering
      \includegraphics[width=\linewidth]{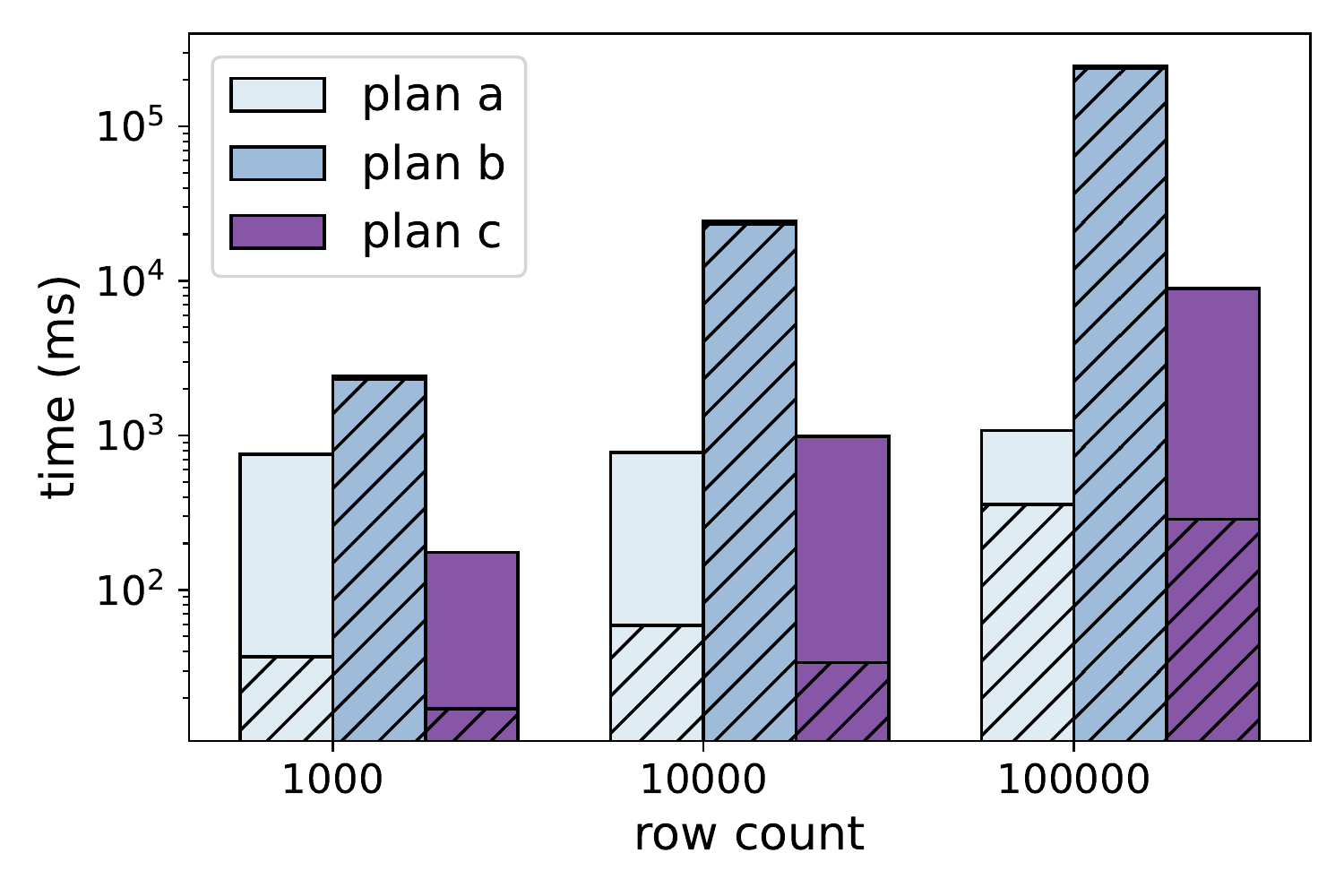}
      \caption{Row count of R: 10K.}\label{fig:tablejoin2}
    \end{subfigure}
    \caption{\small{Execution sub-plans and calibration results for cross-engine \textit{ExecuteSQL}. The part with '//' denote execution time of \textit{ExecuteSQL} operator.}}
    \end{figure}

For relation operators, we test the \textit{ExecuteSQL} operator. Based on the locations of the tables in the query, there are different candidate  execution sub-plans for this operator. For example, if all tables involved are AWESOME tables generated from the upstream operators, then there are two candidate plans: (a) store all relations to in-memory SQLite,  and execute the query in SQLite; (b) store all relations in PostgreSQL, and  execute the query in PostgreSQL. If there are both AWESOME tables and PostgreSQL tables involved in the query,   the two candidate plans are illustrated in Figure~\ref{fig:executeSQL}: (a) as the left dashed rectangle shows, we store AWESOME tables to PostgreSQL, then execute the query in PostgreSQL; (b) as the right dashed rectangle shows, we store AWESOME tables to SQLite and select the columns needed from PostgreSQL tables and store them to SQLite, then the query will be executed in SQLite. 

\noindent\textbf{Datasets.} We design a set of graph datasets and relation datasets  which are used for  graph- and relation-related operators respectively. We present the statistics in Table~\ref{tab:synthetic}. 
\begin{table}[b]
    \centering
    \footnotesize
    \caption{Parameters of synthetic datasets for cost model.}~\label{tab:synthetic}
    \begin{tabular}{ccc}
    \toprule
    &\textbf{Parameter} & \textbf{Value} \\ \midrule
        \multirow{4}{*}{\textbf{graph dataset 1}} &edge size & 500, 1k, $\cdots$, 800k \\
        &avg. density & 2 \\
        &node property & value: String\\
        &keyword size & 50, 100, 500, 1k, 2k\\\hline
        \multirow{3}{*}{\textbf{graph dataset 2}} & node size & 5k, 10k, $\cdots$, 500k \\
        & node property & tweet: String \\
        &keyword size & 50, 100, 500, 1000\\\midrule
        \multirow{2}{*}{\textbf{relation dataset}} & PostgreSQL table row count & 100, 1k, 10k, 100k\\
        & Awesome table row count & 100, 1k, 10k, 100k\\ \bottomrule
    \end{tabular}
\end{table}

For graph datasets, there are two types of graphs: The first type of datasets is used to test operators including \textit{CreateGraph}, \textit{PageRank} and the Type I Cypher queries: We created several property graphs  with different edge sizes, and to simplify the model we kept the density of graphs as a constant value 2; each node (or edge) has a value property which is a unigram and we make sure each node's (or edge's) property is unique, then we created keywords lists with different sizes  from the values set as the predicates. The second dataset is designed for the Type II Cypher queries: We created graphs with different node sizes and each node has a \textit{tweet} property whose value  is a tweet text collected from Twitter; All the unigrams are collected from these tweets and after removing the most and the least frequent words, we randomly selected words to create different sizes of keywords lists which will be used to do text search. 


\begin{figure}[t]
    \begin{subfigure}{.24\textwidth}
      \centering
      \includegraphics[width=\linewidth]{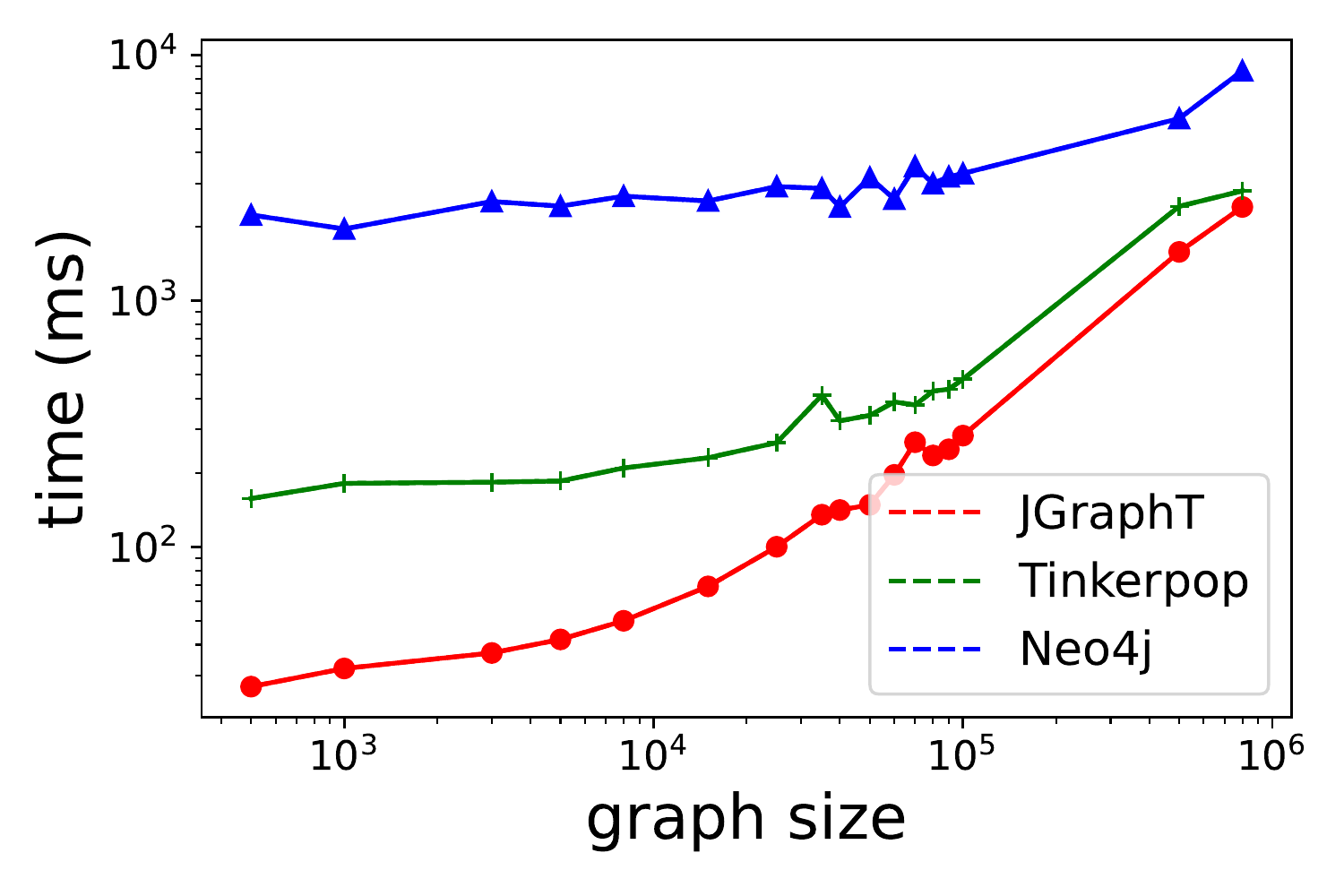}
      \caption{Graph Creation}
    \end{subfigure}%
    \begin{subfigure}{.24\textwidth}
      \centering
      \includegraphics[width=\linewidth]{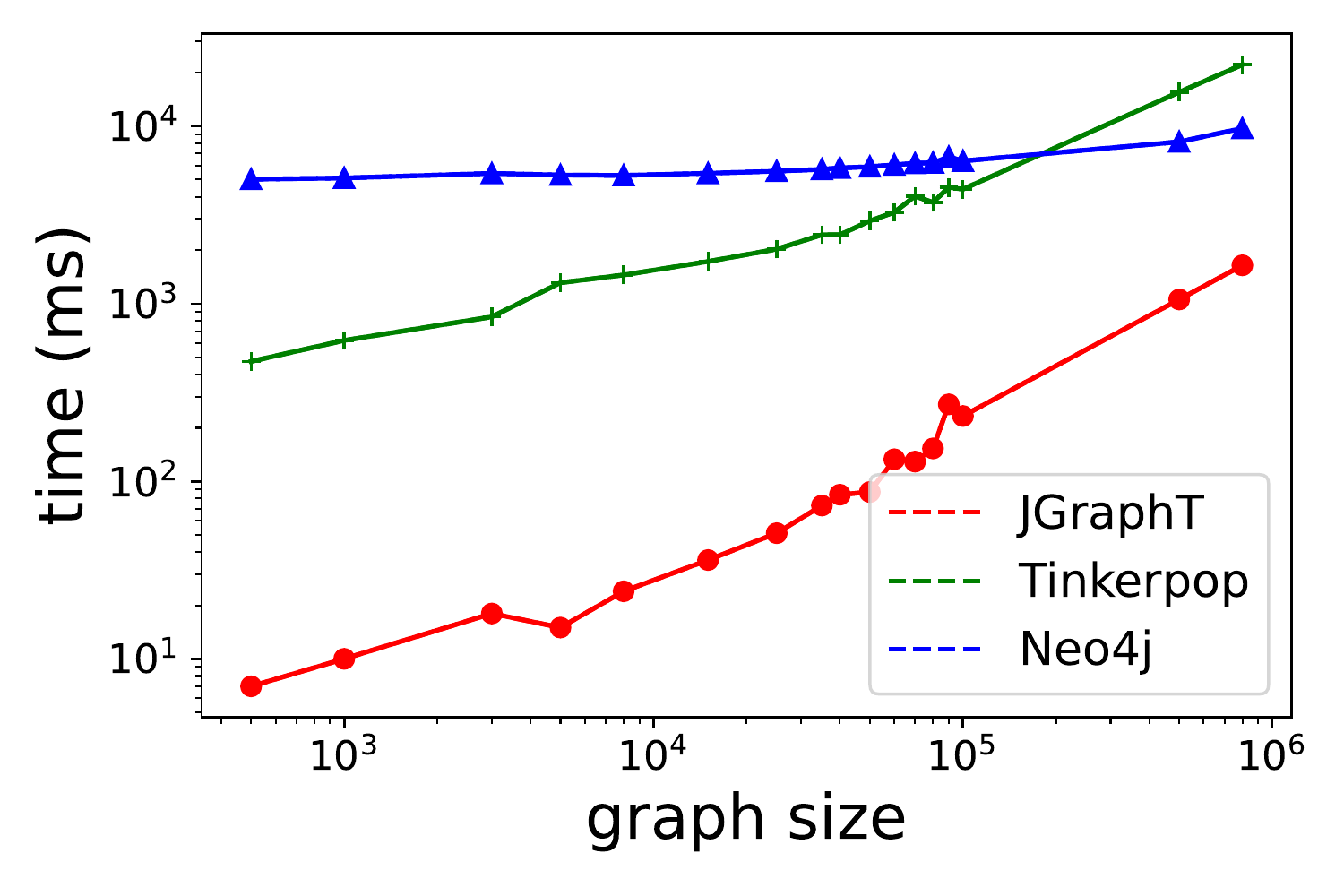}
      \caption{Page Rank}
    \end{subfigure}
        \begin{subfigure}{.24\textwidth}
      \centering
      \includegraphics[width=\linewidth]{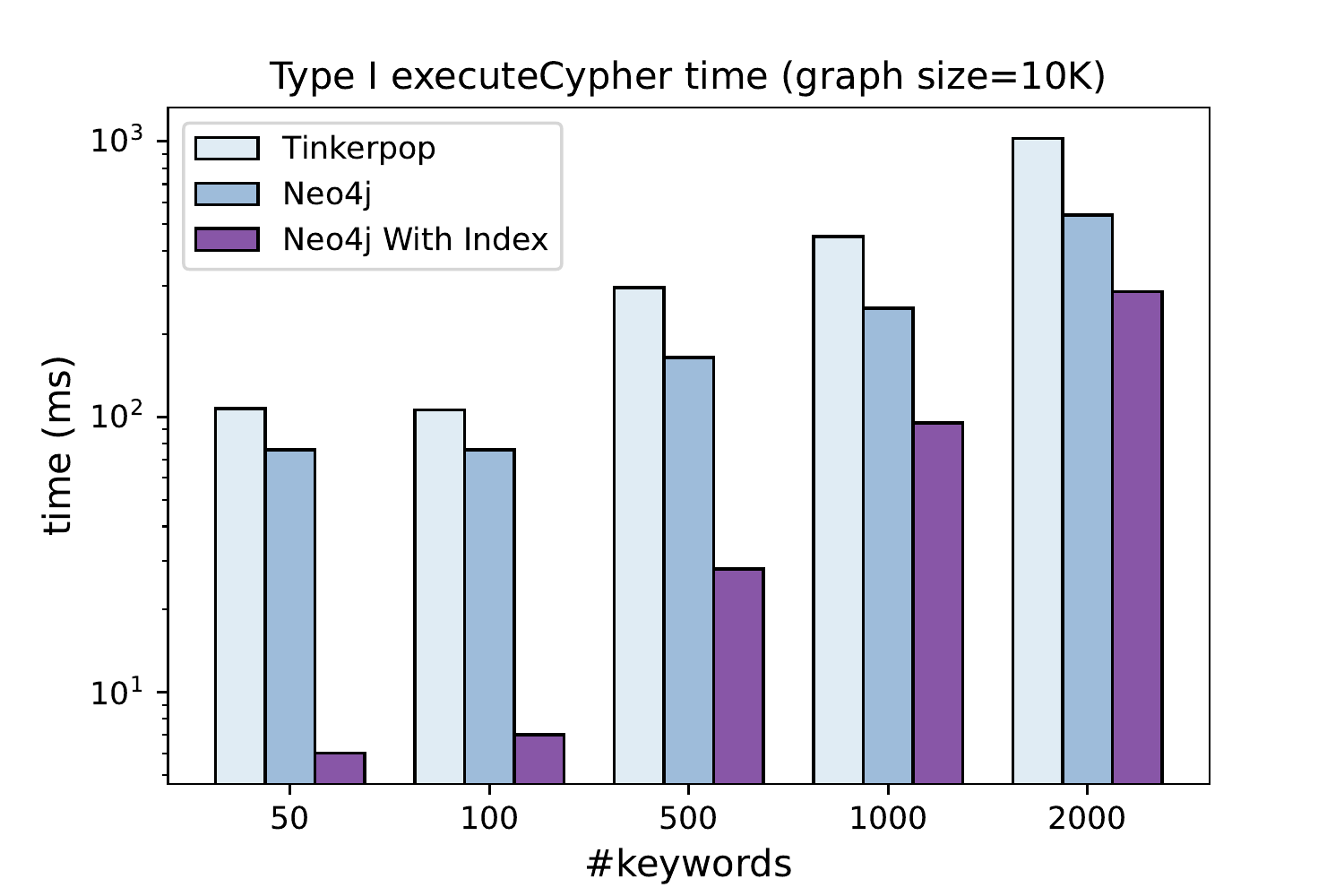}
      \caption{Type I Cypher Query}
    \end{subfigure}%
    \begin{subfigure}{.24\textwidth}
      \centering
      \includegraphics[width=\linewidth]{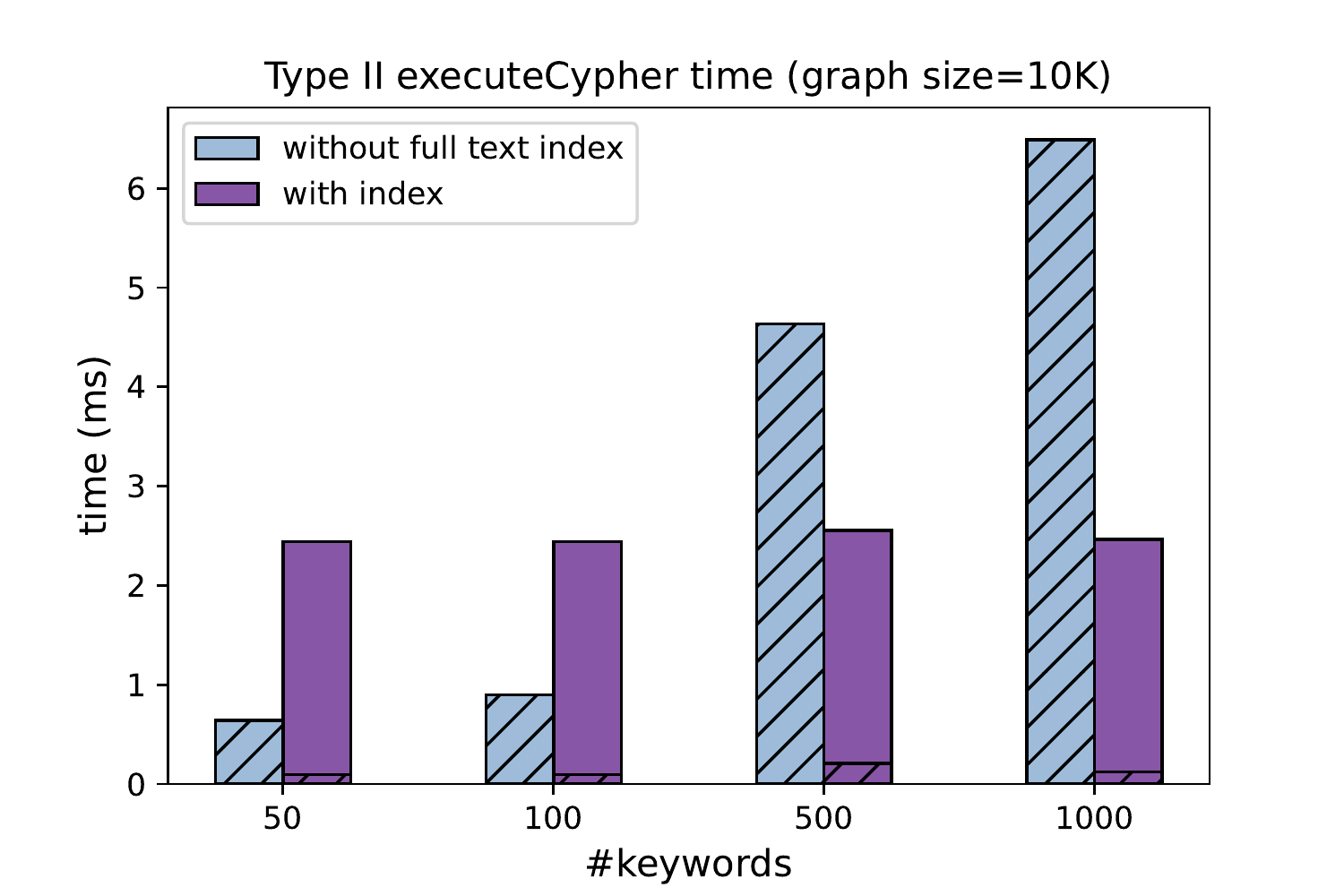}
      \caption{Type II Cypher Query}
    \end{subfigure}
\caption{Calibration results for graph operators}
\label{fig:cali}
\end{figure}

\noindent\textbf{Calibration Results.} We present some  calibration results for some operators/patterns in Figure \ref{fig:cali},  Figure \ref{fig:tablejoin1} and Figure \ref{fig:tablejoin2}. Figure \ref{fig:cali} shows part of the  calibration results for  some graph operators.  
Figure \ref{fig:tablejoin1} and \ref{fig:tablejoin2} show  the calibration results for the  \textit{ExecuteSQL} operator where the query includes a PostgreSQL table and an AWESOME table and  the two sub-plans correspond to the sub-plans in Figure~\ref{fig:executeSQL}.

\vspace{-1.5ex}
\subsection{Training and Cost Estimation}
The individual cost model for each operator is trained based on the calibration results  to minimize the loss function, i.e., mean squared error. 
At run-time, when the input of a virtual node is returned from the upstream operator, the features are collected and passed to the overall cost model (Equation \ref{equ:cost}) to compute the cost for each candidate physical sub-plan. The best sub-plan with the lowest cost will be selected. 

\section{Experiments}\label{sec:exp}
In this section, we first empirically validate whether AWESOME is able to improve efficiency of analytical polystore workloads. 
Then, we drill into how the cost model of AWESOME contributes.

\begin{figure*}[th]
\minipage{0.9\textwidth}
    \includegraphics[width=\linewidth]{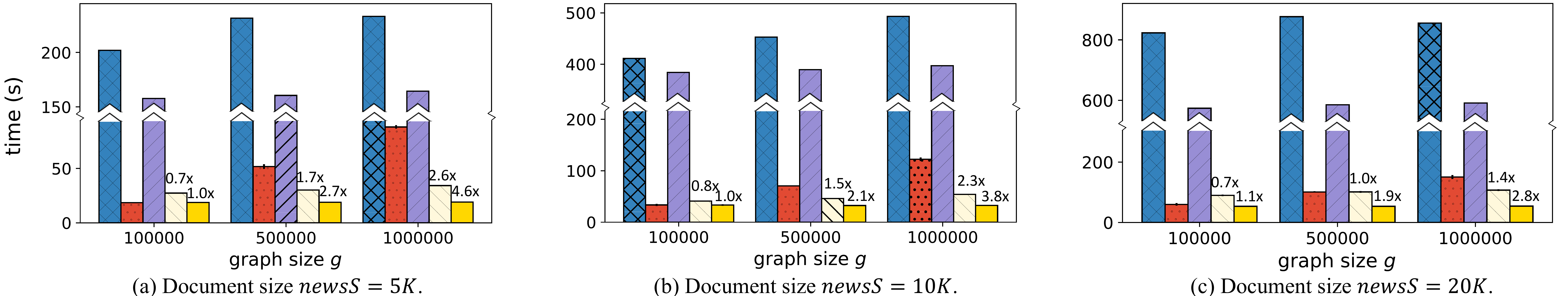}
      \caption{Total execution time for  \textit{PoliSci} w.r.t.~different $newsS$.}
     \label{fig:case2result}
\endminipage\hfill
\minipage{0.9\textwidth}
    \includegraphics[width=\linewidth]{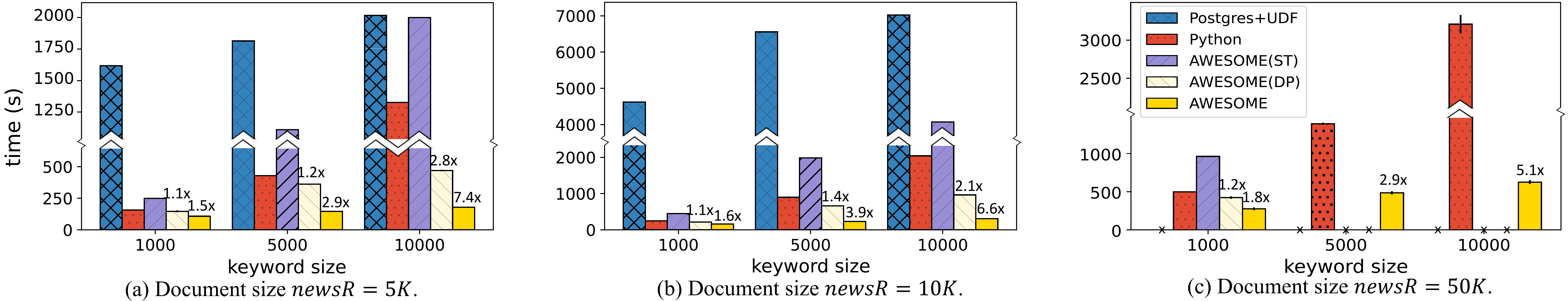}
    \caption{Total execution time for  \textit{NewsAnalysis} w.r.t.~different $newsR$. } \label{fig:case1result}
\endminipage\hfill
\minipage{0.9\textwidth}%
    \includegraphics[width=\linewidth]{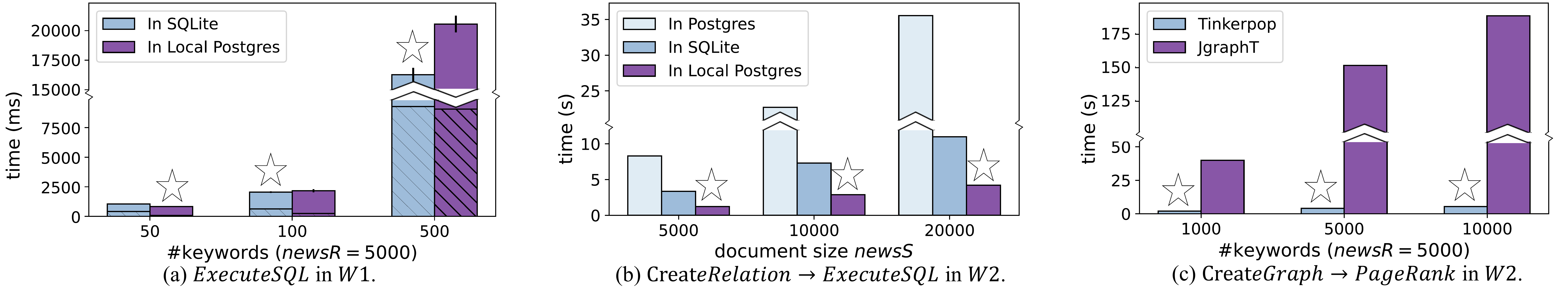}
    \caption{Execution time of different physical plans.}  \label{fig:costmodel}
\endminipage
\end{figure*}

\subsection{Experimental Setup}
We focus on the single-machine multi-cores setting, and the distributed version of AWESOME will be our future work. The experiments were run on CloudLab \cite{duplyakin2019design}, a free and flexible cloud platform for research. 
The machine has 24-core AMD 7402P CPUs, 128 GB memory and 1.6 TB disk space. It runs Ubuntu 18.04. 

\noindent\textbf{Datasets.}
We  collect four real world datasets to run the workloads. 
\begin{itemize}[leftmargin=*]
    \item \textit{Newspaper}: A relation stored in a  PostgreSQL database aliased as News with size around 36 GB. There are over 1M news articles with an average length of 500 words collected from the Chicago Tribune newspaper. 
    \item \textit{SenatorHandler}: A PostgreSQL relation of about 90 United States senators with their names and twitter user names stored in PostgresSQL database aliased as Senator. 
    \item \textit{NewsSolr}: A collection of news stored in the Solr engine with size around 20 GB.
    \item \textit{TwitterG}: Attribute graphs stored in Neo4j that represent the Twitter social networks of different sizes. 
    A node in \textit{TwitterG} is labeled as either ``User'' or ``Tweet''. 
    A User node has a property \textit{userName}, and a Tweet node has a property \textit{tweet} storing  tweet content. A User node  connects to another User node by a directed edge if the first one \textit{mentions} the second, and a User node  connects to a Tweet node by a directed edge if the user \textit{authors} the tweet.
    \end{itemize}
\vspace{-0.5em}
\noindent\textbf{Workloads.}
We evaluate two analytical workloads.
\textbf{W1:} \textit{PoliSci} focuses on the polystore aspect of the system where input data is  stored in heterogeneous data stores. \textbf{W2:} \textit{NewsAnalysis}, a complex text analytical task,  focuses on analytical functions including graph algorithm like PageRank~\cite{page1999pagerank} and NLP functions like LDA~\cite{pritchard2000inference}. 

The illustration and ADIL script of \textit{PoliSci} are shown in Figure~\ref{fig:poliscidiagram} and Figure~\ref{fig:polisciscript}. 
This workload queries on the \textit{NewsSolr}, \textit{SenatorHandler} and \textit{TwitterG} dataset.
For workload \textit{NewsAnalysis}, the corresponding ADIL script is given in  Appendix B. It selects news from \textit{News} dataset and applies LDA model to detect topics from the corpus. 
Then it implements the method in \cite{DBLP:conf/ictir/GollapalliL18} to evaluate the quality of each topic. Specifically, for each topic, from news corpus  a word-neighbor graph is constructed  which consists of  the keywords in the topic and the aggregated PageRank value of these keywords in this graph will be used as a metric to evaluate the quality of the topic.  

\noindent\textbf{Parameter Settings.}
For the \textit{PoliSci} workload, we  change  two parameters: 1) $newsS$, the size of documents $doc$ selected from NewsSolr dataset and is set by  changing $rows=5000$ in the Solr query  to different values. $newsS \in \{5K, 10K, 20K\}$. 2) the  size of the  graph  $TwitterG$: the graph size, denoted by $g$, is the number of nodes in it. $g \in \{ 100K, 500K, 1M\}$. $newsS$ will have impact on the size of $entity$ and thus will influence the \textit{executeSQL} execution time, and $g$ will influence on executing the  two Cypher queries.  
For the \textit{NewsAnalysis} workload, we vary two parameters: $newsR$ is the size of news selected from \textit{Newspaper} relation: $newsR \in \{5K, 10K, 50K\}$; $t$ is the number of keywords in each topic: $t \in \{1K, 5K, 10K\}$ for the end-to-end experiments and $t \in \{50, 100, 500\}$ for the drill down analysis.
$newsR$ and $t$ control the size of the documents  and the size of the word-neighbor graph for each topic.


\noindent\textbf{Compared Methods.}
We implement and compare the following methods. We put the SQL scripts for \textbf{Postgres+UDF} method and Python code for \textbf{Python}  in Appendix A and C. For \textbf{Postgres+UDF} implementation, the scripts have around 400 and 1000 tokens respectively for the two workloads and Python codes have around 380 and 600 tokens for each. While ADIL scripts only have around 100 tokens for each.

\begin{itemize}[leftmargin=*]
    \item \textbf{Postgres+UDF}: It stores all datasets to a single store, Postgres, and  uses pure SQL scripts with user-defined functions written in Python or implemented by MADLIB \cite{hellerstein2012madlib} toolkit.    
    \item \textbf{Python}: The workloads were implemented in Python. To make sure the results are the same as AWESOME's and to make the comparison fair, for analytical functions such as LDA and NER, we use the same toolkits as used by AWESOME. 
    \item \textbf{AWESOME(ST)}: ST stands for single threaded. It does not use any AWESOME features including logical plan optimization, data parallel execution and cost model.
    \item \textbf{AWESOME(DP)}: DP stands for data parallelism. It only applies data parallelism feature. 
    \item \textbf{AWESOME}: It has full AWESOME features including logical optimization, cost model and data parallelism. 
\end{itemize}

\subsection{End-To-End Efficiency}
For AWESOME baselines, the end-to-end execution time is the total execution time from taking a workload as input, parsing, validating, logical planning, physical planning and executing. 
For \textbf{Postgres+UDF} baseline, to make the comparison fair, we exclude the data movement cost, which is the time spent to move data from others stores to Postgres, from the total cost. 
Figure~\ref{fig:case2result} and Figure~\ref{fig:case1result} present the end-to-end execution costs of the four compared methods. The numbers on top of bars denote the speed-up ratio to the \textbf{Python} baseline. The black \textbf{X}s mean that the program either pops out an error or can not finish in 3 hours.

From the results, \textbf{AWESOME(DP)} and \textbf{AWESOME} show great efficiency and scalability when varying the parameters. It dramatically speeds up the execution time over   \textbf{Postgres+UDF}. For \textit{NewsAnalysis}, when $newsR=50K$,  \textbf{Postgres+UDF} pops out a sever connection lost error message when running in-Postgres LDA function. 
Comparing with \textbf{AWESOME(ST)}, data parallelism  achieves significant performance under any parameter setting. 
In the following, we mainly discuss the performance gain of  \textbf{AWESOME(DP)} and \textbf{AWESOME}  over the baseline method \textbf{Python}.

For workload \textit{PoliSci}, when the  graph size $g$ increases, AWESOME achieves an increasingly large speedup over the Python implementation. When $g=100K$, \textbf{AWESOME(DP)} is a little bit slower than \textbf{Python}, but with the cost model, \textbf{AWESOME} has similar  performance as \textbf{Python}; When $g$ is increased to $1M$ and $newsS=5K$, \textbf{AWESOME(DP)} and \textbf{AWESOME} speeds up the execution time by $\sim3$x and $\sim5$x respectively.

For workload \textit{NewsAnalysis}, when $newsR=5K$ and keywords size is large ($t=10K$), \textbf{AWESOME(DP)} and  \textbf{AWESOME}   can achieve up to $\sim3$x  and $\sim7$x speed-up respectively over \textbf{Python}; when $newsR=50K$ and keyword size is  $10K$, \textbf{AWESOME(DP)} pops out an out of memory error because it selected a bad execution plan,  while \textbf{AWESOME} is scalable and can finish in around 10 minutes which is about $5$x speedup over \textbf{Python}.

\noindent\textbf{Why not single DBMS with UDF?} From our experience with implementing the \textbf{Postgres+UDF}, we found that this single DBMS with UDF setting fails to qualify polystore analytical tasks for three reasons: 1) Data movement cost. Users need to write ad-hoc code to move data from various stores to a single store. 2) Programming difficulty. It is not flexible to program with pure SQL. For  workload \textit{NewsAnalysis}, even with  MADLIB toolkit which implements LDA and PageRank UDFs, hundreds of lines of SQL needed to be written (as shown in the Appendix A). 3) Efficiency. The in-DBMS implementation of  analytical functions such as LDA and PageRank  are much less efficient than using the mature packages from Java or Python, and the current in-DBMS implementation of some analytical functions is not scalable as well.  
\vspace{-1em}
\subsection{Drill-Down Analysis}
We present  some detailed evaluation results to explain how \textbf{AWESOME}  achieves a better performance over \textbf{AWESOME(DP)}. 

We take some snippets from each workload and  show the execution time of different  candidate sub-plans for these snippets in Figure \ref{fig:costmodel}. The parts with ``//'' hatch denote execution time for \textit{ExecuteSQL} physical operator.  The bars with stars on top are the best execution plans selected by AWESOME cost model. 


 For workload \textit{PoliSci}, Figure \ref{fig:costmodel} (a) presents the execution time for different sub-plans regarding to the  cross engine \textit{ExecuteSQL} logical operator where one table ($SenatorHandler$ table) is from PostgreSQL and another (named entity table) is an AWESOME table. The three execution plans are illustrated in Figure~ \ref{fig:executeSQL}. When the selected  documents size increases, the named entity table's size will increase and  the in local Postgres execution plan will be much more effective than the in remote Postgres one by not moving large size of data to a remote Postgres server. In SQLite execution  is not efficient because even though storing table data to in memory SQLite is fast, SQLite is slow when the two tables join on Text columns.

For workload \textit{NewsAnalysis}, Figure \ref{fig:costmodel} (b) shows the execution time for logical plan  \textit{ExecuteSQL}. The possible physical sub-plans are shown in Node 1 in Figure~\ref{fig:physicalplan}. 
The cost model does not look at each single operator, e.g., \textit{ExecuteSQL}, to decide the best physical operator, instead, it looks at the sub-plan which consists of both creation and query execution to determine the best physical sub-plan.
 We selected some small keywords sizes for this drill down analysis: $t=\{50, 100, 500\}$ since when $t$ gets larger, the In-SQLite execution   always dominates. As (b) shows, the SQL query execution time of local Postgres is  less than that of SQLite,  however, when considering both table creation time and execution time, the in-memory SQLite plan will be chosen for $t=\{100, 500\}$. These results on small snippets demonstrate the effectiveness  of our  cost model.

\begin{figure}[h]
    \begin{subfigure}{.24\textwidth}
      \centering
      \includegraphics[width=\linewidth]{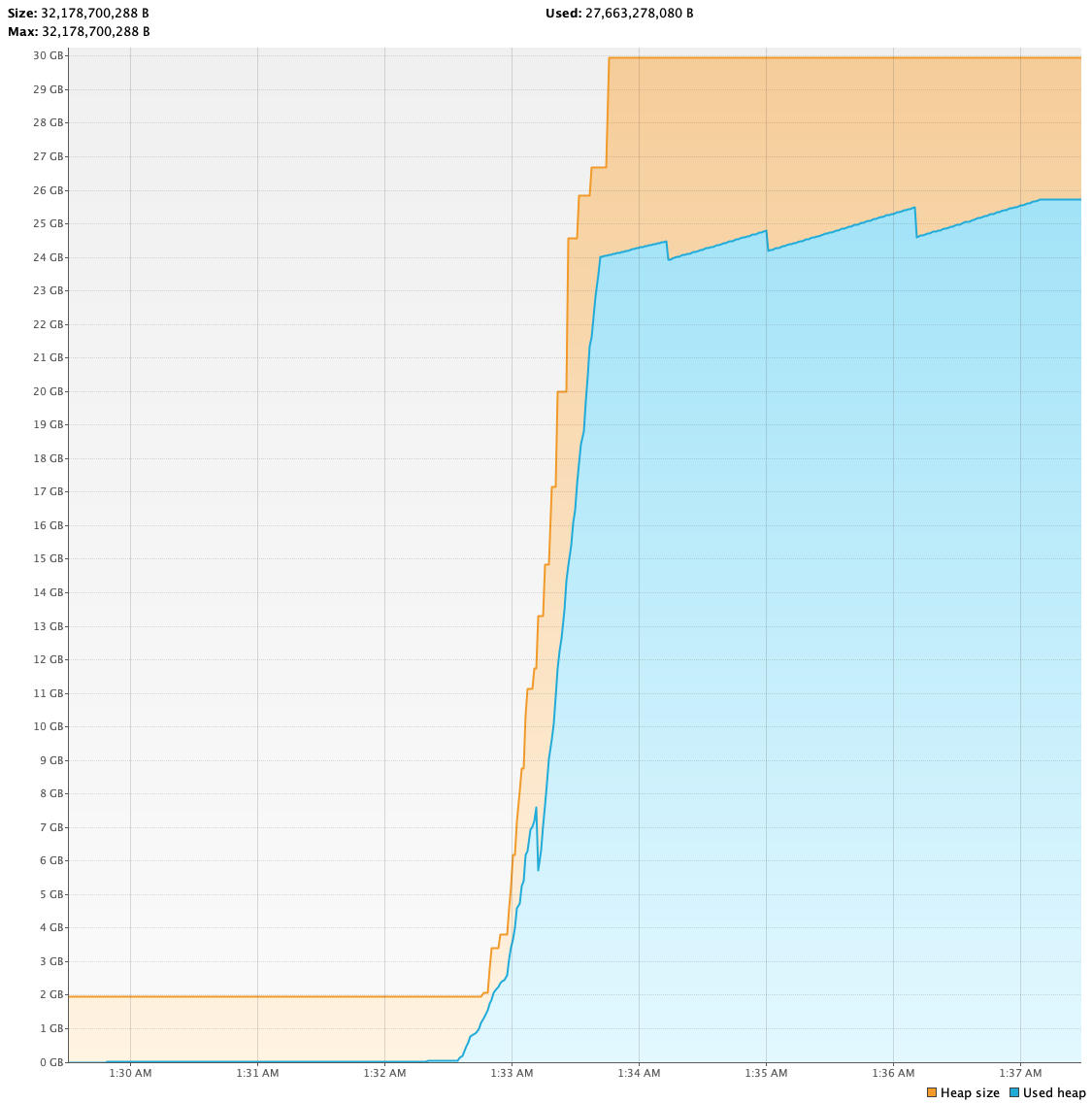}
      \caption{Without buffering mechanism.}
    \end{subfigure}%
    \begin{subfigure}{.24\textwidth}
      \centering
      \includegraphics[width=\linewidth]{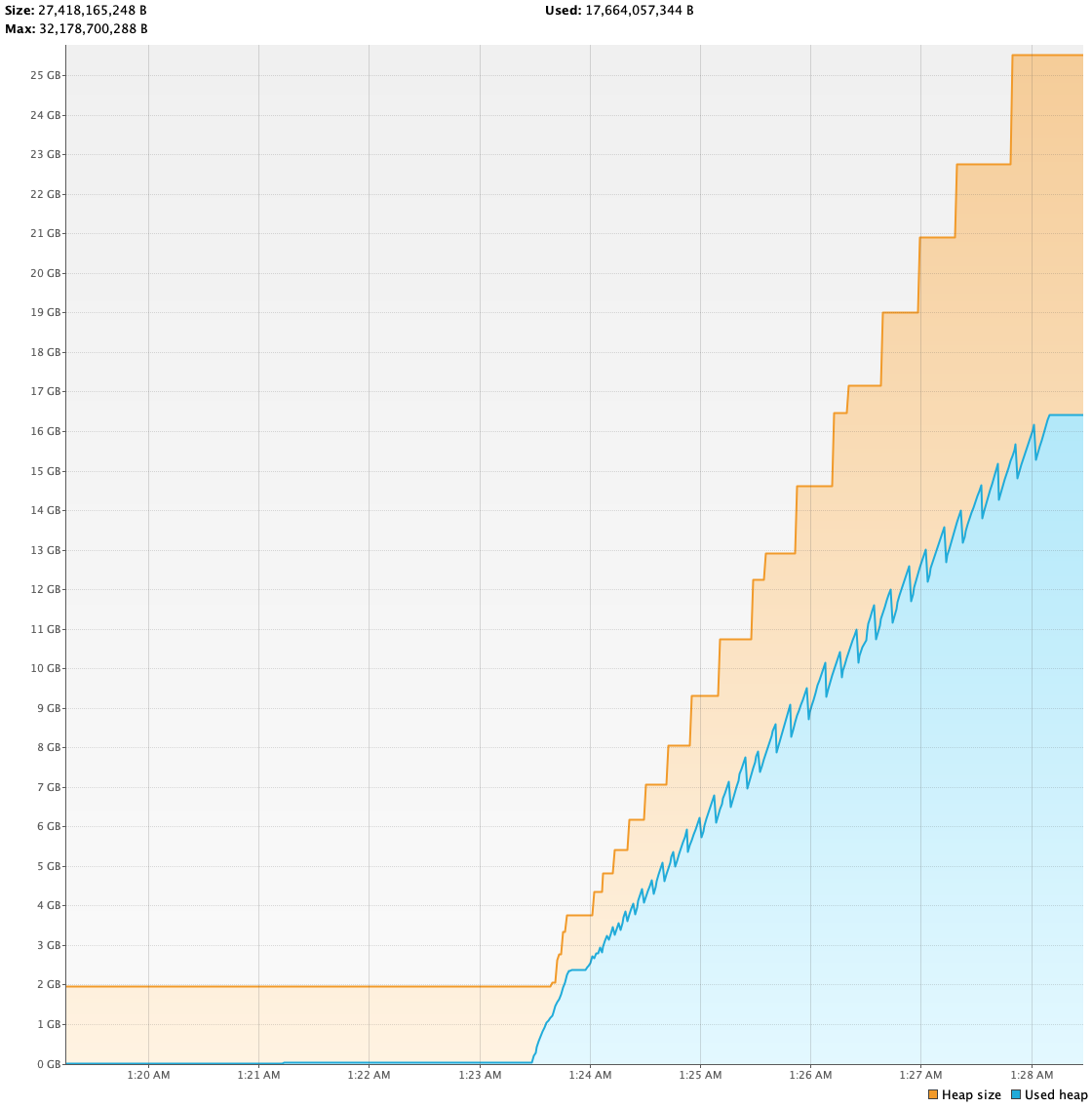}
      \caption{With buffering mechanism.}
    \end{subfigure}
\caption{Memory footprint of executing a chain.}
\label{fig:memory}
\end{figure}

We pick a snippet from \textit{NewsAnalysis} to analyze the effectiveness of buffering mechanism:
\begin{Verbatim}[commandchars=+\[\]]
rawNews := +textbf[executeSQL]("News", "select id as newsid, 
            news as newsText from usnewspaper 
            where src = $src limit 1000");
processedNews := +textbf[tokenize](rawNews.newsText, 
            docid=rawNews.newsid,
            stopwords="stopwords.txt");
\end{Verbatim}
This snippet is  translated to a   chain of physical operators where a buffer mechanism can be added to any two successive operators: \textit{ExecuteSQL} $\to$ \textit{CreateCorpusFromColumn} $\to$ \textit{SplitByPatterns} $\to$ \textit{FilterStopWords}. The number of documents retrieved by the SQL query is set as 1 million  by changing the ``limit'' clause. The memory footprint is presented in Figure \ref{fig:memory}.  Adding buffering mechanism decreases the used heap size from around 27 GB to about 17 GB ($\sim 37\%$ reduction) with a small run-time overhead ($\sim 8\%$ increase).

\vspace{-1ex}
\section{Related Work}
\label{sec:related}
We present the features comparison of selected  prior polystore systems and AWESOME in Table~\ref{tab:relatedwork}. 
Besides the systems shown here, there are some existing work~\cite{lucas2018rheemstudio, kruse2020rheemix, khayyat2015bigdansing, gog2015musketeer, doka2016mix} which focus on integrating multiple data processing platforms such as Spark, Hadoop, GraphX to process heterogeneous data. However, they do not focus on DBMSs, and hence are not discussed in detail in this paper.  
We conclude some important language and system design features for analytical polystores based on our experience working with domain scientists.   As the table suggests, none of the prior polystore systems has all of these features. 
\vspace{-1ex}
\subsection{Polystore Languages}
BigDAWG \cite{duggan2015bigdawg1,she2016bigdawg,duggan2015bigdawg2}, Estocada \cite{alotaibi2019towards,alotaibi2020estocada} and Tatooine \cite{bonaque2016mixed} all support querying backend DBMSs  using native languages.  However, they do not support analytical functions or any control flow logic such as IF-ELSE or For loop.

RheemLatin \cite{kruse2020rheemix,lucas2018rheemstudio,agrawal2018rheem} is an extension from  PigLatin \cite{olston2008pig}. Similar to ADIL, it has its native data models and  grammars; it has Loop operators to support  tasks with iterations.  Users write platform agnostic analysis plan using Rheem keywords which makes it easier to do query optimization but also brings  several deficiencies: 1) For queries against DBMSs, it shows that users can express a SQL query in RheemLatin,  however, it will be hard or impossible to rewrite a Cypher query or Solr query using Rheem keywords. 2)  It does not support analytical functions such LDA or NLP annotations in native. Users can write an analytical function as a combination of RheemLatin keywords, however, it requires a lot of programming efforts and  expert knowledge.

Myria \cite{wang2017myria} provides a hybrid imperative-declarative query language called MyriaL. It consists of a sequence of assignment statements where the left-hand sides  are  relation variables and the right-hand sides can be  a  SQL query, comprehensions, or function calls. It also supports flow control logic such as DO-WHILE structure. However, SQL is the only supported query language.

Hybrid.Poly \cite{podkorytov2017hybrid, podkorytov2019hybrid} provides a hybrid relational
analytical query language stemming from SQL which is
based on their  extended relational model. It uses an SQL-like language where analytical functions (e.g., dot products) are specified in the SELECT clause, but it does not  provide built-in text or graph analytical functions.

\vspace{-1.5ex}
\subsection{Polystore Systems}
All of these polystore systems  support RDBMS, but to the best of our knowledge, none of them  support both graph and  text DBMS. Among all of them, only Hybrid supports in-memory database engines.  We conclude some design details for selected systems.

BigDAWG \cite{duggan2015bigdawg1,she2016bigdawg,duggan2015bigdawg2}
organizes storage engines into a number of islands. An island is composed of a data model, a set of operations and a set of candidate storage engines. It supports heterogeneous data models including relational tables and arrays, and supports cast functions to migrate data from one island with one data model to another island with a different model.  To our best knowledge, BigDAWG does not support graph DBMS or analytical functions.

The Rheemix system \cite{kruse2020rheemix} is a cross-platform system built over diverse engines including PostgreSQL, Spark, Flink and Java Streams. In Rheemix, analytical tasks are specified as a workflow of Rheemix operators (e.g., \texttt{filter, map, groupBy}) that are directly mapped to operators of the underlying systems through operator inflation. Their operators are primitive (fine-granular) operators, which makes optimization easier at the cost of expressiveness.

Hybrid.Poly \cite{podkorytov2017hybrid, podkorytov2019hybrid}, a relationally backed in-memory polystore system,  only uses one consolidated parallel
engine as a back-end to store all data. It decomposes a full analysis into a series of analytical queries where variables are passed by explicitly creating tables of intermediate results. It extends the relational data model with
additional types and operations to support matrix and vector
abstractions and operators over them. Different from AWESOME, it does not support text or graph data models or any on-disk databases.

Our general observation is that, while there is a clear growth in creating and using polystores for different applications, no existing polystore system supports all requisite  features needed by domain scientists to solve a variety of real-world analytical workloads. 
\vspace{-1ex}
\section{Conclusion and Future Work}\label{sec:conclusion}
In this work, we empower an emerging class of large-scale data science workloads over social media data that naturally straddle analytics over relations, graphs, and text.
In contrast to complementary work on polystores that aim for high generality, we build a more specific tri-store system we call AWESOME that offers a succinct domain-specific language on top of standard unistore engines, automatically handles intermediate data, and performs automatic query optimizations. We empirically demonstrate the functionality and efficiency of AWESOME.
As for future work, we plan to pursue new cross-model query optimization opportunities to make AWESOME even faster and also scale to distributed execution.
\begin{acks}
  This work was partly supported by two National Science Foundation Awards (Number \#1909875 and \#1738411). 
\end{acks}

\newpage\clearpage

\bibliographystyle{ACM-Reference-Format}
\bibliography{ref}


\begin{thebibliography}{37}


\ifx \showCODEN    \undefined \def \showCODEN     #1{\unskip}     \fi
\ifx \showDOI      \undefined \def \showDOI       #1{#1}\fi
\ifx \showISBNx    \undefined \def \showISBNx     #1{\unskip}     \fi
\ifx \showISBNxiii \undefined \def \showISBNxiii  #1{\unskip}     \fi
\ifx \showISSN     \undefined \def \showISSN      #1{\unskip}     \fi
\ifx \showLCCN     \undefined \def \showLCCN      #1{\unskip}     \fi
\ifx \shownote     \undefined \def \shownote      #1{#1}          \fi
\ifx \showarticletitle \undefined \def \showarticletitle #1{#1}   \fi
\ifx \showURL      \undefined \def \showURL       {\relax}        \fi
\providecommand\bibfield[2]{#2}
\providecommand\bibinfo[2]{#2}
\providecommand\natexlab[1]{#1}
\providecommand\showeprint[2][]{arXiv:#2}

\bibitem[\protect\citeauthoryear{??}{sol}{2021}]%
        {solr}
 \bibinfo{year}{2021}\natexlab{}.
\newblock \bibinfo{title}{{Apache Solr}}.
\newblock \bibinfo{howpublished}{\url{https://solr.apache.org/}}.
\newblock
\newblock
\shownote{[Online; accessed July 23 2021].}


\bibitem[\protect\citeauthoryear{??}{neo}{2021}]%
        {neo4j}
 \bibinfo{year}{2021}\natexlab{}.
\newblock \bibinfo{title}{{Graph Database Platform | Graph Database Management
  System | Neo4j}}.
\newblock \bibinfo{howpublished}{\url{https://neo4j.com/}}.
\newblock
\newblock
\shownote{[Online; accessed July 23 2021].}


\bibitem[\protect\citeauthoryear{??}{pos}{2021}]%
        {postgres}
 \bibinfo{year}{2021}\natexlab{}.
\newblock \bibinfo{title}{{PostgreSQL: The World's Most Advanced Open Source
  Relational Database}}.
\newblock \bibinfo{howpublished}{\url{https://www.postgresql.org/}}.
\newblock
\newblock
\shownote{[Online; accessed July 23 2021].}


\bibitem[\protect\citeauthoryear{Agrawal, Chawla, Contreras-Rojas, Elmagarmid,
  Idris, Kaoudi, Kruse, Lucas, Mansour, Ouzzani, et~al\mbox{.}}{Agrawal
  et~al\mbox{.}}{2018}]%
        {agrawal2018rheem}
\bibfield{author}{\bibinfo{person}{Divy Agrawal}, \bibinfo{person}{Sanjay
  Chawla}, \bibinfo{person}{Bertty Contreras-Rojas}, \bibinfo{person}{Ahmed
  Elmagarmid}, \bibinfo{person}{Yasser Idris}, \bibinfo{person}{Zoi Kaoudi},
  \bibinfo{person}{Sebastian Kruse}, \bibinfo{person}{Ji Lucas},
  \bibinfo{person}{Essam Mansour}, \bibinfo{person}{Mourad Ouzzani},
  {et~al\mbox{.}}} \bibinfo{year}{2018}\natexlab{}.
\newblock \showarticletitle{RHEEM: enabling cross-platform data processing: may
  the big data be with you!}
\newblock \bibinfo{journal}{\emph{Proceedings of the VLDB Endowment}}
  \bibinfo{volume}{11}, \bibinfo{number}{11} (\bibinfo{year}{2018}),
  \bibinfo{pages}{1414--1427}.
\newblock


\bibitem[\protect\citeauthoryear{Alotaibi, Bursztyn, Deutsch, Manolescu, and
  Zampetakis}{Alotaibi et~al\mbox{.}}{2019}]%
        {alotaibi2019towards}
\bibfield{author}{\bibinfo{person}{Rana Alotaibi}, \bibinfo{person}{Damian
  Bursztyn}, \bibinfo{person}{Alin Deutsch}, \bibinfo{person}{Ioana Manolescu},
  {and} \bibinfo{person}{Stamatis Zampetakis}.}
  \bibinfo{year}{2019}\natexlab{}.
\newblock \showarticletitle{Towards scalable hybrid stores: constraint-based
  rewriting to the rescue}. In \bibinfo{booktitle}{\emph{Proceedings of the
  2019 International Conference on Management of Data}}.
  \bibinfo{pages}{1660--1677}.
\newblock


\bibitem[\protect\citeauthoryear{Alotaibi, Cautis, Deutsch, Latrache,
  Manolescu, and Yang}{Alotaibi et~al\mbox{.}}{2020}]%
        {alotaibi2020estocada}
\bibfield{author}{\bibinfo{person}{Rana Alotaibi}, \bibinfo{person}{Bogdan
  Cautis}, \bibinfo{person}{Alin Deutsch}, \bibinfo{person}{Moustafa Latrache},
  \bibinfo{person}{Ioana Manolescu}, {and} \bibinfo{person}{Yifei Yang}.}
  \bibinfo{year}{2020}\natexlab{}.
\newblock \showarticletitle{ESTOCADA: towards scalable polystore systems}.
\newblock \bibinfo{journal}{\emph{Proceedings of the VLDB Endowment}}
  \bibinfo{volume}{13}, \bibinfo{number}{12} (\bibinfo{year}{2020}),
  \bibinfo{pages}{2949--2952}.
\newblock


\bibitem[\protect\citeauthoryear{Bonaque, Cao, Cautis, Goasdou{\'e}, Letelier,
  Manolescu, Mendoza, Ribeiro, Tannier, and Thomazo}{Bonaque
  et~al\mbox{.}}{2016}]%
        {bonaque2016mixed}
\bibfield{author}{\bibinfo{person}{Rapha{\"e}l Bonaque},
  \bibinfo{person}{Tien~Duc Cao}, \bibinfo{person}{Bogdan Cautis},
  \bibinfo{person}{Fran{\c{c}}ois Goasdou{\'e}}, \bibinfo{person}{Javier
  Letelier}, \bibinfo{person}{Ioana Manolescu}, \bibinfo{person}{Oscar
  Mendoza}, \bibinfo{person}{Swen Ribeiro}, \bibinfo{person}{Xavier Tannier},
  {and} \bibinfo{person}{Micha{\"e}l Thomazo}.}
  \bibinfo{year}{2016}\natexlab{}.
\newblock \showarticletitle{Mixed-instance querying: a lightweight integration
  architecture for data journalism}. In \bibinfo{booktitle}{\emph{VLDB}}.
\newblock


\bibitem[\protect\citeauthoryear{Dasgupta, Coakley, and Gupta}{Dasgupta
  et~al\mbox{.}}{2016}]%
        {gupta:awesome:2016}
\bibfield{author}{\bibinfo{person}{Subhasis Dasgupta}, \bibinfo{person}{Kevin
  Coakley}, {and} \bibinfo{person}{Amarnath Gupta}.}
  \bibinfo{year}{2016}\natexlab{}.
\newblock \showarticletitle{Analytics-driven data ingestion and derivation in
  the AWESOME polystore}. In \bibinfo{booktitle}{\emph{2016 IEEE International
  Conference on Big Data (Big Data)}}. IEEE, \bibinfo{pages}{2555--2564}.
\newblock


\bibitem[\protect\citeauthoryear{Doka, Papailiou, Giannakouris, Tsoumakos, and
  Koziris}{Doka et~al\mbox{.}}{2016}]%
        {doka2016mix}
\bibfield{author}{\bibinfo{person}{Katerina Doka}, \bibinfo{person}{Nikolaos
  Papailiou}, \bibinfo{person}{Victor Giannakouris}, \bibinfo{person}{Dimitrios
  Tsoumakos}, {and} \bibinfo{person}{Nectarios Koziris}.}
  \bibinfo{year}{2016}\natexlab{}.
\newblock \showarticletitle{Mix ‘n’match multi-engine analytics}. In
  \bibinfo{booktitle}{\emph{2016 IEEE International Conference on Big Data (Big
  Data)}}. IEEE, \bibinfo{pages}{194--203}.
\newblock


\bibitem[\protect\citeauthoryear{Duggan, Elmore, Kraska, Madden, Mattson, and
  Stonebraker}{Duggan et~al\mbox{.}}{2015a}]%
        {duggan2015bigdawg1}
\bibfield{author}{\bibinfo{person}{Jennie Duggan}, \bibinfo{person}{Aaron
  Elmore}, \bibinfo{person}{Tim Kraska}, \bibinfo{person}{Sam Madden},
  \bibinfo{person}{Tim Mattson}, {and} \bibinfo{person}{Michael Stonebraker}.}
  \bibinfo{year}{2015}\natexlab{a}.
\newblock \showarticletitle{The bigdawg architecture and reference
  implementation}.
\newblock \bibinfo{journal}{\emph{New England Database Day}}
  (\bibinfo{year}{2015}).
\newblock


\bibitem[\protect\citeauthoryear{Duggan, Elmore, Stonebraker, Balazinska, Howe,
  Kepner, Madden, Maier, Mattson, and Zdonik}{Duggan et~al\mbox{.}}{2015b}]%
        {duggan2015bigdawg2}
\bibfield{author}{\bibinfo{person}{Jennie Duggan}, \bibinfo{person}{Aaron~J
  Elmore}, \bibinfo{person}{Michael Stonebraker}, \bibinfo{person}{Magda
  Balazinska}, \bibinfo{person}{Bill Howe}, \bibinfo{person}{Jeremy Kepner},
  \bibinfo{person}{Sam Madden}, \bibinfo{person}{David Maier},
  \bibinfo{person}{Tim Mattson}, {and} \bibinfo{person}{Stan Zdonik}.}
  \bibinfo{year}{2015}\natexlab{b}.
\newblock \showarticletitle{The bigdawg polystore system}.
\newblock \bibinfo{journal}{\emph{{ACM SIGMOD} Record}} \bibinfo{volume}{44},
  \bibinfo{number}{2} (\bibinfo{year}{2015}), \bibinfo{pages}{11--16}.
\newblock


\bibitem[\protect\citeauthoryear{Duplyakin, Ricci, Maricq, Wong, Duerig, Eide,
  Stoller, Hibler, Johnson, Webb, et~al\mbox{.}}{Duplyakin
  et~al\mbox{.}}{2019}]%
        {duplyakin2019design}
\bibfield{author}{\bibinfo{person}{Dmitry Duplyakin}, \bibinfo{person}{Robert
  Ricci}, \bibinfo{person}{Aleksander Maricq}, \bibinfo{person}{Gary Wong},
  \bibinfo{person}{Jonathon Duerig}, \bibinfo{person}{Eric Eide},
  \bibinfo{person}{Leigh Stoller}, \bibinfo{person}{Mike Hibler},
  \bibinfo{person}{David Johnson}, \bibinfo{person}{Kirk Webb},
  {et~al\mbox{.}}} \bibinfo{year}{2019}\natexlab{}.
\newblock \showarticletitle{The Design and Operation of $\{$CloudLab$\}$}. In
  \bibinfo{booktitle}{\emph{2019 USENIX annual technical conference (USENIX ATC
  19)}}. \bibinfo{pages}{1--14}.
\newblock


\bibitem[\protect\citeauthoryear{Francis, Green, Guagliardo, Libkin, Lindaaker,
  Marsault, Plantikow, Rydberg, Selmer, and Taylor}{Francis
  et~al\mbox{.}}{2018}]%
        {francis2018cypher}
\bibfield{author}{\bibinfo{person}{Nadime Francis}, \bibinfo{person}{Alastair
  Green}, \bibinfo{person}{Paolo Guagliardo}, \bibinfo{person}{Leonid Libkin},
  \bibinfo{person}{Tobias Lindaaker}, \bibinfo{person}{Victor Marsault},
  \bibinfo{person}{Stefan Plantikow}, \bibinfo{person}{Mats Rydberg},
  \bibinfo{person}{Petra Selmer}, {and} \bibinfo{person}{Andr{\'e}s Taylor}.}
  \bibinfo{year}{2018}\natexlab{}.
\newblock \showarticletitle{Cypher: An evolving query language for property
  graphs}. In \bibinfo{booktitle}{\emph{Proceedings of the 2018 International
  Conference on Management of Data}}. \bibinfo{pages}{1433--1445}.
\newblock


\bibitem[\protect\citeauthoryear{Gog, Schwarzkopf, Crooks, Grosvenor, Clement,
  and Hand}{Gog et~al\mbox{.}}{2015}]%
        {gog2015musketeer}
\bibfield{author}{\bibinfo{person}{Ionel Gog}, \bibinfo{person}{Malte
  Schwarzkopf}, \bibinfo{person}{Natacha Crooks}, \bibinfo{person}{Matthew~P
  Grosvenor}, \bibinfo{person}{Allen Clement}, {and} \bibinfo{person}{Steven
  Hand}.} \bibinfo{year}{2015}\natexlab{}.
\newblock \showarticletitle{Musketeer: all for one, one for all in data
  processing systems}. In \bibinfo{booktitle}{\emph{Proceedings of the Tenth
  European Conference on Computer Systems}}. \bibinfo{pages}{1--16}.
\newblock


\bibitem[\protect\citeauthoryear{Gollapalli and Li}{Gollapalli and Li}{2018}]%
        {DBLP:conf/ictir/GollapalliL18}
\bibfield{author}{\bibinfo{person}{Sujatha~Das Gollapalli} {and}
  \bibinfo{person}{Xiaoli Li}.} \bibinfo{year}{2018}\natexlab{}.
\newblock \showarticletitle{Using PageRank for Characterizing Topic Quality in
  {LDA}}. In \bibinfo{booktitle}{\emph{Proceedings of the 2018 {ACM} {SIGIR}
  International Conference on Theory of Information Retrieval, {ICTIR} 2018,
  Tianjin, China, September 14-17, 2018}},
  \bibfield{editor}{\bibinfo{person}{Dawei Song}, \bibinfo{person}{Tie{-}Yan
  Liu}, \bibinfo{person}{Le~Sun}, \bibinfo{person}{Peter Bruza},
  \bibinfo{person}{Massimo Melucci}, \bibinfo{person}{Fabrizio Sebastiani},
  {and} \bibinfo{person}{Grace~Hui Yang}} (Eds.). \bibinfo{publisher}{{ACM}},
  \bibinfo{pages}{115--122}.
\newblock
\urldef\tempurl%
\url{https://doi.org/10.1145/3234944.3234955}
\showDOI{\tempurl}


\bibitem[\protect\citeauthoryear{Guo, Lu, Zhang, Sun, and Yuan}{Guo
  et~al\mbox{.}}{2020}]%
        {guo2020multi}
\bibfield{author}{\bibinfo{person}{Qingsong Guo}, \bibinfo{person}{Jiaheng Lu},
  \bibinfo{person}{Chao Zhang}, \bibinfo{person}{Calvin Sun}, {and}
  \bibinfo{person}{Steven Yuan}.} \bibinfo{year}{2020}\natexlab{}.
\newblock \showarticletitle{Multi-Model Data Query Languages and Processing
  Paradigms}. In \bibinfo{booktitle}{\emph{Proceedings of the 29th ACM
  International Conference on Information \& Knowledge Management}}.
  \bibinfo{pages}{3505--3506}.
\newblock


\bibitem[\protect\citeauthoryear{Hellerstein, R{\'e}, Schoppmann, Wang,
  Fratkin, Gorajek, Ng, Welton, Feng, Li, et~al\mbox{.}}{Hellerstein
  et~al\mbox{.}}{2012}]%
        {hellerstein2012madlib}
\bibfield{author}{\bibinfo{person}{Joe Hellerstein},
  \bibinfo{person}{Christopher R{\'e}}, \bibinfo{person}{Florian Schoppmann},
  \bibinfo{person}{Daisy~Zhe Wang}, \bibinfo{person}{Eugene Fratkin},
  \bibinfo{person}{Aleksander Gorajek}, \bibinfo{person}{Kee~Siong Ng},
  \bibinfo{person}{Caleb Welton}, \bibinfo{person}{Xixuan Feng},
  \bibinfo{person}{Kun Li}, {et~al\mbox{.}}} \bibinfo{year}{2012}\natexlab{}.
\newblock \showarticletitle{The MADlib analytics library or MAD skills, the
  SQL}.
\newblock \bibinfo{journal}{\emph{arXiv preprint arXiv:1208.4165}}
  (\bibinfo{year}{2012}).
\newblock


\bibitem[\protect\citeauthoryear{Khan, Zimmermann, Jha, Gadepally, d’Aquin,
  and Sahay}{Khan et~al\mbox{.}}{2019}]%
        {khan2019one}
\bibfield{author}{\bibinfo{person}{Yasar Khan}, \bibinfo{person}{Antoine
  Zimmermann}, \bibinfo{person}{Alokkumar Jha}, \bibinfo{person}{Vijay
  Gadepally}, \bibinfo{person}{Mathieu d’Aquin}, {and}
  \bibinfo{person}{Ratnesh Sahay}.} \bibinfo{year}{2019}\natexlab{}.
\newblock \showarticletitle{One size does not fit all: querying web
  polystores}.
\newblock \bibinfo{journal}{\emph{IEEE Access}}  \bibinfo{volume}{7}
  (\bibinfo{year}{2019}), \bibinfo{pages}{9598--9617}.
\newblock


\bibitem[\protect\citeauthoryear{Khayyat, Ilyas, Jindal, Madden, Ouzzani,
  Papotti, Quian{\'e}-Ruiz, Tang, and Yin}{Khayyat et~al\mbox{.}}{2015}]%
        {khayyat2015bigdansing}
\bibfield{author}{\bibinfo{person}{Zuhair Khayyat}, \bibinfo{person}{Ihab~F
  Ilyas}, \bibinfo{person}{Alekh Jindal}, \bibinfo{person}{Samuel Madden},
  \bibinfo{person}{Mourad Ouzzani}, \bibinfo{person}{Paolo Papotti},
  \bibinfo{person}{Jorge-Arnulfo Quian{\'e}-Ruiz}, \bibinfo{person}{Nan Tang},
  {and} \bibinfo{person}{Si Yin}.} \bibinfo{year}{2015}\natexlab{}.
\newblock \showarticletitle{Bigdansing: A system for big data cleansing}. In
  \bibinfo{booktitle}{\emph{Proceedings of the 2015 ACM SIGMOD international
  conference on management of data}}. \bibinfo{pages}{1215--1230}.
\newblock


\bibitem[\protect\citeauthoryear{Kruse, Kaoudi, Contreras-Rojas, Chawla,
  Naumann, and Quian{\'e}-Ruiz}{Kruse et~al\mbox{.}}{2020}]%
        {kruse2020rheemix}
\bibfield{author}{\bibinfo{person}{Sebastian Kruse}, \bibinfo{person}{Zoi
  Kaoudi}, \bibinfo{person}{Bertty Contreras-Rojas}, \bibinfo{person}{Sanjay
  Chawla}, \bibinfo{person}{Felix Naumann}, {and}
  \bibinfo{person}{Jorge-Arnulfo Quian{\'e}-Ruiz}.}
  \bibinfo{year}{2020}\natexlab{}.
\newblock \showarticletitle{RHEEMix in the data jungle: a cost-based optimizer
  for cross-platform systems}.
\newblock \bibinfo{journal}{\emph{The VLDB Journal}}  \bibinfo{volume}{29}
  (\bibinfo{year}{2020}), \bibinfo{pages}{1287--1310}.
\newblock


\bibitem[\protect\citeauthoryear{Lu and Holubov{\'a}}{Lu and
  Holubov{\'a}}{2019}]%
        {lu2019multi}
\bibfield{author}{\bibinfo{person}{Jiaheng Lu} {and} \bibinfo{person}{Irena
  Holubov{\'a}}.} \bibinfo{year}{2019}\natexlab{}.
\newblock \showarticletitle{Multi-model databases: a new journey to handle the
  variety of data}.
\newblock \bibinfo{journal}{\emph{ACM Computing Surveys (CSUR)}}
  \bibinfo{volume}{52}, \bibinfo{number}{3} (\bibinfo{year}{2019}),
  \bibinfo{pages}{1--38}.
\newblock


\bibitem[\protect\citeauthoryear{Lu and Da~Xu}{Lu and Da~Xu}{2018}]%
        {lu2018internet}
\bibfield{author}{\bibinfo{person}{Yang Lu} {and} \bibinfo{person}{Li Da~Xu}.}
  \bibinfo{year}{2018}\natexlab{}.
\newblock \showarticletitle{Internet of Things (IoT) cybersecurity research: A
  review of current research topics}.
\newblock \bibinfo{journal}{\emph{IEEE Internet of Things Journal}}
  \bibinfo{volume}{6}, \bibinfo{number}{2} (\bibinfo{year}{2018}),
  \bibinfo{pages}{2103--2115}.
\newblock


\bibitem[\protect\citeauthoryear{Lucas, Idris, Contreras-Rojas,
  Quian{\'e}-Ruiz, and Chawla}{Lucas et~al\mbox{.}}{2018}]%
        {lucas2018rheemstudio}
\bibfield{author}{\bibinfo{person}{Ji Lucas}, \bibinfo{person}{Yasser Idris},
  \bibinfo{person}{Bertty Contreras-Rojas}, \bibinfo{person}{Jorge-Arnulfo
  Quian{\'e}-Ruiz}, {and} \bibinfo{person}{Sanjay Chawla}.}
  \bibinfo{year}{2018}\natexlab{}.
\newblock \showarticletitle{RheemStudio: Cross-platform data analytics made
  easy}. In \bibinfo{booktitle}{\emph{2018 IEEE 34th International Conference
  on Data Engineering (ICDE)}}. IEEE, \bibinfo{pages}{1573--1576}.
\newblock


\bibitem[\protect\citeauthoryear{McCandless, Hatcher, Gospodneti{\'c}, and
  Gospodneti{\'c}}{McCandless et~al\mbox{.}}{2010}]%
        {mccandless2010lucene}
\bibfield{author}{\bibinfo{person}{Michael McCandless}, \bibinfo{person}{Erik
  Hatcher}, \bibinfo{person}{Otis Gospodneti{\'c}}, {and} \bibinfo{person}{O
  Gospodneti{\'c}}.} \bibinfo{year}{2010}\natexlab{}.
\newblock \bibinfo{booktitle}{\emph{Lucene in action}}.
  Vol.~\bibinfo{volume}{2}.
\newblock \bibinfo{publisher}{Manning Greenwich}.
\newblock


\bibitem[\protect\citeauthoryear{Moustaka, Vakali, and Anthopoulos}{Moustaka
  et~al\mbox{.}}{2018}]%
        {moustaka2018systematic}
\bibfield{author}{\bibinfo{person}{Vaia Moustaka}, \bibinfo{person}{Athena
  Vakali}, {and} \bibinfo{person}{Leonidas~G Anthopoulos}.}
  \bibinfo{year}{2018}\natexlab{}.
\newblock \showarticletitle{A systematic review for smart city data analytics}.
\newblock \bibinfo{journal}{\emph{ACM Computing Surveys (cSuR)}}
  \bibinfo{volume}{51}, \bibinfo{number}{5} (\bibinfo{year}{2018}),
  \bibinfo{pages}{1--41}.
\newblock


\bibitem[\protect\citeauthoryear{Olston, Reed, Srivastava, Kumar, and
  Tomkins}{Olston et~al\mbox{.}}{2008}]%
        {olston2008pig}
\bibfield{author}{\bibinfo{person}{Christopher Olston},
  \bibinfo{person}{Benjamin Reed}, \bibinfo{person}{Utkarsh Srivastava},
  \bibinfo{person}{Ravi Kumar}, {and} \bibinfo{person}{Andrew Tomkins}.}
  \bibinfo{year}{2008}\natexlab{}.
\newblock \showarticletitle{Pig latin: a not-so-foreign language for data
  processing}. In \bibinfo{booktitle}{\emph{Proceedings of the 2008 ACM SIGMOD
  international conference on Management of data}}.
  \bibinfo{pages}{1099--1110}.
\newblock


\bibitem[\protect\citeauthoryear{Page, Brin, Motwani, and Winograd}{Page
  et~al\mbox{.}}{1999}]%
        {page1999pagerank}
\bibfield{author}{\bibinfo{person}{Lawrence Page}, \bibinfo{person}{Sergey
  Brin}, \bibinfo{person}{Rajeev Motwani}, {and} \bibinfo{person}{Terry
  Winograd}.} \bibinfo{year}{1999}\natexlab{}.
\newblock \bibinfo{booktitle}{\emph{The PageRank citation ranking: Bringing
  order to the web.}}
\newblock \bibinfo{type}{{T}echnical {R}eport}. \bibinfo{institution}{Stanford
  InfoLab}.
\newblock


\bibitem[\protect\citeauthoryear{Podkorytov and Gubanov}{Podkorytov and
  Gubanov}{2019}]%
        {podkorytov2019hybrid}
\bibfield{author}{\bibinfo{person}{Maksim Podkorytov} {and}
  \bibinfo{person}{Michael Gubanov}.} \bibinfo{year}{2019}\natexlab{}.
\newblock \showarticletitle{Hybrid. Poly: A Consolidated Interactive Analytical
  Polystore System}. In \bibinfo{booktitle}{\emph{2019 IEEE 35th International
  Conference on Data Engineering (ICDE)}}. IEEE, \bibinfo{pages}{1996--1999}.
\newblock


\bibitem[\protect\citeauthoryear{Podkorytov, Soderman, and Gubanov}{Podkorytov
  et~al\mbox{.}}{2017}]%
        {podkorytov2017hybrid}
\bibfield{author}{\bibinfo{person}{Maksim Podkorytov}, \bibinfo{person}{Dylan
  Soderman}, {and} \bibinfo{person}{Michael Gubanov}.}
  \bibinfo{year}{2017}\natexlab{}.
\newblock \showarticletitle{Hybrid. poly: An interactive large-scale in-memory
  analytical polystore}. In \bibinfo{booktitle}{\emph{2017 IEEE International
  Conference on Data Mining Workshops (ICDMW)}}. IEEE, \bibinfo{pages}{43--50}.
\newblock


\bibitem[\protect\citeauthoryear{Pritchard, Stephens, and Donnelly}{Pritchard
  et~al\mbox{.}}{2000}]%
        {pritchard2000inference}
\bibfield{author}{\bibinfo{person}{Jonathan~K Pritchard},
  \bibinfo{person}{Matthew Stephens}, {and} \bibinfo{person}{Peter Donnelly}.}
  \bibinfo{year}{2000}\natexlab{}.
\newblock \showarticletitle{Inference of population structure using multilocus
  genotype data}.
\newblock \bibinfo{journal}{\emph{Genetics}} \bibinfo{volume}{155},
  \bibinfo{number}{2} (\bibinfo{year}{2000}), \bibinfo{pages}{945--959}.
\newblock


\bibitem[\protect\citeauthoryear{She, Ravishankar, and Duggan}{She
  et~al\mbox{.}}{2016}]%
        {she2016bigdawg}
\bibfield{author}{\bibinfo{person}{Zuohao She}, \bibinfo{person}{Surabhi
  Ravishankar}, {and} \bibinfo{person}{Jennie Duggan}.}
  \bibinfo{year}{2016}\natexlab{}.
\newblock \showarticletitle{BigDAWG polystore query optimization through
  semantic equivalences}. In \bibinfo{booktitle}{\emph{2016 IEEE High
  Performance Extreme Computing Conference (HPEC)}}. IEEE,
  \bibinfo{pages}{1--6}.
\newblock


\bibitem[\protect\citeauthoryear{Shrestha and Bhalla}{Shrestha and
  Bhalla}{2020}]%
        {shrestha2020survey}
\bibfield{author}{\bibinfo{person}{Shashank Shrestha} {and}
  \bibinfo{person}{Subhash Bhalla}.} \bibinfo{year}{2020}\natexlab{}.
\newblock \showarticletitle{A Survey on the Evolution of Models of Data
  Integration}.
\newblock \bibinfo{journal}{\emph{International Journal of Knowledge Based
  Computer Systems8 (1 \& 2), June \& December}} (\bibinfo{year}{2020}),
  \bibinfo{pages}{11--16}.
\newblock


\bibitem[\protect\citeauthoryear{Simmons, Armstrong, Soderman, and
  Gubanov}{Simmons et~al\mbox{.}}{2017}]%
        {simmons2017hybrid}
\bibfield{author}{\bibinfo{person}{Mark Simmons}, \bibinfo{person}{Daniel
  Armstrong}, \bibinfo{person}{Dylan Soderman}, {and} \bibinfo{person}{Michael
  Gubanov}.} \bibinfo{year}{2017}\natexlab{}.
\newblock \showarticletitle{Hybrid. media: High velocity video ingestion in an
  in-memory scalable analytical polystore}. In \bibinfo{booktitle}{\emph{2017
  IEEE International Conference on Big Data (Big Data)}}. IEEE,
  \bibinfo{pages}{4832--4834}.
\newblock


\bibitem[\protect\citeauthoryear{Wang, Baker, Balazinska, Halperin, Haynes,
  Howe, Hutchison, Jain, Maas, Mehta, et~al\mbox{.}}{Wang
  et~al\mbox{.}}{2017}]%
        {wang2017myria}
\bibfield{author}{\bibinfo{person}{Jingjing Wang}, \bibinfo{person}{Tobin
  Baker}, \bibinfo{person}{Magdalena Balazinska}, \bibinfo{person}{Daniel
  Halperin}, \bibinfo{person}{Brandon Haynes}, \bibinfo{person}{Bill Howe},
  \bibinfo{person}{Dylan Hutchison}, \bibinfo{person}{Shrainik Jain},
  \bibinfo{person}{Ryan Maas}, \bibinfo{person}{Parmita Mehta},
  {et~al\mbox{.}}} \bibinfo{year}{2017}\natexlab{}.
\newblock \showarticletitle{The Myria Big Data Management and Analytics System
  and Cloud Services.}. In \bibinfo{booktitle}{\emph{CIDR}}. Citeseer.
\newblock


\bibitem[\protect\citeauthoryear{Wolff and Schmidt}{Wolff and Schmidt}{2021}]%
        {wolff2021information}
\bibfield{author}{\bibinfo{person}{Christian Wolff} {and}
  \bibinfo{person}{Thomas Schmidt}.} \bibinfo{year}{2021}\natexlab{}.
\newblock \bibinfo{booktitle}{\emph{Information between Data and Knowledge:
  Information Science and its Neighbors from Data Science to Digital
  Humanities}}. Vol.~\bibinfo{volume}{74}.
\newblock \bibinfo{publisher}{Werner H{\"u}lsbusch}.
\newblock


\bibitem[\protect\citeauthoryear{Wu, Liebman, Stern, Roberts, and Gupta}{Wu
  et~al\mbox{.}}{2019}]%
        {wu2019constructing}
\bibfield{author}{\bibinfo{person}{Xiaohan Wu}, \bibinfo{person}{Benjamin~L
  Liebman}, \bibinfo{person}{Rachel~E Stern}, \bibinfo{person}{Margaret~E
  Roberts}, {and} \bibinfo{person}{Amarnath Gupta}.}
  \bibinfo{year}{2019}\natexlab{}.
\newblock \showarticletitle{On Constructing a Knowledge Base of Chinese
  Criminal Cases}.
\newblock In \bibinfo{booktitle}{\emph{Legal Knowledge and Information
  Systems}}. \bibinfo{publisher}{IOS Press}, \bibinfo{pages}{235--240}.
\newblock


\bibitem[\protect\citeauthoryear{Zheng and Gupta}{Zheng and Gupta}{2019}]%
        {zheng2019social}
\bibfield{author}{\bibinfo{person}{Xiuwen Zheng} {and}
  \bibinfo{person}{Amarnath Gupta}.} \bibinfo{year}{2019}\natexlab{}.
\newblock \showarticletitle{Social network of extreme tweeters: A case study}.
  In \bibinfo{booktitle}{\emph{Proceedings of the 2019 IEEE/ACM International
  Conference on Advances in Social Networks Analysis and Mining}}.
  \bibinfo{pages}{302--306}.
\newblock


\end{thebibliography}

\newpage\clearpage

\appendix
\section{SQL script for two workloads}
\subsection{\textit{NewsAnalysis}}
\begin{Small}
\begin{Verbatim}
create function buildgraphfromtext
(text character varying[], distance integer) 
returns character varying[]
    language plpython2u
as
$$
count = {}
for i in range(len(text)-distance):
    for j in range(1, distance):
        temp = (text[i], text[i+j])
        if temp in count:
            count[temp] += 1
        else:
            count[temp] = 1
result = []
for key in count:
    result.append([key[0], key[1], count[key]])
return result
$$;
create function unnest_2d_1d(anyarray) returns SETOF anyarray
    immutable
    strict
    parallel safe
    language sql
as
$$
SELECT array_agg($1[d1][d2])
FROM   generate_subscripts($1,1) d1
    ,  generate_subscripts($1,2) d2
GROUP  BY d1
ORDER  BY d1
$$;

drop table if exists tokenizednews, graph, topicgraph CASCADE;
drop MATERIALIZED VIEW if exists graphelement;
-- set execution begin time
INSERT INTO timenow(type, starttime, stoptime)
SELECT '5k', now(), clock_timestamp();
---- tokenize and build word neighbor graph
CREATE table tokenizednews as (
select id as docid, news from newspaper 
where  src = ' http://www.chicagotribune.com/' 
order by id limit 5000);
ALTER TABLE tokenizednews ADD COLUMN words TEXT[];
UPDATE tokenizednews SET words =
    regexp_split_to_array(lower(
    regexp_replace(news, E'[,.;\']','', 'g')
    ), E'[\\s+]');
    
create MATERIALIZED VIEW graphelement as (
with temp as (
    select unnest_2d_1d(buildgraphfromtext(words, 5)) as n
    from tokenizednews),
temp2 as (
    select n[1] as word1, n[2] as word2, n[3]::INTEGER as cnt
    from temp)
select word1, word2, sum(cnt) from temp2 group by word1, word2
);    
    
---- LDA
DROP TABLE IF EXISTS news_tf, news_tf_vocabulary,lda_model,
lda_output_data, helper_output_table, 
topicgraph,  pagerank_out, pagerank_out_summary;

SELECT madlib.term_frequency(
    'tokenizednews',    -- input table
    'docid',     -- document id column
    'words',     -- vector of words in document
    'news_tf',   -- output test table with term frequency
    TRUE);       -- TRUE to created vocabulary table
SELECT madlib.lda_train( 
    'news_tf',     -- test table in the form of term frequency
    'lda_model',   -- model table created by LDA training   
    'lda_output_data',  -- readable output data table
    200000,         -- vocabulary size
    10,             -- number of topics
    1000,           -- number of iterations
    1,      -- Dirichlet prior for the per-doc topic multinomial
    0.01    -- Dirichlet prior for the per-topic word multinomial
    );
SELECT madlib.lda_get_topic_desc( 
    'lda_model',    -- LDA model generated in training
    'news_tf_vocabulary',  -- vocabulary table that maps wordid to word
    'helper_output_table',  -- output table for per-topic descriptions
    20000);
INSERT INTO timenow(type, starttime, stoptime) 
SELECT 'LDA', now(), clock_timestamp();

--- create a text network graph   
create table graph as (
    select word1, word2 from graphelement where word1!='' and word2!='' 
    group by word1, word2
);
---- build graph for each one
create table topicgraph as (
    with topicwords as 
        (select word,wordid 
        from helper_output_table 
        where prob > 0 and topicid = 0 
        order by prob desc limit 7000),
    temp as (
        select wordid, word2  from graph, topicwords 
        where word1 = word)
        select temp.wordid as word1, topicwords.wordid as word2, 1 as topic 
        from temp, topicwords 
        where temp.word2=word );
insert into topicgraph(word1, word2, topic) (
    with topicwords as 
        (select word,wordid 
        from helper_output_table 
        where prob > 0 and topicid = 1 
        order by prob desc limit 7000),
    temp as (
        select wordid, word2  from graph, topicwords 
        where word1 = word)
        select temp.wordid as word1, topicwords.wordid as word2, 2 as topic 
        from temp, topicwords 
        where temp.word2=word );
insert into topicgraph(word1, word2,  topic) (
    with topicwords as 
        (select word,wordid 
        from helper_output_table 
        where prob > 0 and topicid = 2 
        order by prob desc limit 7000),
    temp as (
        select wordid, word2  from graph, topicwords 
        where word1 = word)
        select temp.wordid as word1, topicwords.wordid as word2, 3 as topic 
        from temp, topicwords 
        where temp.word2=word );
insert into topicgraph(word1, word2,  topic) (
    with topicwords as 
        (select word,wordid 
        from helper_output_table 
        where prob > 0 and topicid = 3 
        order by prob desc limit 7000),
    temp as (
        select wordid, word2  from graph, topicwords 
        where word1 = word)
        select temp.wordid as word1, topicwords.wordid as word2, 4 as topic 
        from temp, topicwords 
        where temp.word2=word );
insert into topicgraph(word1, word2,  topic) (
    with topicwords as 
        (select word,wordid 
        from helper_output_table 
        where prob > 0 and topicid = 4 
        order by prob desc limit 7000),
    temp as (
        select wordid, word2  from graph, topicwords 
        where word1 = word)
        select temp.wordid as word1, topicwords.wordid as word2, 5 as topic 
        from temp, topicwords 
        where temp.word2=word );
insert into topicgraph(word1, word2,   topic) (
with topicwords as 
        (select word,wordid 
        from helper_output_table 
        where prob > 0 and topicid = 5 
        order by prob desc limit 7000),
    temp as (
        select wordid, word2  from graph, topicwords 
        where word1 = word)
        select temp.wordid as word1, topicwords.wordid as word2, 6 as topic 
        from temp, topicwords 
        where temp.word2=word );
insert into topicgraph(word1, word2,   topic) (
    with topicwords as 
        (select word,wordid 
        from helper_output_table 
        where prob > 0 and topicid = 6 
        order by prob desc limit 7000),
    temp as (
        select wordid, word2  from graph, topicwords 
        where word1 = word)
        select temp.wordid as word1, topicwords.wordid as word2, 7 as topic 
        from temp, topicwords 
        where temp.word2=word );
insert into topicgraph(word1, word2,   topic) (
   with topicwords as 
        (select word,wordid 
        from helper_output_table 
        where prob > 0 and topicid = 7
        order by prob desc limit 7000),
    temp as (
        select wordid, word2  from graph, topicwords 
        where word1 = word)
        select temp.wordid as word1, topicwords.wordid as word2, 8 as topic 
        from temp, topicwords 
        where temp.word2=word ); 
insert into topicgraph(word1, word2,   topic) (
with topicwords as 
        (select word,wordid 
        from helper_output_table 
        where prob > 0 and topicid = 8 
        order by prob desc limit 7000),
    temp as (
        select wordid, word2  from graph, topicwords 
        where word1 = word)
        select temp.wordid as word1, topicwords.wordid as word2, 9 as topic 
        from temp, topicwords 
        where temp.word2=word );
insert into topicgraph(word1, word2,   topic) (
with topicwords as 
        (select word,wordid 
        from helper_output_table 
        where prob > 0 and topicid = 9 
        order by prob desc limit 7000),
    temp as (
        select wordid, word2  from graph, topicwords 
        where word1 = word)
        select temp.wordid as word1, topicwords.wordid as word2, 
        10 as topic 
        from temp, topicwords 
        where temp.word2=word );
--- pagerank for each topic
SELECT madlib.pagerank(
        'news_tf_vocabulary',     -- Vertex table
        'wordid',                 -- Vertix id column
        'topicgraph',               -- Edge table
        'src=word1, dest=word2', -- Comma delimted string of 
        'pagerank_out',       -- Output table of PageRank
         NULL,                 -- Default damping factor (0.85)
         NULL,                 -- Default max iters (100)
         0.00000001,           -- Threshold
        'topic');
INSERT INTO timenow(type, starttime, stoptime) 
SELECT '5k_0', now(), clock_timestamp();
\end{Verbatim}
\end{Small}

\subsection{\textit{PoliSci}}
\begin{Small}
\begin{Verbatim}
create function callner (tname character varying, 
colname character varying, 
filename character varying) 
returns character varying
    language plpython3u
as
$$
import os
import subprocess
with open(filename, 'w') as f:
    for row in plpy.cursor("SELECT " + colname + " FROM "+tname):
        f.write(row[colname]+'\n')
temp_file = filename.split(".")[0]
subprocess.call(['java', '-jar', 
    '/var/lib/postgresql/data/ner/target/NER-1.0-SNAPSHOT.jar', 
    '-i', filename, '-o', temp_file])
return temp_file
$$;
drop table if exists keynews, keyusers, namedentity, timenow;
--- record start time 
INSERT INTO timenow( type, starttime, stoptime) 
    SELECT 'ner_start', now(), clock_timestamp();
create table keynews as (
    select news from newspaper 
    where news @@ to_tsquery('corona|covid|pandemic|vaccine') 
    limit 5000);
select callner('xw_keynews', 'news', 'news.txt');
CREATE TABLE namedentity (
  type text,
  entity text
);
COPY namedentity(type, entity)
FROM 'news'
DELIMITER ','
CSV HEADER;
create table keyusers as (
select distinct t.name as name, t.twittername as twittername
              from twitterhandle t,
                   namedentity e
              where LOWER(e.entity) = LOWER(t.name));
--- record ner time
INSERT INTO timenow( type, starttime, stoptime) 
    SELECT 'ner_end', now(), clock_timestamp();
select text from neo4j_tweet50000, keyusers 
    where text ilike '%' || keyusers.name || '%';
select * from neo4j_user_user50000 
    where userid2 in (
        select userid from neo4j_user50000 
        where username in (select twittername from keyusers));
--- record end time 
INSERT INTO timenow( type, starttime, stoptime) 
    SELECT 'all', now(), clock_timestamp();
\end{Verbatim}
\end{Small}
\newpage

\section{ADIL Script for Workload  \textit{NewsAnalysis}}\label{appendix:script}
\begin{figure}[ht]
\begin{Verbatim}[commandchars=\&\~\!]
/*specify configuration file*/
&textbf~USE! newsDB;
/* main code block */
&textbf~create! analysis NewsAnalysis &textbf~as! (
src :=  "http://www.chicagotribune.com/";
rawNews := &textbf~executeSQL!("News", 
        "select id as newsid, news as newsText 
         from newspaper 
         where src = $src limit 1000");
processedNews := &textbf~preprocess!(rawNews.newsText, 
        docid=rawNews.newsid,
        stopwords="stopwords.txt");
numTop := 10;
DTM, WTM := &textbf~lda!(processedNews, docid=true, topic=numTop);
topicID := [range(0, numberTopic, 1)];
wtmPerTopic := topicID.&textbf~map!(i =>
        WTM where getValue(_:Row, i) > 0.00);
wordsPerTopic := wtmPerTopic.&textbf~map!(i => rowNames(i));
wordsOfInterest := &textbf~union!(wordsPerTopic);
G := &textbf~buildWordNeighborGraph!(processedNews,
        maxDistance=5, splitter=".", 
        words=wordsOfInterest);
relationPerTopic := wordsPerTopic.&textbf~map!(words => 
        (<n:String, m:String, count:Integer>)
        &textbf~executeCypher!(G,  "match (n)-[r]->(m) 
         where n in $words and m in $words 
         return n, m, r.count as count"));
graphPerTopic := relationPerTopic.&textbf~map!(r =>
    &textbf~ConstructGraphFromRelation!(r,
    (:Word {id: r.n})-[:Cooccur{count: r.count}]->
    (:Word{id: r.m})));
scores := graphPerTopic.&textbf~map!(g => 
        &textbf~pageRank!(g, topk=true, num=20));
aggregatePT := scores.&textbf~map!(i => &textbf~sum!(i.pagerank));
/* store a list to rDBMS as a relation*/
&textbf~store!(aggregatePT t, dbName="News", 
  tableName="aggregatePageRankofTopk", 
  columnName=[("id",t.index), ("pagerank",t.value)]);
\end{Verbatim}
    \caption{NewsAnalysis workload written in ADIL script.}\label{fig:firstworkload}
\end{figure}

\newpage \clearpage
\section{Python code for two workloads}
\subsection{\textit{NewsAnalysis}}
\begin{Small}
\begin{Verbatim}
import getopt
import io
import numpy as np
import sys
import time
from multiprocessing import Pool
import math
import networkx as nx
import sqlalchemy as sal
from gensim import corpora
from gensim.models.ldamulticore import LdaMulticore
from gensim.models.wrappers import LdaMallet
from sqlalchemy import text


# tokenize
def tokenize(doc):
    return doc.split(" ")

def build_graph_from_text(docs, dis, words):
    count = {}
    for doc in docs:
        for i in range(len(doc) - dis):
            if doc[i] in words:
                for j in range(1, dis):
                    tempPair = (min(doc[i], doc[i + j]), 
                                max(doc[i], doc[i + j]))
                    if doc[i + j] in words:
                        if tempPair in count:
                            count[tempPair] += 1
                        else:
                            count[tempPair] = 1
    return count


def split(list_a, chunk_size):
    for idx in range(0, len(list_a), chunk_size):
        yield list_a[idx:idx + chunk_size]


# LDA
def LDA(docs):
    id2word = corpora.Dictionary(docs)
    corpus = [id2word.doc2bow(text) for text in docs]
    model = LdaMulticore(corpus=corpus, num_topics=10, 
                         iterations=1000, id2word=id2word,workers=15)
    return model.show_topics(num_words=len(id2word))


def LDA_mallet(docs, threshold):
    path_to_mallet_binary = "/users/Xiuwen/Mallet/bin/mallet"
    id2word = corpora.Dictionary(docs)
    corpus = [id2word.doc2bow(text) for text in docs]
    model = LdaMallet(path_to_mallet_binary, corpus=corpus, num_topics=10, 
                        iterations=1000, id2word=id2word, 
                        random_seed=2, alpha=0.1, workers=96)
    matrix = model.get_topics()
    words = []
    for row in matrix:
        words_ids = np.argsort(row)[-threshold :]
        words.append(set([id2word[w] for w in words_ids]))
    return words

def page_rank(graph_data, num_of_point):
    G = nx.Graph()
    for i in graph_data:
        G.add_edge(i[0], i[1], weight=i[2])
    pr = nx.pagerank(G)
    return sorted(pr.items(), key=lambda val: val[1], 
                  reverse=True)[:num_of_point]


if __name__ == '__main__':
    num_of_docs = ""
    threshold = ""
    argv = sys.argv[1:]
    core = 1
    try:
        opts, args = getopt.getopt(argv, "i:t:c:")
    except getopt.GetoptError:
        print('query1.py -i <size> -t <threshold> -c <cores>')
        sys.exit(2)
    for opt, arg in opts:
        if opt == "-i":
            num_of_docs = arg
        elif opt == "-t":
            threshold = arg
        elif opt == "-c":
            core = int(arg)
    if num_of_docs == "" or threshold == "":
        print('query1.py -i <size> -t <threshold> -c <cores>')
        sys.exit(2)
    start = time.time()
    # read data
    engine = sal.create_engine('postgresql+psycopg2://')
    conn = engine.connect()
    sql = text("select newstext from xw_news_"+num_of_docs)
    result = conn.execute(sql)
    sql_exe = time.time()
    print("sql execution time: " + str(sql_exe - start))
    tokenized_docs = [tokenize(i[0]) for i in result]
    tk_exe = time.time()
    print("tokenize execution time: " + str(tk_exe - sql_exe))
    # read LDA results
    path = "/proj/awesome-PG0/data/"
    if num_of_docs=='5000':
        path = path + "5k/"
    else:
        path = path + "50k/"
    # path = "C://Users//xiuwen//Documents//"
    # get only partial words
    words_index_per_topic = []
    words_file = open(path+'sortedwords.txt', 'r')
    words_lines = words_file.readlines()
    for words in words_lines:
        words_index = [int(i) for i in words.strip().split(", ")][:int(threshold)]
        words_index_per_topic.append(words_index)
    alphabet_file = io.open(path +'alphabet.txt', 'r', encoding='utf-8')
    alphabet = alphabet_file.readline().strip().split(", ")
    words_per_topic = [set([alphabet[i] for i in index]) 
                       for index in words_index_per_topic]
    # get all words
    all_words = list(set.union(*words_per_topic))
    print("size of keywords after union: " + str(len(all_words)))
    lda_exe = time.time()
    print("lda execution time: " + str(lda_exe - tk_exe))
    pool = Pool(processes=core)
    jobs = []
    # split data to the number of cores partitions
    size = int(math.ceil(float(len(tokenized_docs)) / core))
    count_threads = []
    sublists = list(split(tokenized_docs, size))
    print(len(sublists))
    for alist in sublists:
        count_threads.append(pool.apply_async(
            build_graph_from_text, (alist, 5, all_words)))
    pool.close()
    pool.join()
    graph_elements = []
    total_counts = {}
    for c_count in count_threads:
        c_count_map = c_count.get()
        for key in c_count_map:
            if key in total_counts:
                total_counts[key] += c_count_map[key]
            else:
                total_counts[key] = c_count_map[key]
    for key in total_counts:
        graph_elements.append([key[0], key[1], total_counts[key]])
    print (graph_elements[:10])
    bg_exe = time.time()
    print("bg execution time: " + str(bg_exe - lda_exe))
    graph_data_per_topic = []
    # get graph data for each topic
    for i in range(10):
        words_in_this_topic = words_per_topic[i]
        temp_graph = [g for g in graph_elements if g[0] 
                      in words_in_this_topic and g[1] in words_in_this_topic]
        graph_data_per_topic.append(temp_graph)
    bsg_exe = time.time()
    print("bg execution time: " + str(bsg_exe - bg_exe))

    # get pagerank for each topic
    pagerank_all_topics = [page_rank(i, 20) for i in graph_data_per_topic]
    print(pagerank_all_topics)
    end = time.time()
    print("pr execution time: " + str(end - bsg_exe))
    print(end - start)
\end{Verbatim}
\end{Small}

\subsection{\textit{PoliSci}}
\begin{Small}
\begin{Verbatim}
import time
import sqlalchemy as sal
from sner import Ner
import getopt
import sys

if __name__ == '__main__':
    num_of_docs = ""
    tweet = ""
    argv = sys.argv[1:]
    core = 1
    try:
        opts, args = getopt.\
            getopt(argv, "i:t:c:")
    except getopt.GetoptError:
        print('query2.py -i <size> -t <tweet>')
        sys.exit(2)
    for opt, arg in opts:
        if opt == "-i":
            num_of_docs = arg
        elif opt == "-t":
            tweet = arg
    if num_of_docs == "" or tweet == "":
        print('query1.py -i <size> -t <threshold> -c <cores>')
        sys.exit(2)

    start = time.time()
    # sql query without full text search index
    sql = "select news from usnewspaper where news " \
          "@@ to_tsquery('corona|covid|pandemic|vaccine') limit " \
          + num_of_docs
    engine = sal.create_engine('postgresql+psycopg2://')
    conn = engine.connect()
    result = conn.execute(sql)
    docs = [i[0] for i in result]
    # print([ner(i[0]) for i in result])
    print("full text search cost: " + str(time.time() - start))
    # NER
    nes = []
    tagger = Ner(host='localhost', port=9299)
    for d in docs:
        try:
            en = tagger.get_entities(d)
            nes.extend(en)
        except:
            continue
    key_nes = set([i[0].lower() for i in nes if i[1] != 'O'])
    # get senators
    sql = "select name, twittername from twitterhandle"
    conn = engine.connect()
    result = conn.execute(sql)
    senators_name_tn = [[i[0].lower(), i[1]] for i in result]
    # get userid-username
    sql = "select userid, username from xw_neo4j_user"+tweet
    conn = engine.connect()
    result = conn.execute(sql)
    user_id_name = [[i[0], i[1]] for i in result]
    # get user-tweet network
    sql = "select text from xw_neo4j_tweet"+tweet
    conn = engine.connect()
    result = conn.execute(sql)
    texts = [i[0].lower() for i in result]
    # get user-user network
    sql = "select userid1, userid2 from xw_neo4j_user_user"+tweet
    conn = engine.connect()
    result = conn.execute(sql)
    users_users = [[i[0], i[1]] for i in result]
    # get key users name and id
    key_names = set()
    key_users = set()
    for i in key_nes:
        for s in senators_name_tn:
            name = [i for i in s[0].lower().split(" ") if len(i) > 2]
            if i in name:
                key_names.add(i)
                key_users.add(s[1])
    key_users_ids = []
    for i in key_users:
        for j in user_id_name:
            if i == j[1]:
                key_users_ids.append(j[0])
        # get tweets that contain key users names
    key_tweets = []
    for t in texts:
        for i in key_names:
            if i in t.split(" "):
                key_tweets.append(t)
                break
    # get users that mention key user id
    second_key_users = []
    for i in key_users_ids:
        for j in users_users:
            if j[1] == i:
                second_key_users.append(j[0])
    print("total cost: " + str(time.time() - start))
    print(len(second_key_users))
    print(len(key_tweets))
\end{Verbatim}
\end{Small}

\newpage\clearpage
\section{Calibration results.}\label{appendix:benchmark}

\begin{figure}[ht]
    \centering
    \begin{subfigure}[b]{0.23\textwidth}
        \centering
        \includegraphics[width=\textwidth]{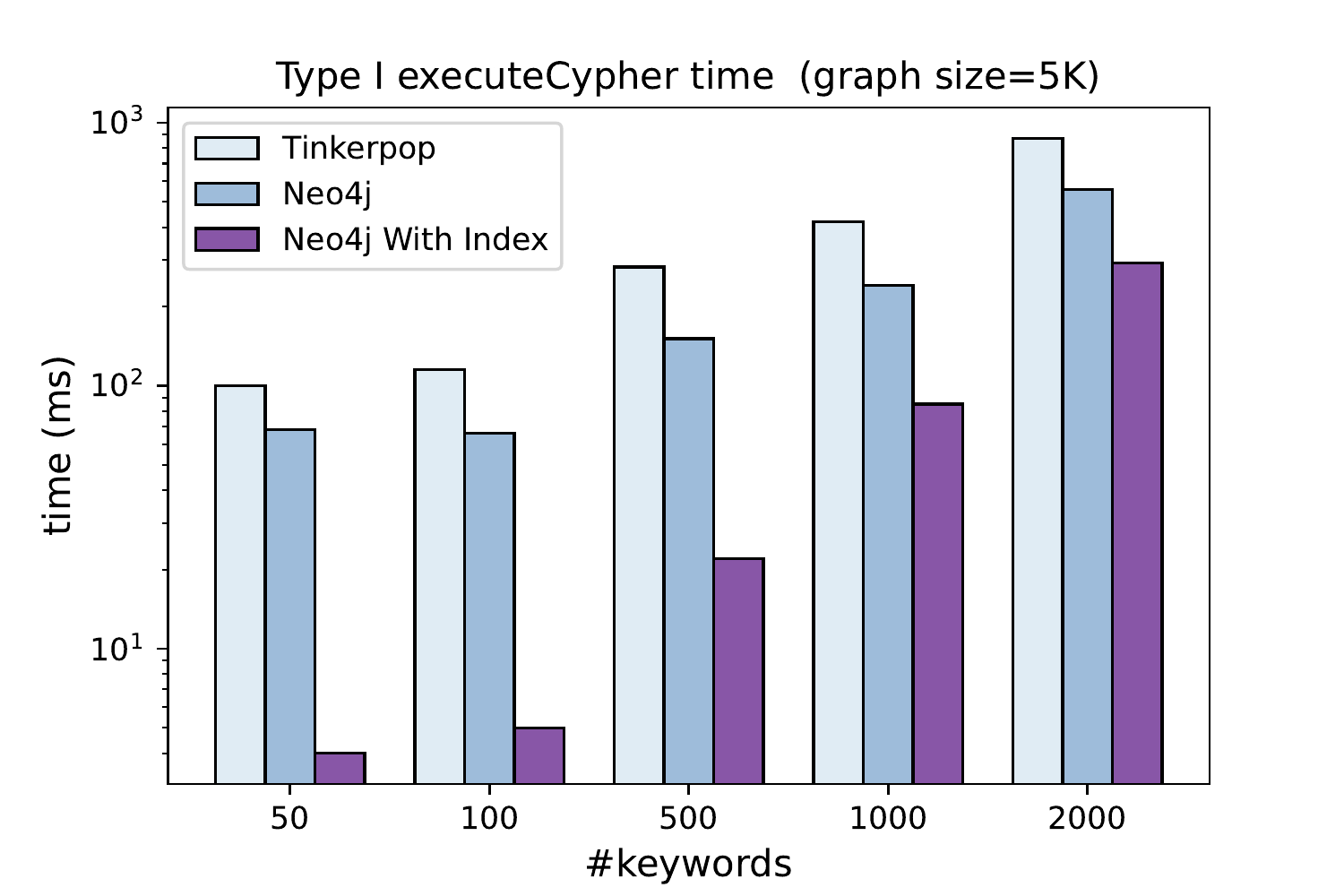}
        \caption{Graph size $=5K$.}
    \end{subfigure}
     \begin{subfigure}[b]{0.23\textwidth}
        \centering
        \includegraphics[width=\textwidth]{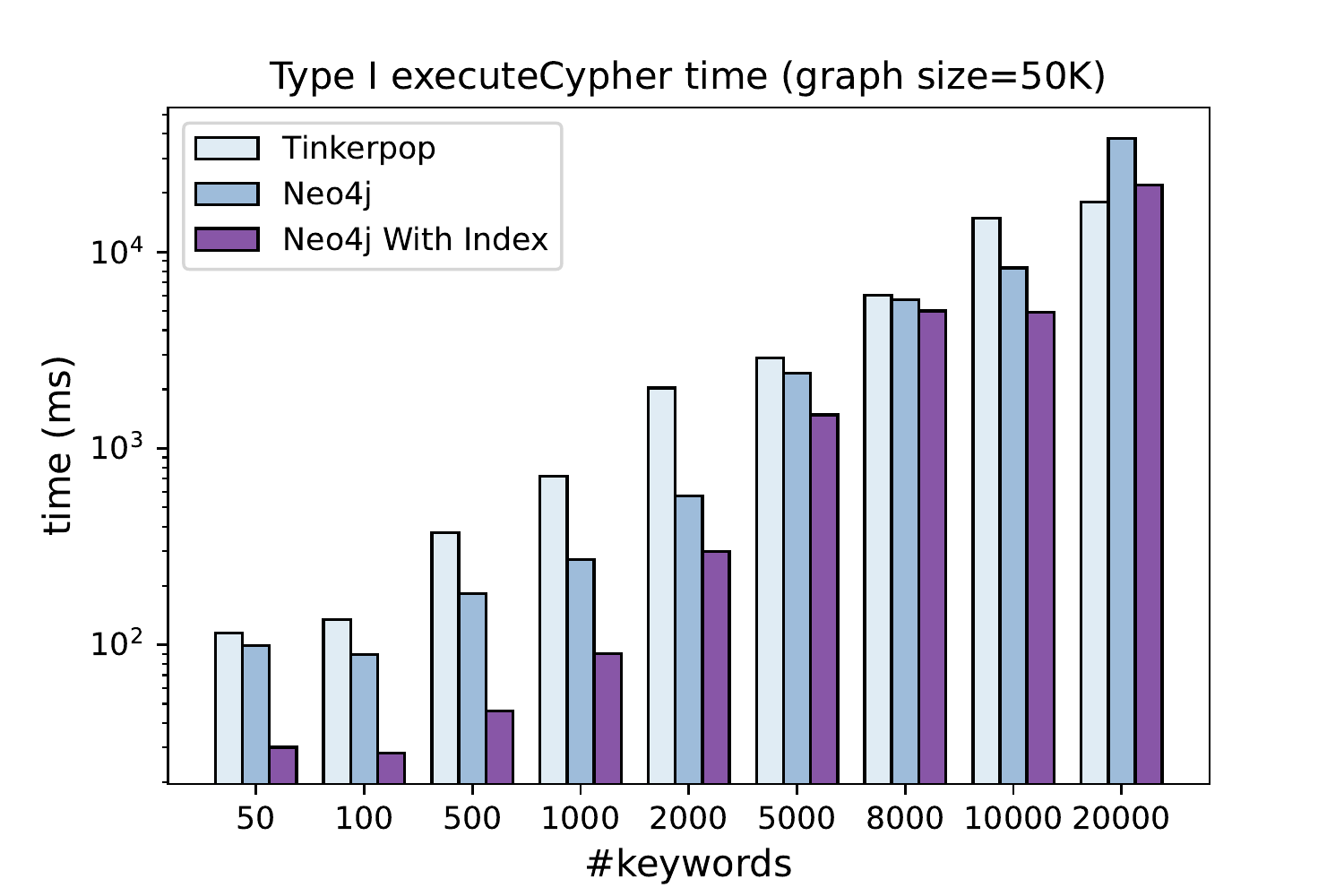}
        \caption{Graph size $=50K$.}
    \end{subfigure}
    \hfill
     \begin{subfigure}[b]{0.23\textwidth}
        \centering
        \includegraphics[width=\textwidth]{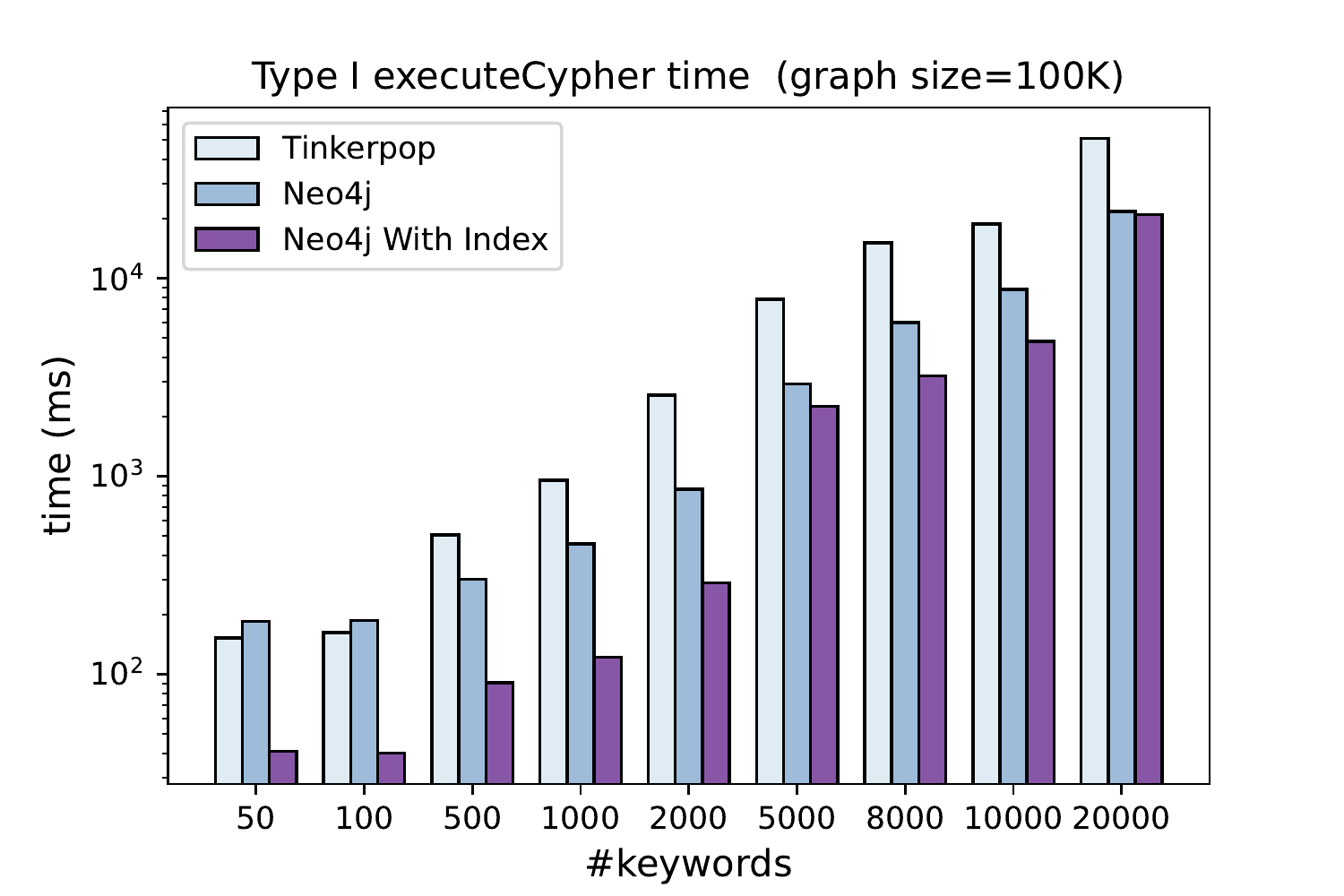}
        \caption{Graph size $=100K$.}
    \end{subfigure}
    \begin{subfigure}[b]{0.23\textwidth}
        \centering
        \includegraphics[width=\textwidth]{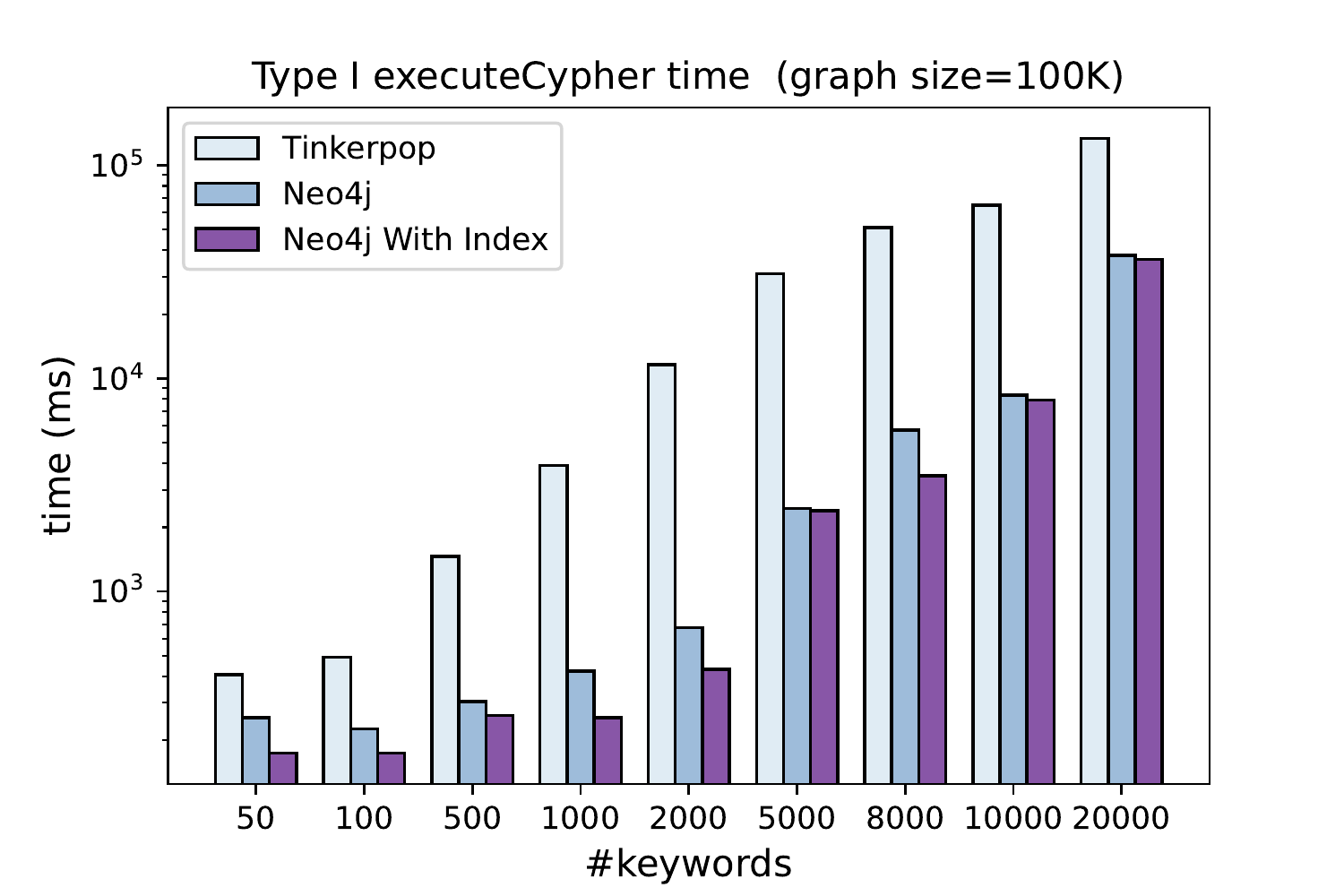}
        \caption{Graph size $=500K$.}
    \end{subfigure}
    \caption{Calibration results for Type I Cypher query w.r.t.~different graph sizes and \#keywords.}
    \label{fig:benchmarktype1}
\end{figure}

\begin{figure}[ht]
    \centering
    \begin{subfigure}[b]{0.23\textwidth}
        \centering
        \includegraphics[width=\textwidth]{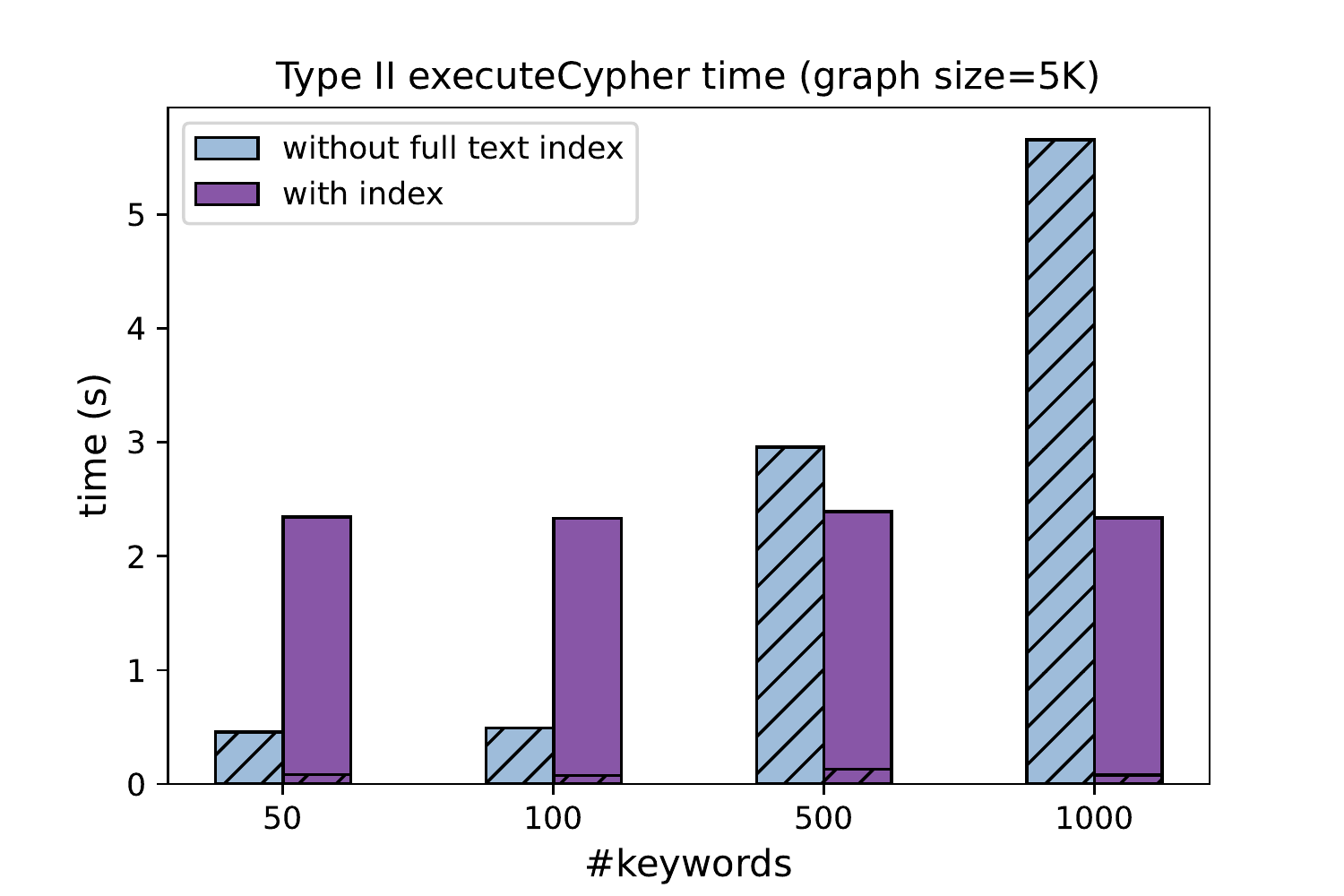}
        \caption{Graph size $=5K$.}
    \end{subfigure}
     \begin{subfigure}[b]{0.23\textwidth}
        \centering
        \includegraphics[width=\textwidth]{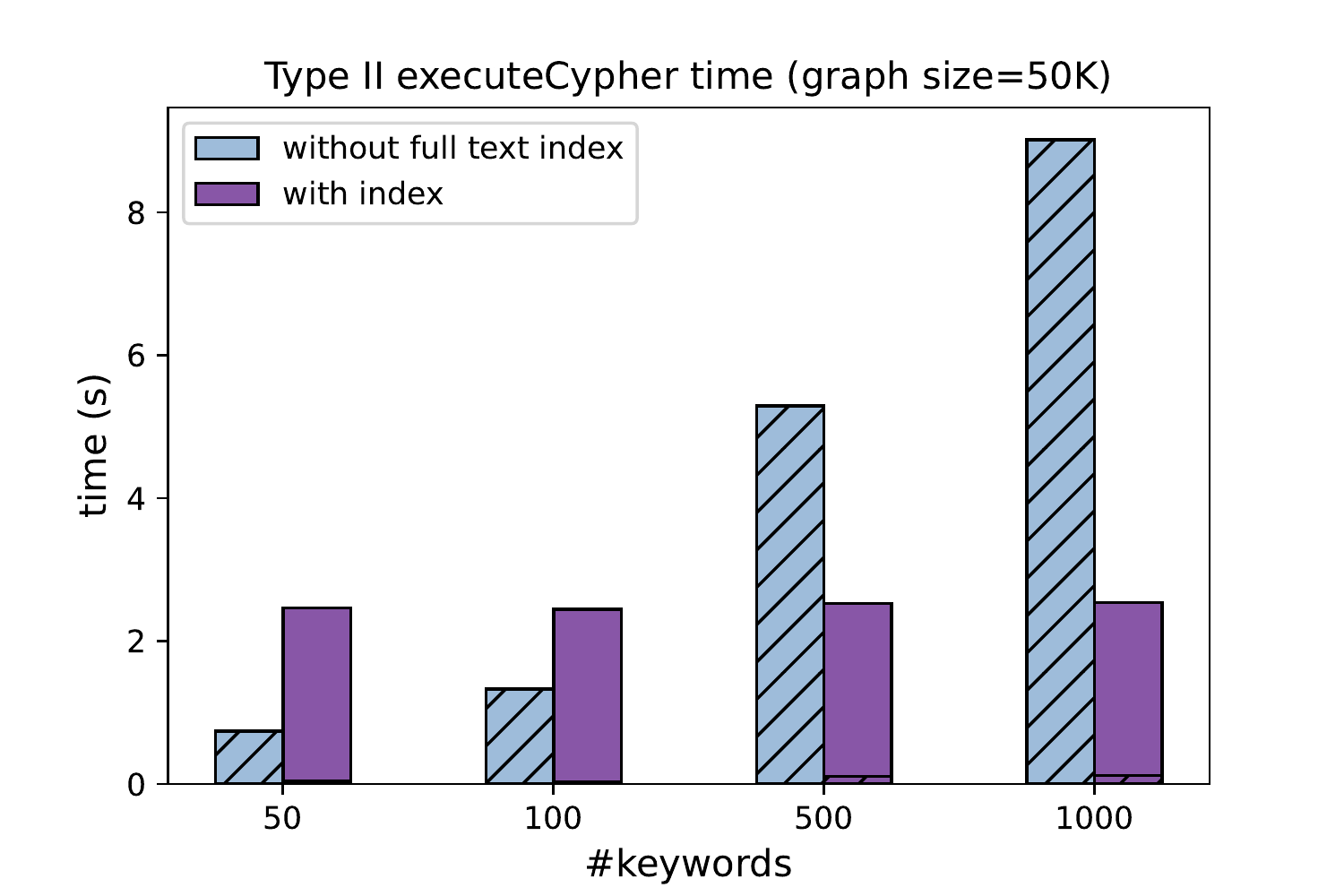}
        \caption{Graph size $=50K$.}
    \end{subfigure}
    \hfill
     \begin{subfigure}[b]{0.23\textwidth}
        \centering
        \includegraphics[width=\textwidth]{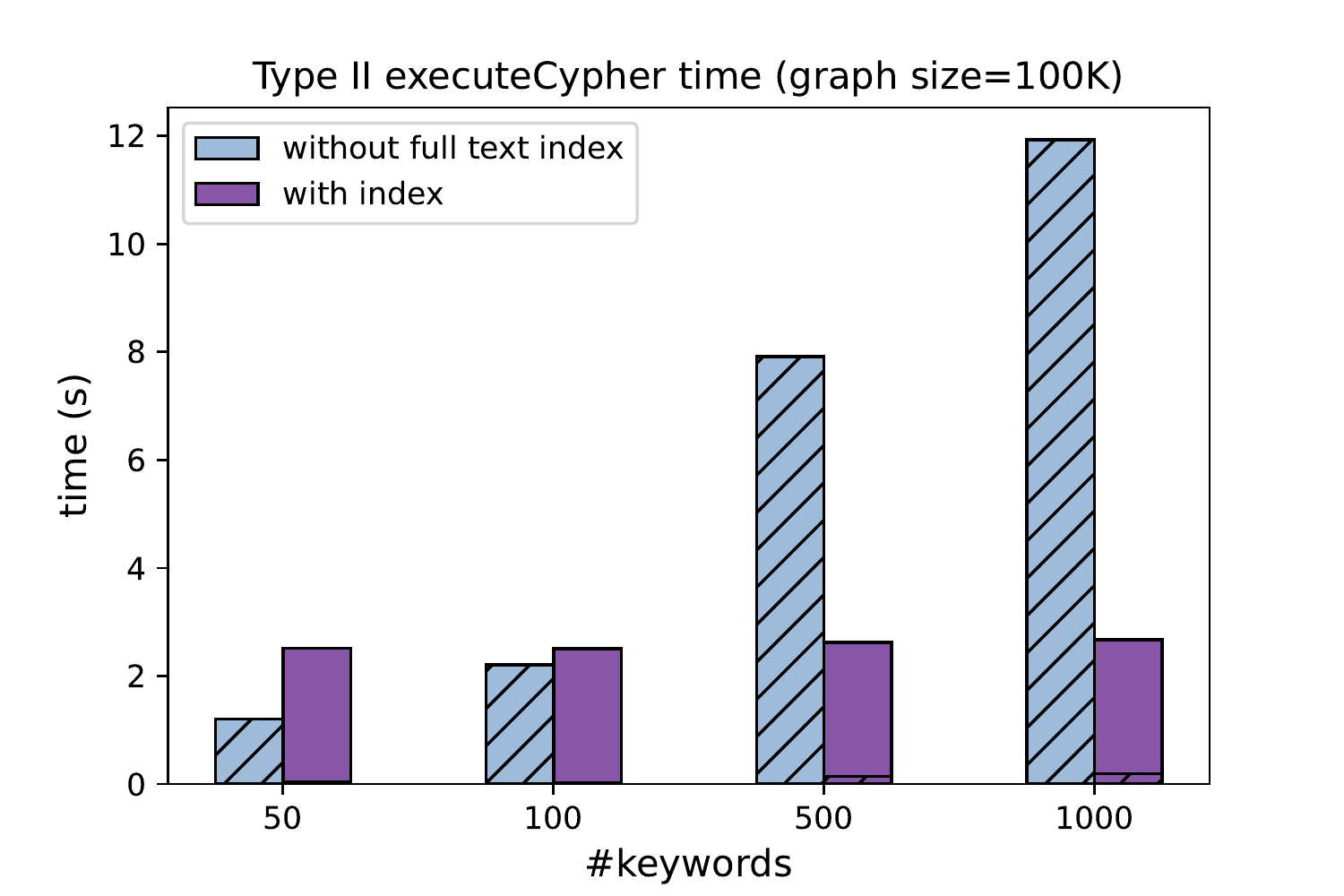}
        \caption{Graph size $=100K$.}
    \end{subfigure}
    \hfill
     \begin{subfigure}[b]{0.23\textwidth}
        \centering
        \includegraphics[width=\textwidth]{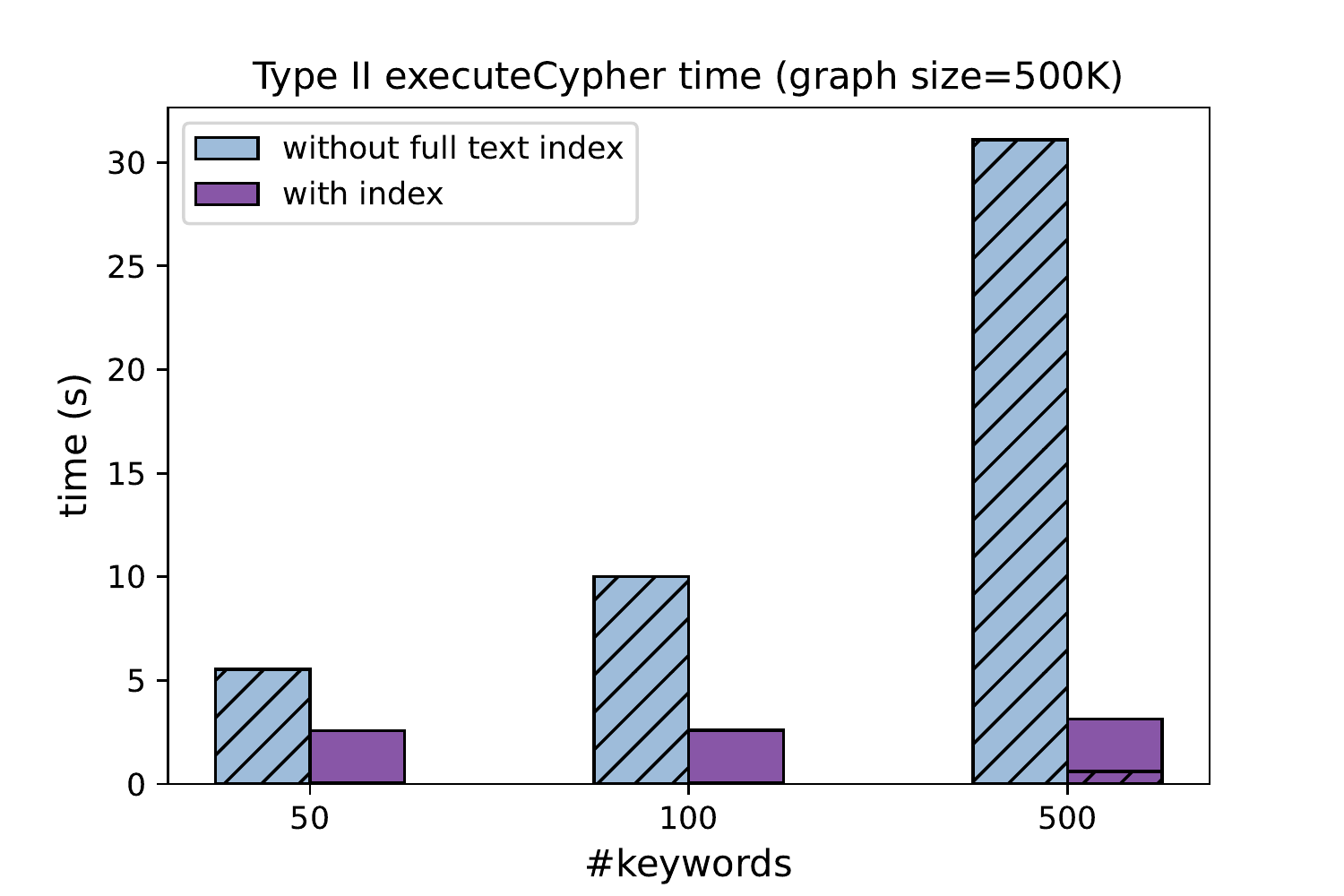}
        \caption{Graph size $=500K$.}
    \end{subfigure}
    \caption{Calibration results for Type II Cypher query w.r.t.~different graph sizes and \#keywords.}
    \label{fig:benchmarktype2}
\end{figure}

\begin{figure}[ht]
    \centering
    \begin{subfigure}[b]{0.23\textwidth}
        \centering
        \includegraphics[width=\textwidth]{figures/tablejoin100.pdf}
        \caption{Row count of R: 100.}
    \end{subfigure}
     \begin{subfigure}[b]{0.23\textwidth}
        \centering
        \includegraphics[width=\textwidth]{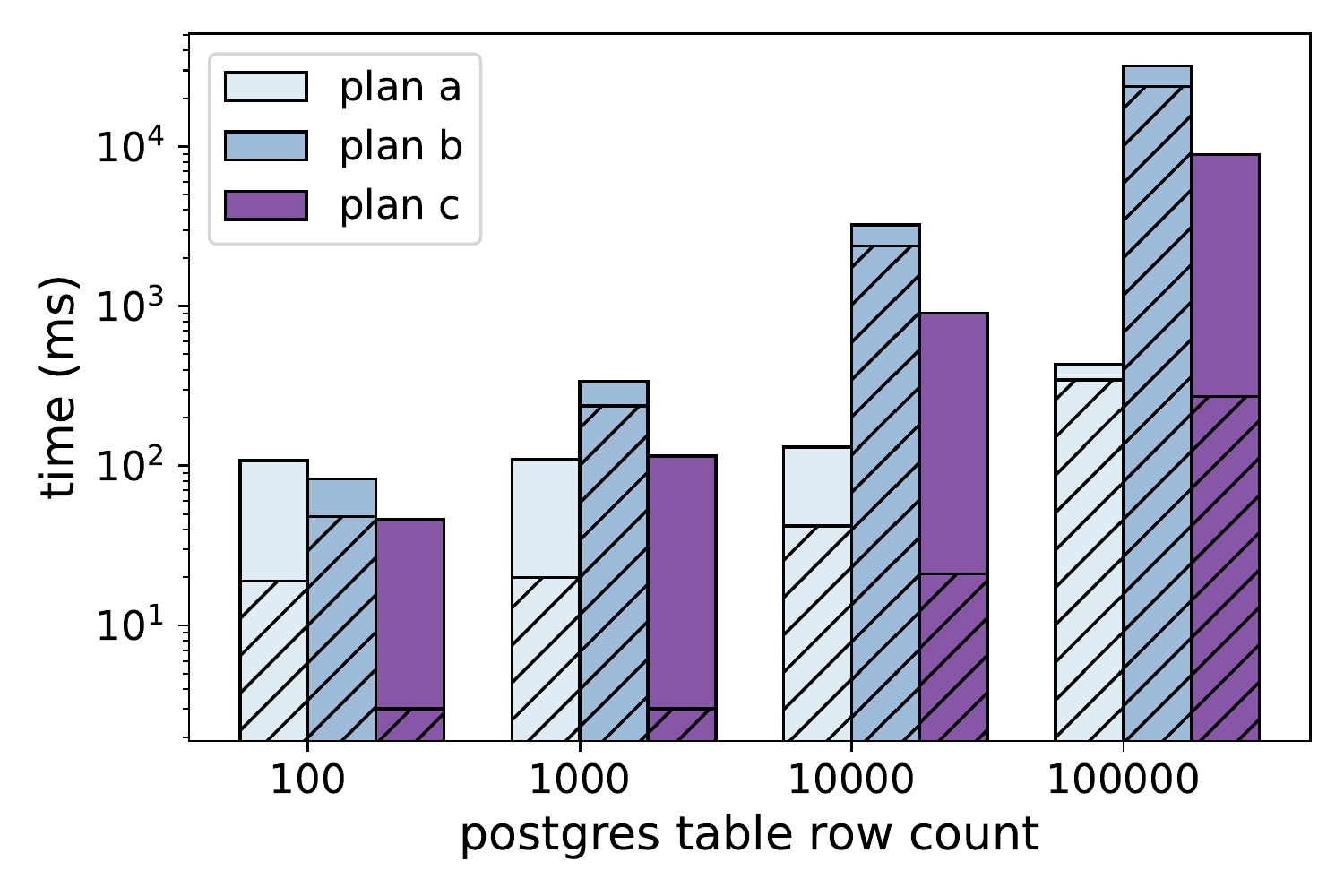}
        \caption{Row count of R: 1K.}
    \end{subfigure}
    \hfill
     \begin{subfigure}[b]{0.23\textwidth}
        \centering
        \includegraphics[width=\textwidth]{figures/tablejoin10k.pdf}
        \caption{Row count of R: 10K.}
    \end{subfigure}
    \hfill
     \begin{subfigure}[b]{0.23\textwidth}
        \centering
        \includegraphics[width=\textwidth]{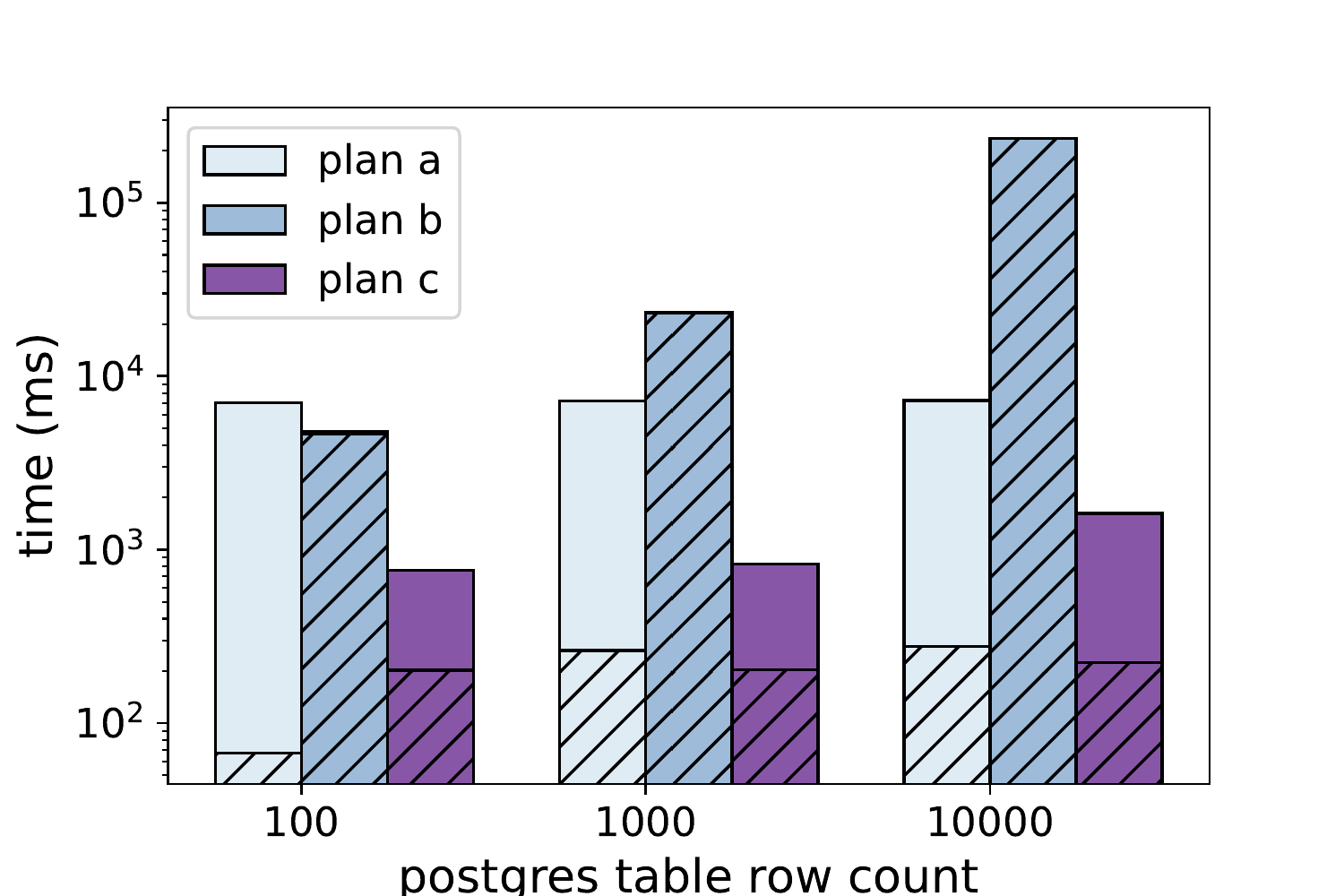}
        \caption{Row count of R: 100K.}
    \end{subfigure}
    \caption{Calibration results for cross engine \textit{executeSQL}}
    \label{fig:benchmark}
\end{figure}

\newpage\clearpage
\onecolumn
\section{AWESOME logical and physical operators.}\label{appendix:ope}
\begin{table*}[h] 
\centering
\caption{AWESOME logical and physical operators.}\label{tab:operators}
\small
\begin{tabular}{lllll} 
\toprule
Types& Logical Operators & Physical Operators  & DataParallelCap  & BufferingCap \\ 
\midrule
Query & \begin{tabular}[c]{@{}l@{}}ExecuteCypher \\ExecuteSQL \\ExecuteSolr\end{tabular} &   \begin{tabular}[c]{@{}l@{}}ExecuteCypherInNeo4j ~\\ExecuteCypherInMeomory ~\\ExecuteSQLInPostgres ~\\ExecuteSQLInSQLite \\ExecuteSolr\\FetchDBMSResults \end{tabular}  & \begin{tabular}[c]{@{}l@{}}ST \\ST \\ST\\ST \\ST \\ST \end{tabular} & \begin{tabular}[c]{@{}l@{}}B \\ B \\B \\B \\ B\\SO \end{tabular} 
\\ 
\midrule
Graph Operations      & \begin{tabular}[c]{@{}l@{}}BuildWordNeighborGraph \\BuildGraphFromRelation \\BuildGraph \\PageRank\\ComputeNodeDegrees\\ ComputeKNeighbors \end{tabular} & 
\begin{tabular}[c]{@{}l@{}}CollectGraphElementsFromDocs ~\\CollectGraphElementsFromRelation ~\\CreateNeo4jGraph ~\\CreateInMemoryGraph ~\\PageRankInNeo4j ~\\PageRankInMemory\\ComputeNodeDegrees\\ ComputeKNeighbors\end{tabular} &  
\begin{tabular}[c]{@{}l@{}} PR \\PR \\ ST \\ ST \\ST\\ ST\\PR \\PR    \end{tabular}
& \begin{tabular}[c]{@{}l@{}}SS \\SS \\SI \\SI\\SO\\SO\\SO\\SO  \end{tabular}
\\ 
\midrule
Relation Operations   & GetColumns                                          & \begin{tabular}[c]{@{}l@{}}GetColumns\\RecordsToList\end{tabular}  &\begin{tabular}[c]{@{}l@{}}ST\\ST\end{tabular}   
&\begin{tabular}[c]{@{}l@{}}SS\\SS\end{tabular}
\\
\midrule
Text Operations       & \begin{tabular}[c]{@{}l@{}}Tokenize\\LDA\\SVD\\TopicModel\\PhraseExtraction\\NER\\\end{tabular}                                  & \begin{tabular}[c]{@{}l@{}}CreatDocumentsFromRecord\\CreatDocumentsFromList\\FilterStopWords\\SplitByPatterns\\CreateTextMatrix\\PhraseExtraction\\NER\\LDA\\SVD\end{tabular}   
& \begin{tabular}[c]{@{}l@{}}PR\\PR\\PR\\PR\\ST\\PR\\PR\\ST\\ST\end{tabular} 
& \begin{tabular}[c]{@{}l@{}}SS\\SS\\SS\\SS\\SI\\SS\\SS\\B\\B\end{tabular} 
\\ 
\midrule
MappedMatrix Operations & \begin{tabular}[c]{@{}l@{}}TopicModel\\ GetValue\\ColumnKeys\\RowKeys\end{tabular}                             & \begin{tabular}[c]{@{}l@{}}LDA\\SVD\\GetValueByIndex\\ GetValueByKeys\\ColumnKeys\\RowKeys\end{tabular} &
\begin{tabular}[c]{@{}l@{}}ST\\ST\\ST\\ST\\ST\\ST\end{tabular} &  
\begin{tabular}[c]{@{}l@{}}B\\B\\B\\B\\B\\B\end{tabular} 
\\ 
\midrule
Other Functions       
& \begin{tabular}[c]{@{}l@{}}Sum\\Range\end{tabular}        & \begin{tabular}[c]{@{}l@{}}SumList\\SumColumn\\SumMatrix\\SumVector\\Range\end{tabular}
& \begin{tabular}[c]{@{}l@{}} PR\\PR\\PR\\PR\\ST \end{tabular}
& \begin{tabular}[c]{@{}l@{}} SI\\SI\\SI\\SI\\SO \end{tabular}
\\ \midrule
Data Movement & Store 
& \begin{tabular}[c]{@{}l@{}}ListToPostgres\\InMemoryRelationToPostgres\\InMemoryGraphToNeo4j\\RecordsToPostgres\\GraphElementsToNeo4j\\ListToCSV\\InMemoryRelationToCSV\\RecordsToCSV\end{tabular} 
& \begin{tabular}[c]{@{}l@{}} ST\\ST\\ST\\ST\\ST\\ST\\ST\\ST \end{tabular}
& \begin{tabular}[c]{@{}l@{}} SI\\B\\B\\SI\\SI\\SI\\B\\SI \end{tabular}
\\
\bottomrule
\end{tabular}
\end{table*}

\end{document}